\documentclass[review]{elsarticle}

\usepackage[colorlinks,citecolor=blue,linktoc=all,linkcolor=cyan]{hyperref}
\usepackage{graphicx}

\usepackage[T1]{fontenc}
\usepackage{dsfont}               % use mathds instead of mathbb for outline fonts
\usepackage{mathrsfs}             % provides mathscr without overwriting mathcal
\usepackage{slashed}              % For Dirac slash notation.
\usepackage{amsmath}
\usepackage{amssymb}
\usepackage{amsbsy}
\usepackage{amsfonts}
\usepackage{bm}

\newcommand{\lea}{\raisebox{-.3ex}{\small $ \
\stackrel{\textstyle <}{\sim} $ }}

\numberwithin{equation}{section}
\numberwithin{table}{section}
\numberwithin{figure}{section}

\journal{Progress in Particle and Nuclear Physics}

\usepackage{titlesec}
\usepackage{sectsty}
\titleformat{\section}{\normalfont\Large\bfseries}{\thesection}{1em}{}
\titleformat{\subsection}{\normalfont\large\bfseries}{\thesubsection}{1em}{}
\titleformat{\subsubsection}{\normalfont\normalsize\bfseries}{\thesubsubsection}{1em}{}
\newcommand{\be}{\begin{eqnarray}}
\newcommand{\ee}{\end{eqnarray}}

\begin{document}
	
	\begin{frontmatter}

		\title{Recent advances in chiral EFT based nuclear forces and their applications}

		%authors, affiliations, corresponding author mention 
	
		\author[mymainaddress]{R. Machleidt}
		%\ead{machleid@uidaho.edu}
		
		\author[mymainaddress]{F. Sammarruca\corref{mycorrespondingauthor} }
		\cortext[mycorrespondingauthor]{Corresponding author}
		\ead{fsammarr@uidaho.edu}
		\address[mymainaddress]{Department of Physics, University of Idaho, Moscow, Idaho 83844, USA}
%		\address[mysecondaryaddress]{Second institution (if applicable)}
		
		\begin{abstract}
		During the past two decades, 
chiral effective field theory has evolved into a powerful tool to derive 
nuclear forces from first principles.
 Nearly all two-nucleon interactions have been worked out up to  sixth order of chiral perturbation theory, while, with few exceptions, three-nucleon forces, which
 play a subtle, but crucial role in microscopic nuclear structure calculations,
  have been derived up to fifth order.
  We review the current status of these forces as well as their applications
  in nuclear many-body systems.
  While the {\it ab initio} description of light nuclei is generally very successful, we
  point out and analyze problems encountered with medium-mass nuclei. 
We also survey the construction of equations of state for symmetric nuclear matter and neutron-rich matter based on chiral forces. A focal point is the symmetry energy and its impact on neutron skins and systems of astrophysical relevance. The physics of neutron-rich systems, from nuclei to compact stars, is essentially determined by the density dependence of the symmetry energy. We review the status of predictions in comparison with latest empirical constraints, with particular attention to those extracted from parity-violating electron scattering.
		\end{abstract}
		
		
\end{frontmatter}

\newpage
\begin{center}
\vspace*{4cm}
\LARGE
{\it In loving memory of Ruprecht Machleidt.} \\
{\it His legacy, both human and scientific, will last for decades to come.}
\end{center}

	\newpage
	
	\thispagestyle{empty}
	\tableofcontents
	
	%to begin the line numbers: 
	%\linenumbers

	%beginning of the core of the manuscript
	\newpage
	
\section{Introduction}

\begin{table}\centering
\caption{Abbreviations and acronyms used in this article.}
\smallskip
%\footnotesize
\scriptsize
\begin{tabular}{ll}
\hline
\hline
   Abbreviation/Acronym   &  Explanation  \\
\hline
ADC(n) & Algebraic diagrammatic construction up to order $n$ \\
AFDMC & Auxiliary-field diffusion Monte Carlo (simulations) \\
AV18 & Argonne $v_{18}$ 2NF~\cite{WSS95}\\
BHF & Brueckner Hartree-Fock (approach) \\
CC & Coupled cluster (method) \\
ChPT & Chiral perturbation theory  \\
CMS & Center-of-mass system  \\
CREX & Calcium radius experiment \\
dof & degrees of freedom \\
EFT & Effective field theory \\
EoS & Equation of state \\
FHNC & Fermi hypernetted chain (method) \\
FRIB   & Facility for Rare Isotope Beams \\
GFMC & Green's function Monte Carlo (method) \\
HF & Hartree-Fock (approximation) \\
HH & Hyperspherical harmonics (method) \\
HI & Heavy ion \\
IANM & Isospin-asymmetric nuclear matter \\
IM-SRG & In-medium similarity renormalization group (method)  \\
$\Lambda_b$ & Breakdown scale \\
LEC & Low-energy constant \\
LENPIC & Low Energy Nuclear Physics International Collaboration \\
LO & Leading order \\
LS & Lippmann-Schwinger \\
MBPT & Many-body perturbation theory \\
NCCI & No-core configuration interaction (method) \\
NCSM & No-core shell model \\
NLO & Next-to-leading order \\
NM & Neutron matter \\
$NN$ & Nucleon-nucleon \\
NNLO, N$^2$LO, N2LO & Next-to-next-to-leading order \\
N$^3$LO, ... & Next-to-next-to-next-to-leading order, ... \\
PREX & Lead (Pb) radius experiment \\
QCD & Quantum chromodynamics \\
QMC & Quantum Monte Carlo (methods)\\
RG & Renormalization group (method) \\
RMF & Relativistic mean field (approach) \\
SCGF & Self-consistent Green's functions (theory) \\
SNM & Symmetric nuclear matter \\
SPP  & Single-particle potential \\
SRG & Similarity renormalization group (method) \\
$T_{\rm lab}$ & Laboratory energy \\
TM & Tucson-Melbourne 3NF~\cite{Coo79,CH01}\\
UV14 & Urbana $v_{14}$ 2NF~\cite{LP81}\\
UV & Urbana V 3NF~\cite{CPW83}\\
UIX & Urbana IX 3NF~\cite{Pud95}\\
VMC & Variational Monte Carlo (method) \\
1PE & One-pion exchange \\
2PE & Two-pion exchange \\
3PE & Three-pion exchange \\
2NF & Two-nucleon force \\
3NF & Three-nucleon force \\
4NF & Four-nucleon force \\
\hline
\hline
\end{tabular}
\label{tab_acro}
\end{table}

%\newpage

The theory of nuclear forces has a long history.
The first serious attempt towards a theory was
launched 
  in 1935, when the Japanese physicist Yukawa~\cite{Yuk35} suggested that nucleons would exchange quanta between each other to create the force. Yukawa constructed his theory in analogy to the theory of the electromagnetic interaction where the exchange of a (massless) photon is the cause of the force. However, in the case of the nuclear force, Yukawa assumed that the mass of the ``force-makers'' was
between the masses of the electron and the proton (which is why these particles were eventually named ``mesons''). The mass of the mesons limits the effect of the force to a finite range, since the uncertainty principal allows massive virtual particles to travel only a finite distance. The meson predicted by Yukawa was finally found in 1947 in cosmic ray and in 1948 in the laboratory and called the pion. Yukawa was awarded the Nobel Prize in 1949. In the 1950's and 60's more mesons were found in accelerator experiments and the meson theory of nuclear forces was extended to include many mesons. These models became known as one-boson-exchange models, which is a reference to the fact that the different mesons are exchanged singly in this model. The one-boson-exchange model is very successful in explaining essentially all properties of the nucleon-nucleon interaction at low energies~\cite{BS69,Erk74,HM75,Mac89,Mac01}. In the 1970's and 80's, meson models were developed that went beyond the simple single-particle exchange mechanism. These models included, in particular, the explicit exchange of two pions with all its complications. Well-known representatives of the latter kind are the Paris~\cite{Lac80} and the Bonn potentials~\cite{MHE87}.

Since these meson models were quantitatively very successful, it appeared that they were the solution of the nuclear force problem. However, with the realization (in the 1970's) that the fundamental theory of strong interactions is quantum chromodynamics (QCD) and not meson theory, all ``meson theories'' had to be viewed as models, and the attempts to derive the nuclear force from first principals had to start all over again.

The problem with a derivation of nuclear forces from QCD is two-fold. First, each nucleon consists of three valence quarks, quark-antiquark pairs, and gluons such that the system of two nucleons is a complicated many-body problem. Second, the force between quarks, which is created by the exchange of gluons, has the feature of being very strong at the low energy-scale that is characteristic of nuclear physics. This extraordinary strength makes it difficult to find converging expansions. Therefore, during the first round of new attempts, QCD-inspired quark models became popular. The positive aspect of these models is that they try to explain nucleon structure (made up from three constituent quarks) and nucleon-nucleon interactions (six quarks) on an equal footing. Some of the gross features of the two-nucleon force, like the ``hard core'' are explained successfully in such models. However, from a critical point of view, it must be noted that these quark-based approaches are yet another set of models and not a theory. Alternatively, one may try to solve the six-quark problem with brute computing power, by putting the six-quark system on a four dimensional lattice of discrete points which represents the three dimensions of space and one dimension of time. This method has become known as lattice QCD and is making progress~\cite{HALQCD,NPLQCD}. However, such calculations are computationally very expensive and cannot be used as a standard nuclear physics tool.

Around 1980/90, a major breakthrough occurred when nobel laureate Steven Weinberg applied the concept of an effective field theory (EFT) to low-energy QCD~\cite{Wei79,Wei90,Wei91}. He simply wrote down the most general theory that is consistent with all the properties of low-energy QCD, since that would make this theory equivalent to low-energy QCD. A particularly important property is the so-called chiral symmetry, which is spontaneously broken. Massless 
spin-$\frac12$ fermions possess the property of 
chirality, which means that their spin and momentum are either parallel (``right-handed'') or anti-parallel (``left-handed'') and remain so forever. Since the quarks, which nucleons are made of (``up'' and ``down'' quarks), are almost 
massless, approximate chiral symmetry is a given. Naively, this symmetry should have the consequence that one finds in nature hadrons of the same mass, but with opposite parities (``parity doublets''). However, this is not the case and such failure is termed a spontaneous breaking of the symmetry. According to a theorem first proven by Goldstone, the spontaneous breaking of a symmetry creates a particle, here, the pion. Thus, the pion becomes the main player in the production of the nuclear force. The interaction of pions with nucleons is weak as compared to the interaction of gluons with quarks. Therefore, pion-nucleon processes can be calculated without problem. Moreover, this effective field theory can be expanded in powers of momentum over ``scale,'' where scale denotes the ``chiral symmetry breaking scale''
or ``breakdown scale.'' This scheme is also known as chiral perturbation theory (ChPT)~\cite{GL84,GL85,GSS88} and allows to calculate the various terms that make up the nuclear potential systematically power by power, or order by order~\cite{ORK96}. Another advantage of the chiral EFT approach is its ability to generate not only the force between two nucleons, but also many-nucleon forces, on the same footing~\cite{Wei92,Kol94}.
In modern theoretical nuclear physics, the chiral EFT approach has gained great popularity and is applied with outstanding success~\cite{ME11,EHM09,HKK19}.

The purpose of this article is twofold: first,  
to review the developments of chiral nuclear interactions up to the highest order
reached so far (which is the sixth power) and, second, to discuss  applications of these forces in the nuclear many-body problem, with emphasis on medium-mass nuclei and neutron-rich systems.

\section{Nuclear forces from chiral EFT: Overview
\label{sec_overview}}

Given an appropriate energy scale, an EFT consists of all interactions consistent with the symmetries that govern
the degrees of freedom relevant at that scale.
For the problem under consideration, 
pertinent degrees of freedom are 
pions (Goldstone bosons), nucleons, and $\Delta(1232)$ isobars.
We start out with just pions and nucleons and will discuss the inclusion of the
$\Delta(1232)$ isobar later.

\subsection{Chiral effective Lagrangians 
\label{sec_Lpi} }

Schematically, we can write the effective Lagrangian as                   
\begin{equation}
{\cal L}
=
{\cal L}_{\pi\pi} 
+
{\cal L}_{\pi N} 
+
{\cal L}_{NN} 
 + \, \ldots \,,
\end{equation}
where ${\cal L}_{\pi\pi}$
deals with the dynamics among pions, 
${\cal L}_{\pi N}$ 
describes the interaction
between pions and a nucleon,
and ${\cal L}_{NN}$  contains two-nucleon contact interactions
which consist of four nucleon-fields (four nucleon legs) and no
meson fields.
The ellipsis stands for terms that involve two nucleons plus
pions and three or more
nucleons with or without pions, relevant for nuclear
many-body forces.
Because pion interactions must      
vanish at zero momentum transfer and in the limit of    
$m_\pi \rightarrow 0$, namely the chiral limit, the                         
Lagrangian is expanded in powers of derivatives
or pion masses:
\begin{eqnarray}
{\cal L}_{\pi\pi} 
 & = &
{\cal L}_{\pi\pi}^{(2)} 
+
{\cal L}_{\pi\pi}^{(4)}
 + \ldots \,, \\
{\cal L}_{\pi N} 
 & = &
{\cal L}_{\pi N}^{(1)} 
+
{\cal L}_{\pi N}^{(2)} 
+
{\cal L}_{\pi N}^{(3)} 
+
{\cal L}_{\pi N}^{(4)} 
+ \ldots , 
\label{eq_LpiN}
\\
\label{eq_LNN}
{\cal L}_{NN} &  = &
{\cal L}^{(0)}_{NN} +
{\cal L}^{(2)}_{NN} +
{\cal L}^{(4)}_{NN} + 
\ldots \,,
\end{eqnarray}
where the superscript refers to the number of derivatives or 
pion mass insertions (chiral dimension)
and the ellipsis stands for terms of higher dimensions.
We use the heavy-baryon formulation of the Lagrangians, originally proposed by Jenkins and Manohar~\cite{JM91} to study heavy quark systems.
The explicit expressions we use can be found in Refs.~\cite{ME11,KGE12}.

\subsection{Chiral perturbation theory and power counting
\label{sec_chpt}}

From these Lagrangians, an infinite number of Feynman diagrams can be generated, which seems to make the theory unmanageable. The way out of this dilemma is
to design a scheme that makes it possible to organize the
diagrams according to their importance. 
Chiral perturbation theory (ChPT) provides such scheme. 

Nuclear potentials are defined by the irreducible types of these
graphs.
By definition, an irreducible graph is a diagram that
cannot be separated into two
by cutting only nucleon lines.
These graphs are then analyzed in terms of powers of 
$Q$ with $Q=p/\Lambda_b$, 
where $p$ is generic for a momentum (nucleon three-momentum
or pion four-momentum) or a pion mass and $\Lambda_b \sim m_\rho \sim$ 0.7 GeV is the 
breakdown scale~\cite{Fur15}. Determining the power $\nu$ has become know
as power counting.

Following the Feynman rules of covariant perturbation theory,
a nucleon propagator is $p^{-1}$,
a pion propagator $p^{-2}$,
each derivative in any interaction is $p$,
and each four-momentum integration $p^4$.
This is also known as naive dimensional analysis or Weinberg counting.

Since we use the heavy-baryon formalism, we encounter terms which include factors of
$p/M_N$, where $M_N$ denotes the nucleon mass.
We count the order of such terms by the rule
$p/M_N \sim (p/\Lambda_b)^2$,
for reasons explained in Refs.~\cite{Wei91,ORK96,Epe06}.

Applying some topological identities, one obtains
for the power of a connected irreducible diagram
involving $A$ nucleons~\cite{ME11,Wei90},
\begin{equation} \nu = -2 +2A - 2C + 2L 
+ \sum_i \Delta_i \, ,
\label{eq_nu} 
\end{equation}
with the `index of the interaction' defined by 
\begin{equation}
\Delta_i  \equiv   d_i + \frac{n_i}{2} - 2  \, .
\label{eq_Deltai}
\end{equation}
In the above equations: 
$C$ represents the number of individually connected parts of the diagram while
$L$ is the number of loops;               
moreover, for each vertex $i$,    
$d_i$ indicates how many derivatives or pion masses are present 
and $n_i$ is the number of nucleon fields.                  
The summation extends over all vertices present in that particular diagram.
Notice also that chiral symmetry implies $\Delta_i \geq 0$. 
Interactions among pions have at least two derivatives
($d_i\geq 2, n_i=0$), while 
interactions between pions and a nucleon have one or more 
derivatives  
($d_i\geq 1, n_i=2$). Finally, pure contact interactions
among nucleons ($n_i=4$)
have $d_i\geq0$.
In this way, a low-momentum expansion based on chiral symmetry 
can be constructed.                   

Naturally,                                            
the powers must be bounded from below for the expansion
to converge. This is in fact the case, 
with $\nu \geq 0$.

To further illustrate the power formula Eq.~(\ref{eq_nu}), let us apply it to diagrams shown in 
Fig.~\ref{fig_hi}. From the LO row, we pick the one-pion exchange diagram (second diagram).
Each vertex in this disgram is a small dot which consists of one derivative ($d_i = 1$)
and two nucleon legs/fields ($n_i=2$); thus, for the small-dot vertices, we have
\begin{equation}
\Delta_i = 1+\frac{2}{2} -2 = 0 \,.
\end{equation}
Moreover, $A=2$, $C=1$, and $L=0$, and so Eq.~(\ref{eq_nu}) results into
\begin{equation}
\nu = -2 + 2\times 2 -2\times 1 + 2\times 0 + \sum_i 0 = 0 \, ,
\end{equation}
as it should for LO. Moving on to NLO, let us pick one triangular diagram from the second row
of Fig.~\ref{fig_hi}. All three small-dot vertices in this diagram have $\Delta_i = 0$.
Furthermore, $A=2$, $C=1$, and $L=1$ (one loop).
Hence,
\begin{equation}
\nu = -2 + 2\times 2 -2\times 1 + 2\times 2 + \sum_i 0 = 2 \, ,
\end{equation}
as required for NLO.
Alternatively, one can also calculate the power of a diagram `from scratch', i.~e., without the help of Eq.~(\ref{eq_nu}). For the NLO triangular diagram that we just discussed, this goes like this:
Each vertex contains one derivative, each meson propagator is (-2), the nucleon propagator is (-1),
and the loop integration is 4; thus,
\begin{equation}
\nu = 1+1+1-2-2-1+4=2 \,,
\end{equation}
in agreement with what we obtained from Eq.~(\ref{eq_nu}).

By the way, the power formula 
Eq.~(\ref{eq_nu}) also
allows to predict
the leading orders of connected multi-nucleon forces.
Consider a $m$-nucleon irreducibly connected diagram
($m$-nucleon force) in an $A$-nucleon system ($m\leq A$).
The number of separately connected pieces is
$C=A-m+1$. Inserting this into
Eq.~(\ref{eq_nu}) together with $L=0$ and 
$\sum_i \Delta_i=0$ yields
$\nu=2m-4$. Thus, two-nucleon forces ($m=2$) appear
at $\nu=0$, three-nucleon forces ($m=3$) at
$\nu=2$ (but they happen to cancel at that order),
and four-nucleon forces at $\nu=4$ (they don't cancel).
More about this in the next sub-section.

For later purposes, we note that for an irreducible 
$NN$ diagram ($A=2$, $C=1$), the
power formula collapses to the very simple expression
\begin{equation}
\nu =  2L + \sum_i \Delta_i \,.
\label{eq_nunn}
\end{equation}

To summarize, at each order                            
$\nu$ we only have a well defined, finite number of diagrams, 
which renders the theory feasible from a practical standpoint.
The magnitude of what has been left out at order $\nu$ can be estimated (in a 
simple way) from 
$Q^{\nu+1}$ (see Sec.~\ref{sec_uncert}, below). The ability to calculate observables (in 
principle) to any degree of accuracy gives the theory 
its predictive power.

\subsection{The ranking of nuclear forces}

\begin{figure}[t]\centering
%\vspace*{-0.5cm}
\scalebox{0.70}{\includegraphics{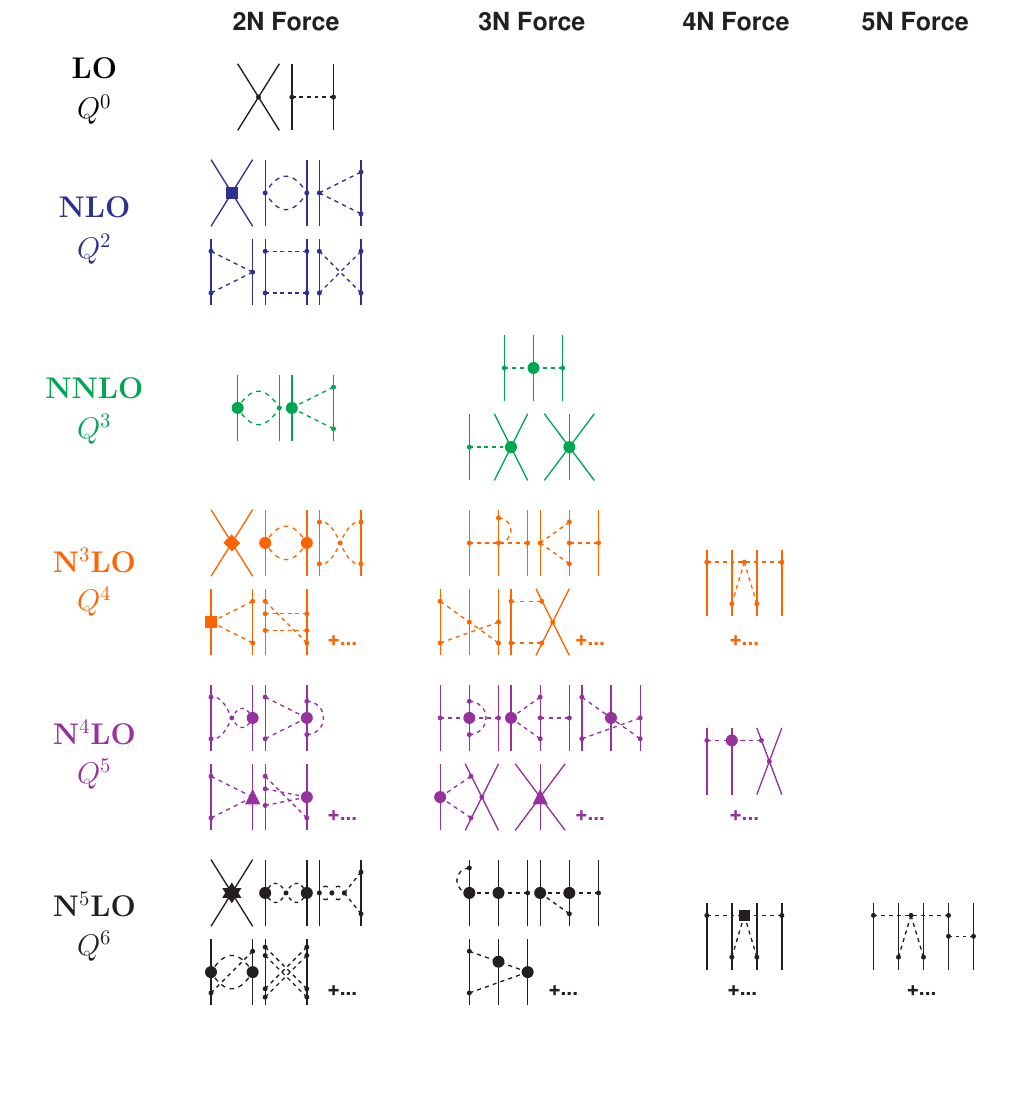}}
\vspace*{-1.0cm}
\caption{Hierarchy of nuclear forces in ChPT. Solid lines
represent nucleons and dashed lines pions. 
Small dots, large solid dots, solid squares, triangles, diamonds, and stars
denote vertices of index $\Delta_i = \, $ 0, 1, 2, 3, 4, and 6, respectively. 
Further explanations are
given in the text.}
\label{fig_hi}
\end{figure}

  \begin{table}[t]\centering
\caption{Some basic constants relevant to work reviewed in this article~\cite{PDG}}
\label{tab_basic}
\smallskip
%\begin{tabular*}{\textwidth}{@{\extracolsep{\fill}}cc}
\begin{tabular}{lcl}
\hline 
\hline 
\noalign{\smallskip}
  Quantity            &  \hspace{1cm} & Value \\
\hline
\noalign{\smallskip}
Charged-pion mass $m_{\pi^\pm}$ && 139.5704 MeV \\
Neutral-pion mass $m_{\pi^0}$ && 134.9768 MeV \\
Average pion-mass $\bar{m}_\pi$ && 138.0392 MeV \\
Proton mass $M_p$ && 938.2721 MeV \\
Neutron mass $M_n$ && 939.5654 MeV \\
Average nucleon-mass $\bar{M}_N$ && 938.9183 MeV \\
$\Delta$-isobar mass $M_{\Delta}$ && 1232 MeV \\
 $\Delta M \equiv  M_\Delta - \bar{M}_N$ && 293.0817 MeV \\
Nucleon axial coupling constant  $g_A$ && 1.29 \\
$\pi N \Delta$ axial coupling constant  $h_A$ && 1.40 \\
Pion-decay constant $f_\pi$ && 92.2 MeV \\
Conversion constant $\hbar c$ && 197.32698 MeV fm\\
\hline
\hline
\noalign{\smallskip}
\end{tabular}
\end{table}

As shown in Fig.~\ref{fig_hi}, nuclear forces appear in ranked
orders in accordance with 
the power counting scheme. 

The lowest power is $\nu = 0$, also known as the leading order (LO).
At LO we have only two contact contributions with no momentum dependence 
($\sim Q^0$). They are signified by the 
four-nucleon-leg diagram 
with a small-dot vertex shown in the first row of 
Fig.~\ref{fig_hi}.
Besides this, we have the
static one-pion exchange (1PE), also shown 
in the first row of                 
Fig.~\ref{fig_hi}.              
Its charge-independent version is given in momentum space by
\begin{equation}
V_{1\pi} ({\vec p}~', \vec p) = -
%V_{1\pi} (\vec q) = - 
\frac{g_A^2}{4f_\pi^2}
\: 
\frac{\vec \sigma_1 \cdot \vec q \,\, \vec \sigma_2 \cdot \vec q  }{q^2 + m_\pi^2} \:
\bm{\tau}_1 \cdot \bm{\tau}_2 \,,
\label{eq_1PEq}
\end{equation}
where ${\vec p}\,'$ and $\vec p$ denote the final and initial nucleon momenta in the 
center-of-mass system, 
respectively. Moreover, $\vec q = {\vec p}\,' - \vec p \, $ 
($q \equiv |\vec q |$)
is the momentum transfer, 
 and $\vec \sigma_{1,2}$ and $\bm{\tau}_{1,2}$ are the spin 
and isospin operators of nucleon 1 and 2, respectively. Parameters
$g_A$, $f_\pi$, and $m_\pi$ denote the axial-vector coupling constant,
pion-decay constant, and the pion mass, respectively; see Table~\ref{tab_basic}
for their values.  

Fourier transform yields the corresponding position-space version of 1PE:
\begin{equation}
\widetilde V_{1\pi}(\vec r) =\frac{g_A^2 \, m_\pi^2}{48 \pi f_\pi^2} \; 
\frac{e^{-x}}{r} 
\left[
\vec\sigma_1 \cdot \vec \sigma_2
+
 \left(1+\frac{3}{x}+\frac{3}{x^2} \right) \, S_{12}(\hat r)
 \right]
 \bm{\tau}_1 \cdot \bm{\tau}_2 \,
 \label{eq_1PEr}
\end{equation}
with $\vec r$ the relative distance between the two nucleons and
$x = m_\pi \, r$ ($r \equiv |\vec r|$).
\\
Moreover,
\begin{equation}
S_{12}(\hat r) = 3 \vec \sigma_1 \cdot \hat r \,\,\: \vec \sigma_2 \cdot \hat r - \vec \sigma_1 \cdot  \vec \sigma_2 
\end{equation}
denotes the standard position-space spin-tensor operator 
with $\hat r = \vec r/r $.
In Eq.~(\ref{eq_1PEr}), a $\delta$-function term has been omitted, since it can be absorbed into the LO contact terms.

In spite of its simplicity, this rough description                      
contains some of the main attributes of the $NN$ force. 
Through the 1PE it generates the tensor component of the force
known to be crucial for the two-nucleon bound state (deuteron).
It also predicts correctly 
$NN$ phase parameters for high partial waves.               
The two terms which result from a partial-wave expansion of the LO contact terms
impact states of zero orbital angular momentum ($S$-waves) and produce attraction at 
short- and intermediate-range.                              

Notice that there are no terms with power 
$\nu=1$, as they would violate parity conservation 
and time-reversal invariance.

The next order is then
$\nu=2$, next-to-leading order, or NLO. The two-pion exchange (2PE) makes its first appearance at this order,
and thus it is referred to as the 
``leading 2PE''. As is well known from decades of nuclear physics, 
this contribution is essential for a realistic account of the intermediate-range attraction.    
However, the leading 2PE has insufficient strength, for the following reason: 
the loops present in the diagrams which involve pions 
carry the power $\nu=2$ [cf.\ Eq.~(\ref{eq_nunn})],
and so only                                  
$\pi NN$ and $\pi \pi NN$ vertices with $\Delta_i = 0$ are allowed at this order. 
These vertices are known to be weak.
Moreover, seven new contacts appear at this order which 
impact $L = 0$ and $L = 1$ states. (As always, two-nucleon contact terms are indicated 
by four-nucleon-leg diagrams and a vertex of appropriate shape, in this case a solid square.) 
At this power, the appropriate operators                                  
include  central,
spin-spin, spin-orbit, and tensor terms, namely all the spin
operator structures needed for a realistic description of the 
two-nucleon force (2NF), although the medium-range attraction still lacks 
sufficient strength.

\begin{figure}[t]\centering
%\vspace*{-0.5cm}
\scalebox{0.7}{\includegraphics{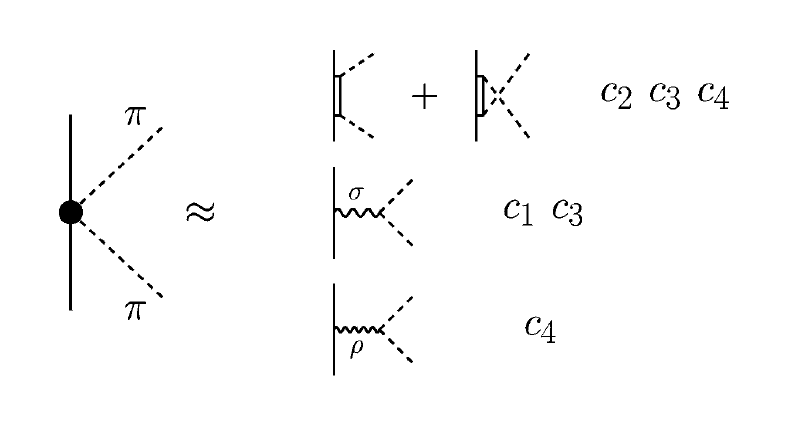}}
\vspace*{-0.5cm}
\caption{Interpretation of the second-order seagull graph (large solid dot)
in terms of resonance exchanges. The $c_i$ refer to the LECs used 
in the ${\cal L}_{\pi N}^{(2)}$ Lagrangian.}
\label{fig_ressat}
\end{figure}

At the next order, 
$\nu=3$ or next-to-next-to-leading order (NNLO), 
the 2PE contains the subleading         
$\pi\pi NN$ seagull vertices with two derivatives~\cite{KBW97}.                   
These vertices,
denoted by a large solid dot
in Fig.~\ref{fig_hi}, 
simulate correlated 2PE
and intermediate $\Delta(1232)$-isobar contributions (cf.\ Fig.~\ref{fig_ressat}).
Consistent with what the meson theory of                               
the nuclear forces~\cite{Lac80,MHE87} has shown since a long time 
concerning the importance of these effects,            
at this order the 2PE finally provides medium-range     
attraction of realistic strength, bringing the description of the $NN$ force
to an almost quantitative level. 
No new contacts become available at NNLO. 

An important advantage of ChPT is that it generates                
two- and many-nucleon forces 
on an equal footing.
Thus, three-nucleon forces (3NFs) appear for the first time at NLO,                          
but their net contribution vanishes at this order~\cite{Wei92}.
The first non-zero 3NF contribution is found 
at NNLO~\cite{Kol94,Epe02}. It is therefore easy to understand why  
3NF are very weak as compared to the 2NF which contributes already at 
$Q^0$.

For $\nu =4$, or next-to-next-to-next-to-leading
order (N$^3$LO), we display some representative diagrams in 
Fig.~\ref{fig_hi}. There is a large attractive one-loop 2PE contribution (the bubble diagram with two large solid dots), which slightly over-estimates the 2NF attraction
at medium range. 
Two-pion-exchange graphs with two loops are seen at this order, together with 
three-pion exchange (3PE), which was determined to be very weak 
at N$^3$LO~\cite{Kai00a,Kai00b}.
The most important feature at this order is the presence 
of 15 additional contacts $\sim Q^4$, signified 
by the four-nucleon-leg diagram in the figure with the diamond-shaped vertex. 
These contacts impact states with orbital angular momentum up to $L = 2$, 
and are the reason for the                            
 quantitative description of the
two-nucleon force (up to approximately 300 MeV
in terms of laboratory energy) 
at this order~\cite{ME11,EM03}.
More 3NF diagrams show up 
at N$^3$LO~\cite{Ber08,Ber11}, as well as the first contributions to 
four-nucleon forces (4NF)~\cite{Epe07}.
We then see that forces involving more and more nucleons appear for the
first time at higher and higher orders, which 
gives theoretical support to the fact that          
2NF $\gg$ 3NF $\gg$ 4NF
\ldots.

Further 2PE and 3PE occur at N$^4$LO (fifth order). The contribution to the 2NF 
at this order has been first calculated  by Entem {\it et al.}~\cite{Ent15a}. It turns out to be moderately repulsive, thus
compensating for the attractive surplus generated at N$^3$LO by the bubble diagram with two solid dots. The long- and intermediate-range 3NF contributions at this order have been evaluated~\cite{KGE12,KGE13}, but not yet applied in nuclear structure calculations. From the analyses in Refs.~\cite{KGE12,KGE13}, one may expect these contributions to be of some importance, as the subleading contributions to the 2P1PE and ring topologies can be understood in terms of $\Delta(1232)$ isobar excitations. Moreover, a new set of 3NF contact terms appears~\cite{GKV11}
that had a successful application in the context of the so-called `$A_y$ puzzle'
of nucleon-deuteron scattering~\cite{Gir19}.
The N$^4$LO 4NF has not been derived yet. Because it contains the subleading $\pi\pi N N$ seagull vertex (large solid dot), interpreted in terms of resonance exchanges, see Fig.~\ref{fig_ressat}, we speculate that this contribution may be non-negligible.

Finally turning to N$^5$LO (sixth order): The dominant 2PE and 3PE contributions to the 2NF have been derived by Entem {\it et al.} in Ref.~\cite{Ent15b}, which represents
the most advanced investigation conducted in chiral EFT for the $NN$ system. The effects are small indicating the desired trend towards convergence of the chiral expansion for the 2NF. 
Moreover, a new set of 26 $NN$ contact terms $\sim Q^6$ occurs that contributes up to $F$-waves (represented by the $NN$ diagram with a star in Fig.~\ref{fig_hi})
bringing the total number of $NN$ contacts to 50~\cite{EM03a}.
The three-, four-, and five-nucleon forces of this order have not yet been evaluated.

\begin{figure}[t]\centering
\vspace*{-1.5cm}
%\hspace*{-1.7cm}
\scalebox{0.5}{\includegraphics{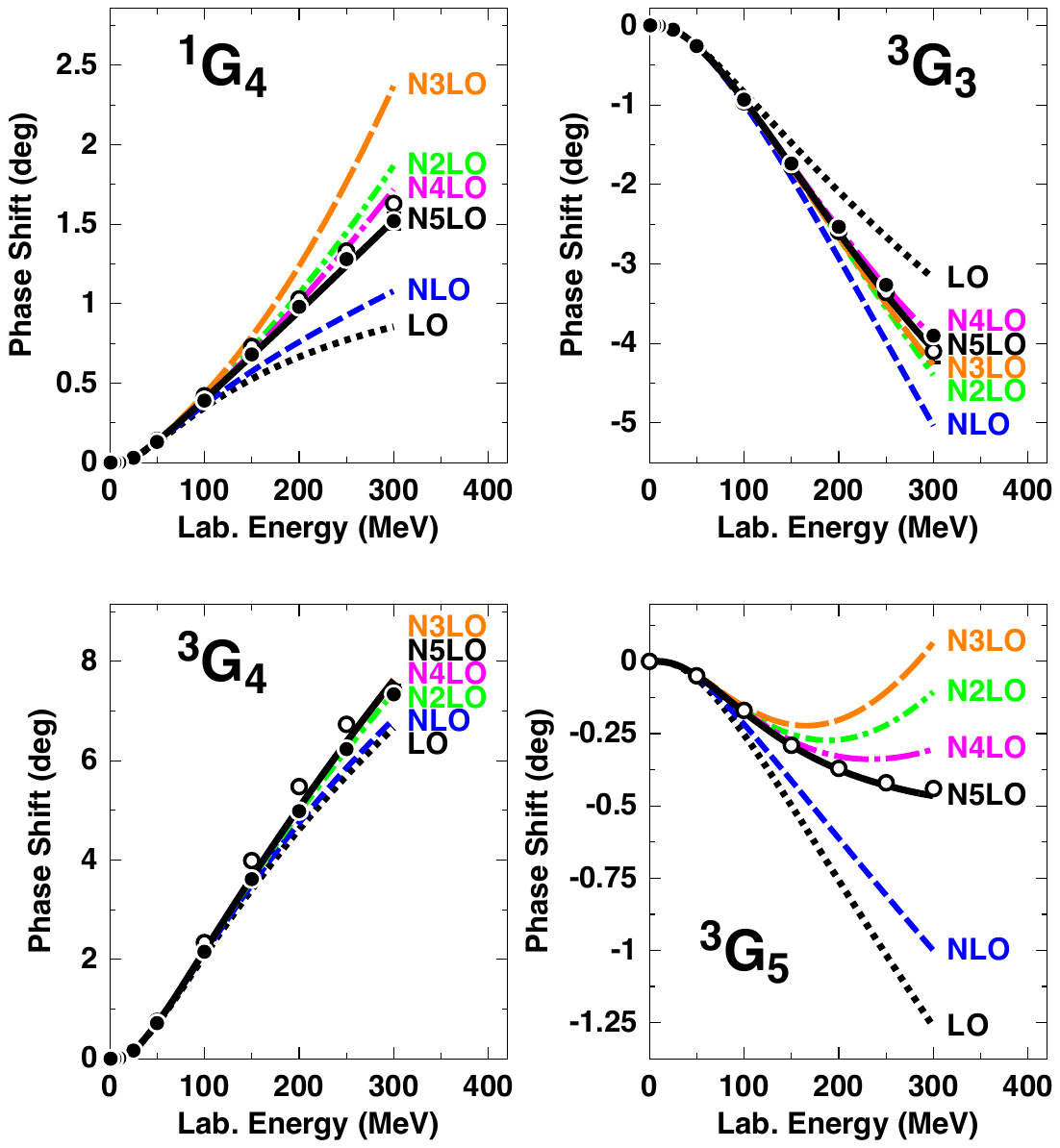}}
\vspace*{-2.0cm}
\caption{Phase-shifts of neutron-proton scattering in $G$ waves at all orders of ChPT
from LO to N$^5$LO.  
The filled and open circles represent the results from the Nijmegen multi-energy $np$ phase-shift analysis~\cite{Sto93} and the GWU single-energy $np$ analysis SP07~\cite{SP07}, respectively.}
\label{fig_ph6a}
\end{figure}

To summarize, we show in Fig.~\ref{fig_ph6a} the contributions to the phase shifts
of peripheral $NN$ scattering through 
 all orders from LO to N$^5$LO
 as obtained from a perturbative calculation.
   Note that the difference between the LO prediction 
(one-pion-exchange, dotted line) and the data (filled and open circles) is to 
be provided by two- and three-pion exchanges, i.e. the intermediate-range part
of the nuclear force. How well that is accomplished is a crucial test for
any theory of nuclear forces.  NLO produces only a small contribution, 
but N$^2$LO creates substantial intermediate-range attraction (most clearly 
seen in $^1G_4$, $^3G_5$). In fact, N$^2$LO is the largest 
contribution among all orders. This is due to the one-loop $2\pi$-exchange 
 triangle diagram which involves one $\pi\pi NN$-contact vertex (large solid dot). 
 As discussed,
the one-loop $2\pi$-exchange at N$^2$LO is attractive and 
 describes the intermediate-range attraction of the
nuclear force about right. At N$^3$LO, more one-loop 2PE is added by the 
bubble diagram with two large solid dots, a contribution that seemingly is 
overestimating the attraction. This attractive surplus is then compensated by
the prevailingly repulsive two-loop $2\pi$- and $3\pi$-exchanges that occur 
at N$^4$LO and N$^5$LO. 

In this context, it is worth noting that also in conventional meson 
theory~\cite{MHE87} the one-loop models for the 2PE contribution always show 
some excess of attraction (cf.  Fig.~2 of Ref.~\cite{EM02a} and Fig.~10 of Ref.~\cite{ME11}). 
The same is 
true for the dispersion theoretic approach pursued by the Paris 
group (see, e.~g., the predictions for $^1D_2$, $^3D_2$, and $^3D_3$
in Fig.~8 of Ref.~\cite{Vin79} which are all too attractive). 
In conventional meson theory, this
attraction is reduced by heavy-meson exchanges ($\rho$-, $\omega$-, and $\pi\rho$-exchange) 
which, however, have no place in chiral effective field theory (as a 
finite-range contribution). Instead, in the latter approach, two-loop 
$2\pi$- and $3\pi$-exchanges provide the corrective action.

\subsection{Adding the $\Delta$ degree of freedom \label{sec_delta}}

\begin{table}
\caption{The $\pi N$ LECs as determined in
the Roy-Steiner-equation analysis of $\pi N$ scattering conducted in  Refs.~\cite{Hof15,Hof16,Sie17}.
The given orders of the chiral expansion refer to the $NN$ system. 
The $c_i$, $\bar{d}_i$, 
and $\bar{e}_i$ are the LECs of the second, third, and fourth order $\pi N$ Lagrangian~\cite{KGE12}, Eq.~(\ref{eq_LpiN}), and are 
 in units of GeV$^{-1}$, GeV$^{-2}$, and GeV$^{-3}$, respectively.
The uncertainties in the last digits are given in parentheses after the values.}
\label{tab_lecs}
\smallskip
\begin{tabular*}{\textwidth}{@{\extracolsep{\fill}}crrrrrr}
\hline 
\hline 
\noalign{\smallskip}
  & \multicolumn{2}{c}{NNLO}
   & \multicolumn{2}{c}{N$^3$LO}
    & \multicolumn{2}{c}{N$^4$LO} \\
    \cline{2-3}       \cline{4-5}            \cline{6-7}
    & $\Delta$-less &   $\Delta$-full 
      & $\Delta$-less &   $\Delta$-full 
         & $\Delta$-less &   $\Delta$-full    \\
\hline
\noalign{\smallskip}
$c_1$ & --0.74(2) & --0.74(2) & --1.07(2) & --1.25(3)  & --1.10(3) & --1.11(3) \\
$c_2$ & 1.81(3) & --0.49(17)  & 3.20(3) & 1.37(16)  & 3.57(4) & 1.52(21) \\
$c_3$ & --3.61(5) & --0.65(22) & --5.32(5) & --2.41(23) & --5.54(6) & --1.99(30) \\
$c_4$ & 2.44(3) & 0.96(11) & 3.56(3) & 1.66(14) & 4.17(4) & 1.88(19) \\
$\bar{d}_1 + \bar{d}_2$ & --- & --- & 1.04(6) & 0.11(11) & 6.18(8) & 1.75(42)  \\
$\bar{d}_3$ & --- & --- & --0.48(2) & --0.81(3) & --8.91(9) & --3.61(48) \\
$\bar{d}_5$ & --- & --- & 0.14(5) & 0.80(7) & 0.86(5) & 1.52(7) \\
$\bar{d}_{14} - \bar{d}_{15}$ & --- & --- & --1.90(6) & --1.04(12) & --12.18(12) & --4.32(79) \\
$\bar{e}_{14}$ & --- & --- & --- & --- & 1.18(4) & 1.67(6) \\
$\bar{e}_{17}$ & --- & --- & --- & --- & --0.18(6) & --0.44(6) \\
\hline
\hline
\noalign{\smallskip}
\end{tabular*}
\end{table}

The lowest excited state of the nucleon is the
$\Delta(1232)$ resonance or isobar 
(a $\pi$-$N$ $P$-wave resonance with both spin and isospin 3/2)
with an excitation energy of $\Delta M=M_\Delta - \bar{M}_N = 293$ MeV.
Because of its strong coupling to the $\pi$-$N$ system and low excitation energy,
it is an important ingredient for models of pion-nucleon scattering in the $\Delta$-region
and pion production from the
two-nucleon system at intermediate energies, where the particle production
proceeds prevailingly through the formation of $\Delta$ isobars~\cite{Mac89}.
At low energies, the more sophisticated conventional models for the 2$\pi$-exchange contribution to the
$NN$ interaction include the virtual excitation of $\Delta$'s, which in these models accounts
for about 50\% of the intermediate-range attraction of the nuclear force---as demonstrated
by the Bonn potential~\cite{MHE87,HM77}. 

Because of its relatively small excitation energy, it is not clear from the outset if, 
in an EFT, the $\Delta$ should be taken into account explicitly or integrated out as
a ``heavy'' degree of freedom. If it is included, then $\Delta M \sim m_\pi$ is considered
as another small expansion parameter, besides the pion mass and small external momenta.
This scheme has become known as the small scale expansion (SSE)~\cite{HHK98}.
Note, however, that this extension is of phenomenological character, since
$\Delta M$ does not vanish in the chiral limit.

In the chiral EFT discussed so far in this article 
(also known as the ``$\Delta$-less'' theory),
the effects due to $\Delta$ isobars
are taken into account implicitly. Note that the dimension-two $\pi N$ LECs, the $c_i$,
have unnaturally large values (cf.\ Table~\ref{tab_lecs}). The reason for this is that
the $\Delta$-isobar (and some meson resonances) contribute considerably
to the $c_i$---a mechanism that has become known as
resonance saturation~\cite{BKM97}, Fig.~\ref{fig_ressat}.
Therefore, the explicit inclusion of the $\Delta$ (``$\Delta$-full'' theory) will take strength out of these
LECs and move this strength to a lower order~\cite{KGW98,KEM07,EKM08,KGE18}. 
As a consequence, the convergence of the expansion improves, which
is another motivation for introducing explicit $\Delta$-degrees of freedom.
We observed that, in the $\Delta$-less theory, the subleading 2PE and 3PE contributions
to the 2NF are larger than the leading ones. The promotion of large contributions
by one order in the $\Delta$-full theory fixes this problem. 

The LECs of the $\pi N$ Lagrangian are usually extracted in the analysis of $\pi$-$N$
scattering data and clearly come out differently in the $\Delta$-full theory as compared
to the $\Delta$-less one. While in the $\Delta$-less theory, the magnitude of the LECs
$c_3$ and $c_4$ is about 3-5 GeV$^{-1}$ (cf.\ Table~\ref{tab_lecs}),
they turn out to be around 1 GeV$^{-1}$ in the $\Delta$-full theory~\cite{Sie17}.

\begin{figure}[t]\centering
\vspace*{-0.5cm}
\scalebox{0.55}{\includegraphics{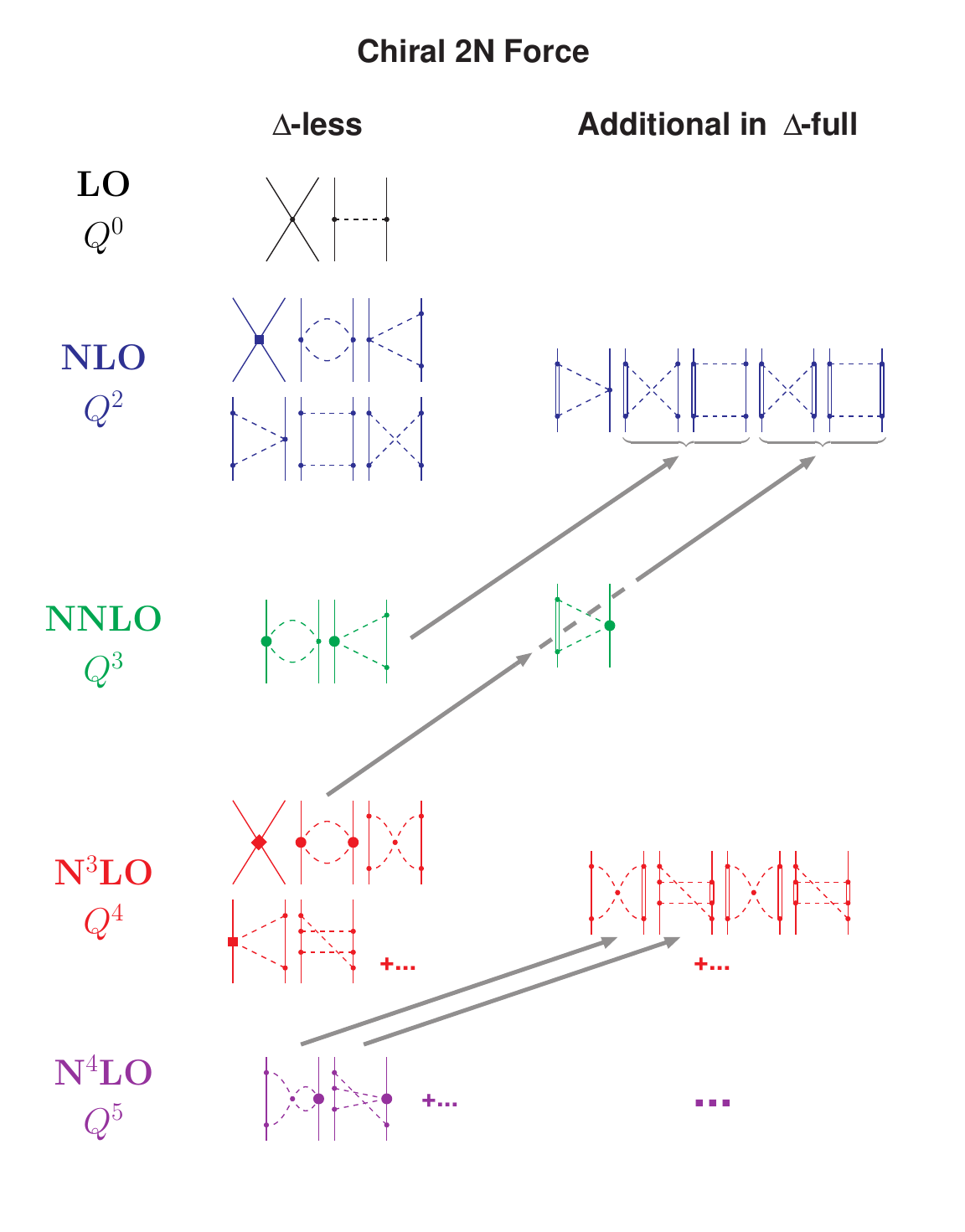}}
\vspace*{-1.0cm}
\caption{Chiral 2NF without and with $\Delta$-isobar degrees of freedom.
Arrows indicate the shift of strength when explicit $\Delta$'s are added to the theory.
Note that the $\Delta$-full theory consists of the diagrams involving $\Delta$'s
{\it plus} the $\Delta$-less ones. Double lines represent $\Delta$-isobars; remaining notation
as in Fig.~\ref{fig_hi}.}
\label{fig_delta_2nf}
\end{figure}

\begin{figure}[t]\centering
\vspace*{-0.5cm}
\scalebox{0.55}{\includegraphics{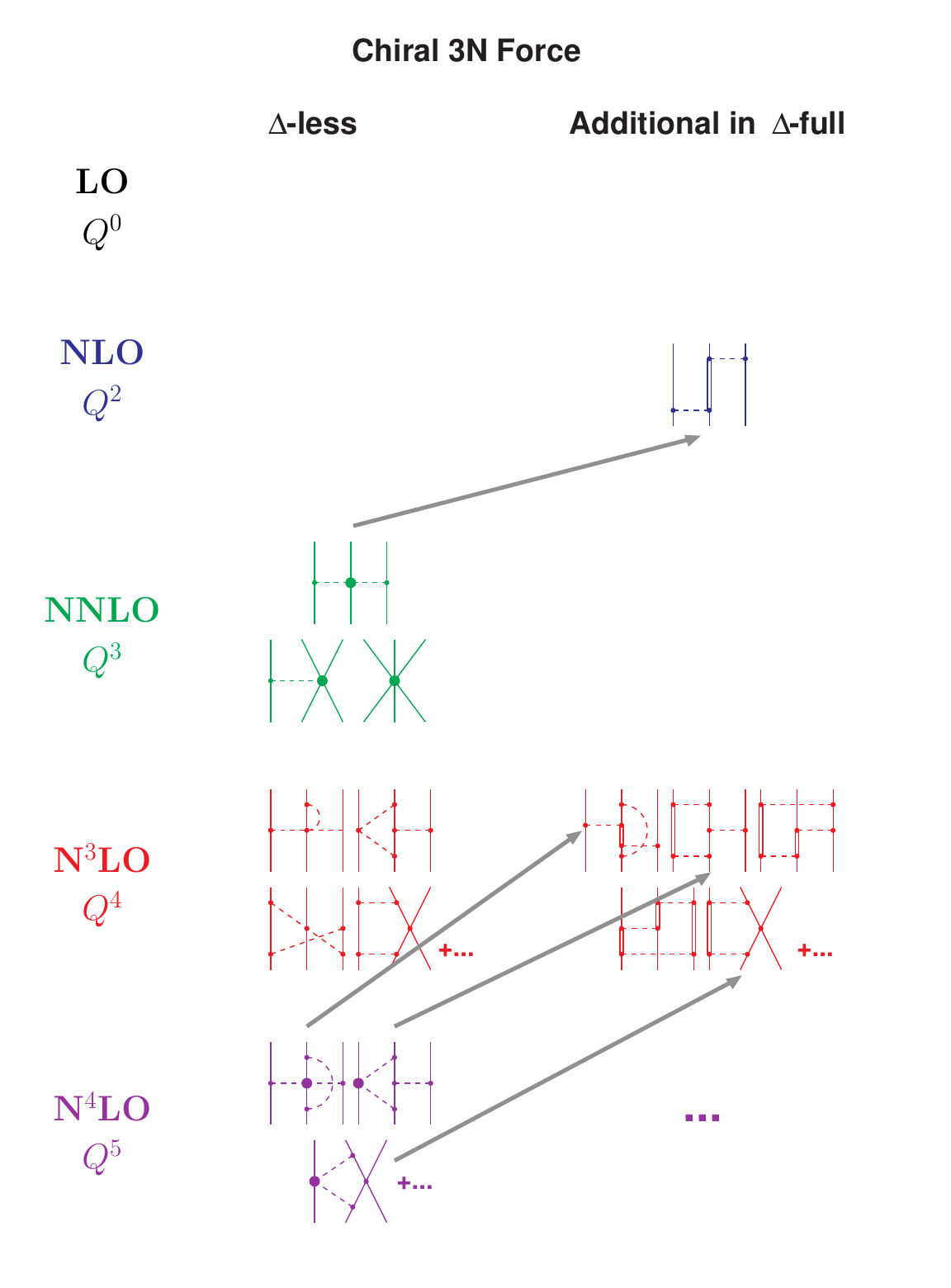}}
\vspace*{-0.5cm}
\caption{The 3NF without and with $\Delta$-isobar degrees of freedom.
Arrows indicate the shift of strength when explicit $\Delta$'s are added to the theory.
Note that the $\Delta$-full theory consists of the diagrams involving $\Delta$'s
{\it plus} the $\Delta$-less ones. Double lines represent $\Delta$-isobars; remaining notation
as in Fig.~\ref{fig_hi}.}
%\vspace*{-1.0cm}
\label{fig_delta_3nf}
\end{figure}

In the 2NF, the virtual excitation of $\Delta$-isobars requires at least one loop and, thus,
the contribution occurs first at $\nu=2$ (NLO), see Fig.~\ref{fig_delta_2nf}.
The $\Delta$ contributions to the 2PE at NLO were first evaluated in Refs.~\cite{ORK96}
using time-ordered perturbation theory and later by Kaiser {\it et al.}~\cite{KGW98} in
covariant perturbation theory. The NNLO contributions have been worked out by
Krebs {\it et al.}~\cite{KEM07}, who verified  the consistency between the $\Delta$-full and $\Delta$-less 
theories.
 
The studies of Refs.~\cite{KGW98,KEM07,NEM21} confirm that
a large amount of the intermediate-range attraction of the 2NF is shifted from NNLO to NLO
with the explicit introduction of the $\Delta$-isobar. 
However, it is also found that
the NNLO 2PE potential of the $\Delta$-less theory
provides a very good approximation to the NNLO potential in the $\Delta$-full theory.

The $\Delta$ isobar also changes the 3NF scenario~\cite{EKM08,KGE18}, see Fig.~\ref{fig_delta_3nf}. 
The leading 2PE 3NF is promoted to NLO. However, substantial 3NF contributions are expected at N$^3$LO from one-loop
diagrams with one, two, or three
intermediate $\Delta$-excitations,
which correspond to diagrams of order N$^4$LO, N$^5$LO, and N$^6$LO, respectively, 
in the $\Delta$-less theory. 
3NF loop-diagrams with one and two $\Delta$'s 
are included in the Illinois force~\cite{Pie01}
in a simplified way.

To summarize, the inclusion of explicit $\Delta$ degrees of freedom does
certainly improve the convergence of the chiral expansion by shifting sizable
contributions from NNLO to NLO. On the other hand, at NNLO
the results for the $\Delta$-full and $\Delta$-less theory are essentially the
same. Note that the $\Delta$-full theory consists of the diagrams involving
$\Delta$'s plus all diagrams of the $\Delta$-less theory. Thus, the $\Delta$-full
theory is much more involved. 

The situation could, however, change at N$^3$LO where potentially large contributions
enter the picture. It may be more efficient to calculate these terms in the $\Delta$-full
theory, because in the $\Delta$-less theory they are spread out
over N$^3$LO, N$^4$LO and, in part, N$^5$LO. 
These higher order contributions are a crucial test for the convergence of
the chiral expansion of nuclear forces and represent a challenging topic for the
future.
The Bochum group has embarked on such program and has, recently, 
calculated the 2PE 3NF diagrams involving intermediate $\Delta$-isobar
excitations~\cite{KGE18}.

\subsection{The short-range $NN$ force \label{sec_ct}}

In previous sections, we mainly discussed the pion-exchange contributions to the $NN$
interaction.
They
describe the long- and intermediate-range
parts of the nuclear force, which are governed by chiral symmetry and rule the peripheral partial waves (cf.\ Fig.~\ref{fig_ph6a}).
However, for a ``complete'' nuclear force, we have to 
describe correctly all partial waves, including the lower ones.
In fact, in calculations of $NN$ observables at low energies (cross
sections, analyzing powers, etc.),
the partial
waves with $L\leq 2$ are the most important ones, generating
the largest contributions.
The same is true for microscopic nuclear structure
calculations.
The lower partial waves are dominated by the
dynamics at short distances.
Therefore, we need to look now more closely into the short-range part
of the $NN$ potential.

In conventional meson theory~\cite{Mac89,MHE87}, the short-range nuclear force
is described by the exchange of heavy mesons, notably the
$\omega(782)$. 
Qualitatively, the short-distance behavior of the $NN$ 
potential is obtained
by Fourier transform of the propagator of
a heavy meson,
\begin{equation}
\int d^3q \frac{e^{i{\vec q} \cdot {\vec r}}}{m^2_\omega
+ {\vec q}^2} \;
\sim \;
 \frac{e^{-m_\omega r}}{r} \; .
\end{equation}

ChPT is an expansion in small momenta $p$, too small
to resolve structures like a $\rho(770)$ or $\omega(782)$
meson, because $p \ll \Lambda_b \approx m_{\rho,\omega}$.
But the latter relation allows us to expand the propagator 
of a heavy meson into a power series,
\begin{equation}
\frac{1}{m^2_\omega + p^2} 
\approx 
\frac{1}{m^2_\omega} 
\left( 1 
- \frac{p^2}{m^2_\omega}
+ \frac{p^4}{m^4_\omega}
-+ \ldots
\right)
,
\label{eq_power}
\end{equation}
where the $\omega$ is representative
for any heavy meson of interest.
The above expansion suggests that it should be 
possible to describe the short distance part of
the nuclear force simply in terms of powers of
$p/m_\omega$, which fits in well
with our over-all 
power expansion since $p/m_\omega \approx p/\Lambda_b$.
In Fig.~\ref{fig_hi}, such terms are denoted by four-nucleon-leg diagrams, like the first
diagram in the first and the second row of the figure. Since in these diagrams, the nucleons come infinitely close to each other (no meson-exchanges in-between them that would keep them apart), these diagrams 
 are dubbed contact terms.

Contact terms play an important role in renormalization.
Regularization
of the loop integrals that occur in multi-pion exchange diagrams
typically generates polynomial terms
with coefficients that are, in part, infinite or scale
dependent (cf.\ Appendix B of Ref.~\cite{ME11}). Contact terms absorb infinities
and remove scale dependences. We also note that contact interactions can be described, qualitatively, using resonance saturation~\cite{EP+02}.

Due to parity, only even powers of $Q$
are allowed.
Thus, the expansion of the contact potential is
formally given by
\begin{equation}
V_{\rm ct} =
V_{\rm ct}^{(0)} + 
V_{\rm ct}^{(2)} + 
V_{\rm ct}^{(4)} + 
V_{\rm ct}^{(6)} 
+ \ldots \; ,
\label{eq_ct}
\end{equation}
where the supersript denotes the power or order.
The contact terms of the various orders are given below.

\paragraph{Zeroth order (LO) $NN$ contact potential.}
\begin{equation}
V_{\rm ct}^{(0)}(\vec{p'},\vec{p}) =
C_S +
C_T \, \vec{\sigma}_1 \cdot \vec{\sigma}_2 
\label{eq_ct0}
\end{equation}
and, in terms of partial waves, we have 
\be
V_{\rm ct}^{(0)}(^1 S_0)          &=&  \widetilde{C}_{^1 S_0} =
4\pi\, ( C_S - 3 \, C_T ) \,,
\nonumber \\
V_{\rm ct}^{(0)}(^3 S_1)          &=&  \widetilde{C}_{^3 S_1} =
4\pi\, ( C_S + C_T ) \,.
\label{eq_ct0_pw}
\ee

\paragraph{Second order (NLO) $NN$ contact potential.}
\be
V_{\rm ct}^{(2)}(\vec{p'},\vec{p}) &=&
C_1 \, q^2 +
C_2 \, k^2 
\nonumber 
\\ &+& 
\left(
C_3 \, q^2 +
C_4 \, k^2 
\right) \vec{\sigma}_1 \cdot \vec{\sigma}_2 
\nonumber 
\\
&+& C_5 \left( -i \vec{S} \cdot (\vec{q} \times \vec{k}) \right)
\nonumber 
\\ &+& 
 C_6 \, ( \vec{\sigma}_1 \cdot \vec{q} )\,( \vec{\sigma}_2 \cdot 
\vec{q} )
\nonumber 
\\ &+& 
 C_7 \, ( \vec{\sigma}_1 \cdot \vec{k} )\,( \vec{\sigma}_2 \cdot 
\vec{k} ) 
\label{eq_ct2}
\ee
with 
$\vec k = (\vec{p'} + \vec{p})/2$ and
$\vec S = (\vec{\sigma}_1 + \vec{\sigma}_2 )/2$ the total spin.

Partial-wave decomposition yields
\be
V_{\rm ct}^{(2)}(^1 S_0)          &=&  C_{^1 S_0} ( p^2 + {p'}^2 ) 
\nonumber \\
V_{\rm ct}^{(2)}(^3 P_0)          &=&  C_{^3 P_0} \, p p'
\nonumber \\
V_{\rm ct}^{(2)}(^1 P_1)          &=&  C_{^1 P_1} \, p p' 
\nonumber \\
V_{\rm ct}^{(2)}(^3 P_1)          &=&  C_{^3 P_1} \, p p' 
\nonumber \\
V_{\rm ct}^{(2)}(^3 S_1)          &=&  C_{^3 S_1} ( p^2 + {p'}^2 ) 
\nonumber \\
V_{\rm ct}^{(2)}(^3 S_1- ^3 D_1)  &=&  C_{^3 S_1- ^3 D_1}  p^2 
\nonumber \\
V_{\rm ct}^{(2)}(^3 D_1- ^3 S_1)  &=&  C_{^3 S_1- ^3 D_1}  {p'}^2 
\nonumber \\
V_{\rm ct}^{(2)}(^3 P_2)          &=&  C_{^3 P_2} \, p p'   \,,
\label{eq_ct2_pw}
\ee
which obviously contributes up to $P$ waves.

\paragraph{Fourth order (N$^3$LO) $NN$ contact potential.}
\be
V_{\rm ct}^{(4)}(\vec{p'},\vec{p}) &=&
D_1 \, q^4 +
D_2 \, k^4 +
D_3 \, q^2 k^2 +
D_4 \, (\vec{q} \times \vec{k})^2 
\nonumber 
\\ &+& 
\left(
D_5 \, q^4 +
D_6 \, k^4 +
D_7 \, q^2 k^2 +
D_8 \, (\vec{q} \times \vec{k})^2 
\right) \vec{\sigma}_1 \cdot \vec{\sigma}_2 
\nonumber 
\\ &+& 
\left(
D_9 \, q^2 +
D_{10} \, k^2 
\right) \left( -i \vec{S} \cdot (\vec{q} \times \vec{k}) \right)
\nonumber 
\\ &+& 
\left(
D_{11} \, q^2 +
D_{12} \, k^2 
\right) ( \vec{\sigma}_1 \cdot \vec{q} )\,( \vec{\sigma}_2 
\cdot \vec{q})
\nonumber 
\\ &+& 
\left(
D_{13} \, q^2 +
D_{14} \, k^2 
\right) ( \vec{\sigma}_1 \cdot \vec{k} )\,( \vec{\sigma}_2 
\cdot \vec{k})
\nonumber 
\\ &+& 
D_{15} \left( 
\vec{\sigma}_1 \cdot (\vec{q} \times \vec{k}) \, \,
\vec{\sigma}_2 \cdot (\vec{q} \times \vec{k}) 
\right) .
\label{eq_ct4}
\ee
The rather lengthy partial-wave expressions at this order
are given in Appendix E of Ref.~\cite{ME11}. These contacts affect partial waves up to $D$ waves.

\paragraph{Sixth order (N$^5$LO).}
At sixth order, 26 new contact terms appear, bringing the total number to 50. These terms
as well as their partial-wave decomposition have been worked out in Ref.~\cite{EM03a}.
They contribute up to $F$-waves.
Except for the $F$-wave terms, these contacts have not been used in the construction of $NN$ potentials.

\subsection{Regularization and non-perturbative renormalization}
\label{sec_reno}

Iteration of the potential $\widehat V$ in the Lippmann-Schwinger (LS) equation [Eq.~(\ref{eq_LS}) below]
requires cutting $\widehat V$ off for high momenta to avoid infinities.
This is consistent with the fact that ChPT
is a low-momentum expansion which
is valid only for momenta $p \ll \Lambda_b \approx 0.7$ GeV.
Therefore, the potential $\widehat V$
is multiplied
with a regulator function $f(p',p)$,
\begin{equation}
{\widehat V}(\vec{ p}~',{\vec p})
\longmapsto
{\widehat V}(\vec{ p}~',{\vec p}) \, f(p',p) \,.
\end{equation}
One choice for $f$ is the nonlocal function
\begin{equation}
f(p',p) = \exp[-(p'/\Lambda)^{2n}-(p/\Lambda)^{2n}] \,,
\label{eq_f}
\end{equation}
such that
\begin{equation}
{\widehat V}(\vec{ p}~',{\vec p}) \, f(p',p) 
\approx
{\widehat V}(\vec{ p}~',{\vec p})
\left\{1-\left[\left(\frac{p'}{\Lambda}\right)^{2n}
+\left(\frac{p}{\Lambda}\right)^{2n}\right]+ \ldots \right\} 
\,,
\label{eq_reg_exp}
\end{equation}
with a cutoff parameter 
$\Lambda  < \Lambda_b $.

Equation~(\ref{eq_reg_exp}) provides an indication of the fact that
the exponential cutoff does not necessarily
affect the given order at which 
the calculation is conducted.
For sufficiently large $n$, the regulator introduces contributions that 
are beyond the given order. Assuming a good rate
of convergence of the chiral expansion, such orders are small 
as compared to the given order and, thus, do not
affect the accuracy at the given order.
(In actual calculations, one uses, of course,
the exponential form, Eq.~(\ref{eq_f}),
and not the expansion Eq.~(\ref{eq_reg_exp}).)

It is pretty obvious that results for the $T$-matrix may
depend sensitively on the regulator and its cutoff parameter.
This is acceptable if one wishes to build models.
For example, the meson models of the past~\cite{Mac89,MHE87} 
always depended sensitively on the choices for the
cutoff parameters, and they were
welcome as additional fit parameters to further improve the reproduction of the $NN$ data.
However, the EFT approach wishes to be more fundamental
in nature and not just another model.

In field theories, divergent integrals are not uncommon and methods have
been devised for how to deal with them.
One regulates the integrals and then removes the dependence
on the regularization parameters (scales, cutoffs)
by renormalization. In the end, the theory and its
predictions do not depend on cutoffs
or renormalization scales.

Renormalizable quantum field theories, like QED,
have essentially one set of prescriptions 
that takes care of renormalization through all orders. 
In contrast, 
EFTs are renormalized order by order, i.~e., each order comes with the contact terms
needed to renormalize that order. 
Note that this applies only to perturbative calculations. 
The $NN$ {\it potential} is calculated perturbatively and hence properly renormalized.

 However, the story is different for the $NN$ {\it amplitude} ($T$-matrix) that results from a solution of the  LS equation [Eq.~(\ref{eq_LS}), below],  which is a {\it nonperturbative} resummation of the potential.
This resummation is necessary in {\it nuclear} EFT because
nuclear physics is characterized by bound states which
are nonperturbative in nature.
EFT power counting may be different for nonperturbative processes as
compared to perturbative ones. Such difference may be caused by the infrared
enhancement of the reducible diagrams generated in the LS equation.

Weinberg's discussion in Refs.~\cite{Wei90,Wei91} may suggest that the contact terms
introduced to renormalize the perturbatively calculated
potential, based upon naive dimensional analysis (``Weinberg counting''),
may also be sufficient to renormalize the nonperturbative
resummation of the potential in the LS equation.

Weinberg's alleged assumption may not be correct as first pointed out by Kaplan, Savage, and Wise (KSW)~\cite{KSW96} who, therefore, suggested to treat 1PE 
perturbatively---a prescrition which, however, has convergence problems~\cite{FMS00}.
The KSW critique resulted in a flurry of publications on the renormalization of the $NN$ amplitude,
 and we refer the interested reader to 
section 4.5 of Ref.~\cite{ME11} for an account of the first phase of discussion.
However, even today, no generally accepted solution to this problem has emerged and some more recent proposals can be found in 
Refs.~\cite{HKK19,NTK05,Bir06,LY12,Lon16,Val11,Val11a,Val16,Val17,EGM17,Kon17,Epe18,Kol20,Val19,EO21}.

Concerning the construction of quantitative $NN$ potential
(by which we mean $NN$ potentials suitable for use in contemporary many-body nuclear methods), 
only Weinberg counting
has been used with success during the past 
25 years~\cite{ORK96,EM03,EGM00,EGM05,EKM15a,EKM15b,Eks13,Gez14,Pia15,Pia16,PAA15,Car16,RKE18,Eks15,Eks18,EMN17}.

In spite of the criticism, Weinberg counting may be perceived as not unreasonable by the following argument.
For a successful EFT (in its domain of validity), one must be able to claim independence of the predictions on the regulator within the theoretical error.
Also,                                         
truncation errors must decrease as we go to higher and higher orders.
These are precisely the goals of renormalization.  

Lepage~\cite{Lep97} has stressed that the cutoff independence should be examined
for cutoffs below the hard scale and not beyond. Ranges of cutoff independence within the
theoretical error are to be identified using Lepage plots~\cite{Lep97}.

In Ref.~\cite{Mar13}, the error of the predictions was quantified by calculating the $\chi^2$/datum 
for the reproduction of the $np$ elastic scattering data
as a function of the cutoff parameter $\Lambda$ of the regulator function
Eq.~(\ref{eq_f}). Predictions by chiral $np$ potentials at 
order NLO and NNLO were investigated applying Weinberg counting 
for the $NN$ contact terms.
It is found that the reproduction of the $np$ data at lab.\ energies below 200 MeV is generally poor
at NLO, while at NNLO the $\chi^2$/datum assumes acceptable values (a clear demonstration of
order-by-order improvement). Furthermore, at NNLO, 
a ``plateau'' of constant low $\chi^2$ for
cutoff parameters ranging from about 450 to 850 MeV can be identified. This may be perceived as cutoff independence
(and, thus, successful renormalization) for the relevant range of cutoff parameters.

Alternatively, one may go for a compromise between Weinberg's prescription of full resummation
of the potential
and Kaplan, Savage, and Wise's~\cite{KSW96} suggestion of perturbative pions---as discussed in
Ref.~\cite{Kol20}: 1PE is resummed only in lower partial waves and all corrections are included in 
distorted-wave perturbation theory. 
However, since current {\it ab initio} calculations are tailored such that they need a potential as input,  the question is if there is a way to reconcile those (low-cutoff) potentials 
with the approch of
partially perturbative pions. A first attempt to address this issue has recently been undertaken
by Valderrama~\cite{Val19}.

\section{Quantitative chiral $NN$ potentials
\label{sec_pot}}

\subsection{Various representations of $NN$ potentials \label{sec_pot1}}

We have now rounded up everything needed for a realistic
nuclear force---long, intermediate, and short ranged 
components---and so we can finally proceed to the lower
partial waves. However, here we encounter another problem.
The two-nucleon system at low angular momentum, particularly
in $S$ waves, is characterized by the
presence of a shallow bound state (the deuteron)
and large scattering lengths.
Thus, perturbation theory does not apply.
In contrast to $\pi$-$\pi$ and $\pi$-$N$,
the interaction between nucleons is not suppressed
in the chiral limit ($Q\rightarrow 0$).
Weinberg~\cite{Wei91} showed that the strong enhancement of the
scattering amplitude arises from purely nucleonic intermediate
states (``infrared enhancement''). He therefore suggested to use perturbation theory to
calculate the $NN$ potential (i.e., the irreducible graphs) and to apply this potential
in a scattering equation 
to obtain the $NN$ amplitude. 
Current chiral $NN$ potential constructions
 follow this prescription.

The potential $V$ as discussed in previous sections is, in principal, an invariant amplitude and, thus, satisfies a relativistic scattering equation, for which 
one may consider the Blankenbecler-Sugar
(BbS) equation~\cite{BS66}, which reads explicitly,
\begin{equation}
{T}({\vec p}~',{\vec p})= {V}({\vec p}~',{\vec p})+
\int d^3p'' \:
{V}({\vec p}~',{\vec p}~'') \:
\frac{M_N^2}{E_{p''}} \:  
\frac{1}
{{ p}^{2}-{p''}^{2}+i\epsilon} \:
{T}({\vec p}~'',{\vec p}) 
\label{eq_bbs2}
\end{equation}
with $E_{p''}\equiv \sqrt{M_N^2 + {p''}^2}$ and $M_N$ the nuclear mass.
The advantage of using a relativistic scattering equation is that it automatically
includes relativistic corrections to all orders. Thus, in the scattering equation,
no propagator modifications are necessary when raising the order to which the
calculation is conducted.

Defining
\begin{equation}
\widehat{V}({\vec p}~',{\vec p})
\equiv 
\sqrt{\frac{M_N}{E_{p'}}}\:  
{V}({\vec p}~',{\vec p})\:
 \sqrt{\frac{M_N}{E_{p}}}
\label{eq_minrel1}
\end{equation}
and
\begin{equation}
\widehat{T}({\vec p}~',{\vec p})
\equiv 
\sqrt{\frac{M_N}{E_{p'}}}\:  
{T}({\vec p}~',{\vec p})\:
 \sqrt{\frac{M_N}{E_{p}}} 
\,,
\label{eq_minrel2}
\end{equation}
the BbS equation collapses into the usual, nonrelativistic
LS equation,
\begin{equation}
 \widehat{T}({\vec p}~',{\vec p})= \widehat{V}({\vec p}~',{\vec p})+
\int d^3p''\:
\widehat{V}({\vec p}~',{\vec p}~'')\:
\frac{M_N}
{{ p}^{2}-{p''}^{2}+i\epsilon}\:
\widehat{T}({\vec p}~'',{\vec p}) \, .
\label{eq_LS}
\end{equation}
Since 
$\widehat V$ 
satisfies Eq.~(\ref{eq_LS}), 
it can be used like a nonrelativistic potential, and 
$\widehat{T}$ 
may be perceived as the conventional nonrelativistic 
$T$-matrix.
The above momentum-space equation is equivalent
to the nonrelativistic Schr\"odinger equation, which one would apply for
the configuration-space versions of the chiral potentials.
Since, for reasons explained in Sec.~\ref{sec_rspace}, below, most practitioners perceive it as desirable to keep configuration-space
potentials local, in all chiral $r$-space potentials the approximation
is applied:
\begin{eqnarray}
\widehat{V}({\vec p}~',{\vec p}) & \approx & {V}({\vec p}~',{\vec p}) \,, \\
\widehat{T}({\vec p}~',{\vec p}) & \approx & {T}({\vec p}~',{\vec p}) \,,
\end{eqnarray}
because the square-root factors in Eqs.~(\ref{eq_minrel1}) and
(\ref{eq_minrel2}) are nonlocal.

Over the past 20 years, a large number of chiral $NN$ potentials have been constructed and Table~\ref{tab_pots} provides an overview of these activities,
which we will discuss now in more detail.

\begin{table}
\caption{Chiral $NN$ potentials published during the past 20 years}
\smallskip
%\begin{tabular}{cccccccccc}
%\begin{tabular*}{\textwidth}{@{\extracolsep{\fill}}llllllllll}
%\footnotesize
\scriptsize
%\tiny
\begin{tabular}{llllllllll}
\hline
\hline
\noalign{\smallskip}
 Year & Authors  & Name &  Order(s) & $\Delta$'s$^a$ & Locality & Cutoff(s) & Max. $T_{lab}$$^b$  & $\chi^2$/datum$^c$ & Ref(s).   \\
\hline
\hline
\noalign{\smallskip}
\multicolumn{3}{l}{\underline{Set 1 (Early Birds):}} \\
2003 & Entem, Machleidt & Idaho & N$^3$LO & No & Nonlocal & 500 MeV & 300 MeV & 1.3 & \cite{EM03} \\
2005 & Epelbaum {\it et al.} & Bochum & N$^3$LO & No & Nonlocal & 450, 600 MeV & 300 MeV & 14.5, 2.3 & \cite{EGM05} \\
%\hline
\noalign{\smallskip}
\multicolumn{3}{l}{\underline{Set 2 (G\"oteborg/Oak Ridge):}} \\
2013 & Ekstr\"om {\it et al.} & NNLO$_{\rm opt}$ & NNLO & No & Nonlocal & 500 MeV & 290 MeV &  9.5  & \cite{Eks13} \\
2015 & Ekstr\"om {\it et al.} & NNLO$_{\rm sat}$ & NNLO & No & Nonlocal & 450 MeV & 35 MeV &  39.0$^d$, 42.7$^{e}$  & \cite{Eks15} \\
2018 & Ekstr\"om {\it et al.} & $\Delta$NNLO & NNLO & Yes & Nonlocal & 450 MeV & 200 MeV &  25.1$^f$  & \cite{Eks18} \\
2020 & Jiang, Ekstr\"om {\it et al.} & $\Delta$NNLO$_{\rm GO}$ & NNLO & Yes & Nonlocal & 394, 450 MeV & 200 MeV & 32.6$^f$, 29.6$^f$   & \cite{Jia20} \\
%\hline
\noalign{\smallskip}
\multicolumn{3}{l}{\underline{Set 3 (Configuration space):}} \\
2014 & Gezerlis {\it et al.} &    & LO-NNLO & No & Local & 1.0-1.2 fm & 250 MeV & 12.2$^g$ & \cite{Gez14} \\
2015 & Piarulli {\it et al.} &    & NNLO/N$^3$LO$^h$ & Yes & Min.\ nonloc.$^i$ & 0.8-1.2 fm & 300 MeV & 1.35(2) & \cite{Pia15} \\
2016 & Piarulli {\it et al.} & Norfolk, NV2  & LO-N(3)LO$^h$ & Yes & Local & 0.8-1.2 fm & 200 MeV & 1.40$^j$ & \cite{Pia16} \\
2023 & Saha {\it et al.} & Idaho & LO-N$^3$LO & No & Local & 1.0-1.2 fm & 200 MeV & 1.45$^k$ & \cite{Sah23} \\
2023 & Somasundaram {\it et al.} & LO-N$^3$LO$_{\rm LA}$ & LO-N$^3$LO & No & Max.\ local$^l$ & 0.6-0.9 fm & 400 MeV &  & \cite{Som23} \\
%\hline
\noalign{\smallskip}
\multicolumn{4}{l}{\underline{Set 4 (Latest high accuracy and high precision):}} \\
2015 & Epelbaum {\it et al.} & LENPIC & LO-N$^4$LO & No & Semilocal$^m$ & 0.8-1.2 fm & 300 MeV &  & \cite{EKM15a,EKM15b} \\
2017 & Entem {\it et al.} & Idaho & LO-N$^4$LO & No & Nonlocal & 450-550 MeV & 300 MeV & 1.15$^n$ & \cite{EMN17} \\
2018 & Reinert {\it et al.} & LENPIC & LO-N$^4$LO$^+$ & No & Semilocal$^m$ & 350-550 MeV & 300 MeV & 1.03$^p$ & \cite{RKE18} \\
2021 & Nosyk {\it et al.} & Idaho & LO-NNLO & Yes & Nonlocal & 394, 450 MeV & 200 MeV & 3.71$^q$ & \cite{NEM21} \\
\hline
\hline
\noalign{\smallskip}
%\end{tabular*}
\end{tabular}
\footnotesize
$^a$ $\Delta(1232)$-isobar excitations included, yes or no?
\\
$^b$ Maximum lab.\ energy up to which $NN$ phase shifts or data are fitted.
\\
$^c$ $\chi^2$/datum for the reproduction of the $NN$ data up to pion-production threshold---unless noted otherwise.
\\
$^d$ $\chi^2$/datum for the $NN$ data up to 100 MeV.
\\
$^{e}$ $\chi^2$/datum for the $NN$ data up to 190 MeV.
\\
$^f$ $\chi^2$/datum for the $NN$ data up to 200 MeV.
\\
$^g$ $\chi^2$/datum for the $NN$ data up to 190 MeV for the NNLO potential with cutoff 1.0 fm.
\\
$^h$ 2PE at NNLO, contacts at N$^3$LO.
\\
$^i$ Minimally nonlocal.
\\
$^j$ $\chi^2$/datum for the $NN$ data up to 200 MeV for the N(3)LO potential with cutoff 0.8 fm.
\\
$^k$ $\chi^2$/datum for the $NN$ data up to 190 MeV for the N$^3$LO potential with cutoff 1.0 fm.
\\
$^l$ Maximally local.
\\
$^m$ Pion-exchanges local, contacts nonlocal.
\\
$^n$ For the N$^4$LO potential with cutoff 500 MeV.
\\
$^p$ For the N$^4$LO$^+$ potential with cutoff 450 MeV.
\\
$^q$ $\chi^2$/datum for the $NN$ data up to 200 MeV for the $\Delta$NNLO potential with cutoff 450 MeV.
\label{tab_pots}
\end{table}

\subsubsection{Momentum space potentials}
\label{sec_mom}

Since ChPT is a low-momentum expansion, the most natural way
to construct a chiral $NN$ potential is in momentum space. 
Therefore, the first quantitative chiral $NN$ 
potentials---the ($\Delta$-less) 
N$^3$LO  ``Early Birds'' potentials~\cite{EM03,EGM05}
(cf.\ Table~\ref{tab_pots})---as well as
the latest high accuracy and high
precision potentials at
N$^4$LO~\cite{EKM15a,EMN17,RKE18} are all represented in 
momentum space. 
Differences between these potentials have mainly  to do with the types of regulators applied. While all Idaho potentials~\cite{EM03,EMN17} and the older Bochum potentials~\cite{EGM05} use the nonlocal
regulator function Eq.~(\ref{eq_f}) for contacts as well as pion-exchanges,
the more recent Bochum potentials~\cite{EKM15a,RKE18} apply local procedures to the pion-exchange contributions to reduce the impact of finite-cutoff artifacts.

Also the $\Delta$-full theory has been invoked for the construction of $NN$ potentials. The G\"oteborg/Oak Ridge group has constructed families of
NNLO potentials of this kind~\cite{Eks18,Jia20}. Unfortunately, these 
$\Delta$-full NNLO potentials by the G\"oteborg/Oak Ridge group
severely lack accuracy as does their $\Delta$-less 
predecessor~\cite{Eks15}---with 
$\chi^2$/datum between 30 and 40, which is unacceptable (cf.\ the  ``Set 2'' of Table~\ref{tab_pots}).
Consequently, their predictions for nuclear many-body
systems are unreliable and even misleading, as discussed in detail in Ref.~\cite{NEM21}, see also Sec.~\ref{sec_recad}, below.
It has also been demonstrated that, at least at NNLO, there is no advantage to the $\Delta$-full
theory as compared to the $\Delta$-less one~\cite{NEM21}.

\subsubsection{Configuration space potentials}
\label{sec_rspace}

Chiral $NN$ potentials have also been constructed in configuration space (position space, ``$r$-space''), see ``Set 3'' of Table~\ref{tab_pots}.
The main motivation for their construction is that
 some {\it ab initio} few- and many-body algorithms, particularly,
the ones known as quantum Monte Carlo (QMC) methods~\cite{Car15,Lyn19}
require local $r$-space potentials as their input.
Variational Monte Carlo (VMC) and Green's Function Monte Carlo (GFMC) techniques
provide reliable solutions of the many-body Schr\H{o}dinger equation for, presently, up to 12
nucleons. Spectra, form factors, transitions, low-energy scattering, and response functions for light
nuclei have been successfully calculated using QMC methods~\cite{PT20}.
A further extension, the Auxiliary Field Diffusion Monte Carlo (AFDMC)~\cite{Car15,Lyn19}, additionally samples
the spin-isospin degrees of freedom, thus, making possible the study of neutron matter (NM).
In summary, QMC techniques have substantially contributed to the progress
in {\it ab initio} nuclear structure of the past 20+ years, and will continue to do so.
Thus, high-quality nuclear interactions suitable for application
by these promising many-body methods are called for.

The first chiral $NN$ potential ever constructed
 was, in fact, represented
in configuration space (a $\Delta$-full NLO potential~\cite{ORK96}).
About 20 years later, a local $\Delta$-less NNLO was developed~\cite{Gez14}.
A more accurate one followed soon after that
was constructed in the hybrid format, NNLO/N$^3$LO~\cite{Pia15,Pia16},
where $\Delta$-full 2PE contributions are included up to NNLO and contact terms 
up to N$^3$LO.
A chiral $r$-space potential up to order
N$^3$LO was recently developed. This construction is consistent in the sense that the same power counting scheme and cutoff procedures are applied at all orders.
The 
contacts and ($\Delta$-less) pion-exchanges are all taken into
account up to N$^3$LO~\cite{Sah23}.
A  maximally local potential at N3LO in $\Delta$-less chiral EFT is under construction by the Los Alamos theory group~\cite{Som23}.
N3LO in delta-less chiral EFT. Our interactions include a
total of 21 contact operators at N3LO, out of which four
are nonlocal.

\begin{figure}\centering
%\vspace*{-0.5cm}
%\hspace*{-0.7cm}
\scalebox{0.42}{\includegraphics{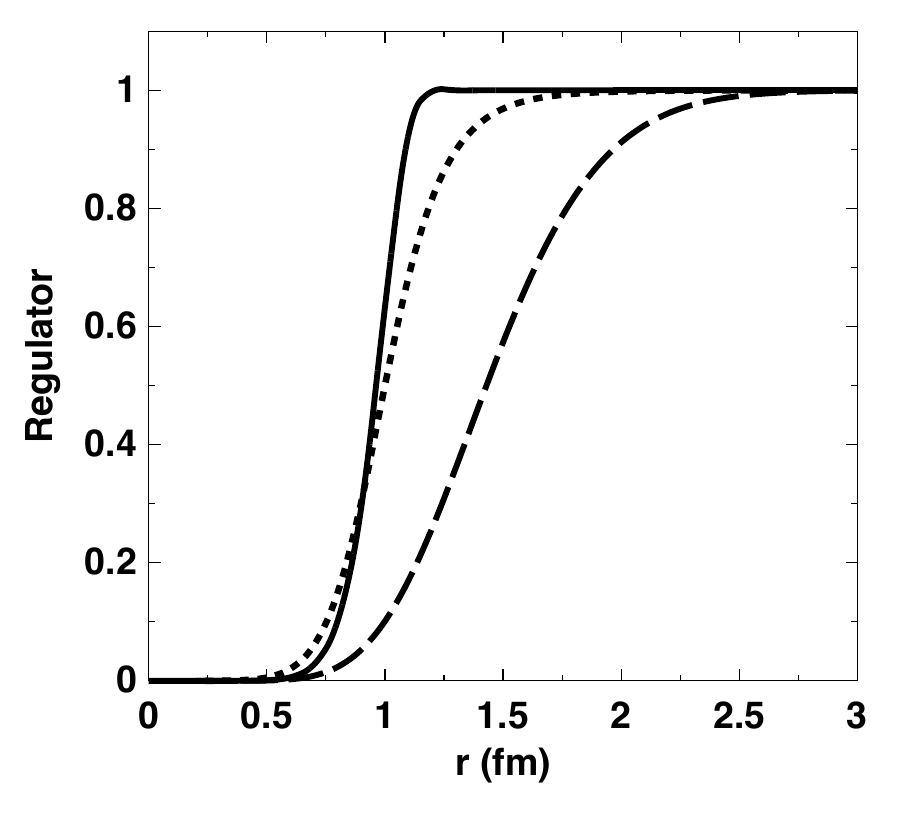}}
\vspace*{-0.3cm}
\caption{Various regulator functions 
applied to the pion-exchange contributions 
of chiral position-space potentials.
 The solid, dashed, and dotted curves represent the regulators 
 $\widetilde f_{1}(r)$, $\widetilde f_{2}(r)$, and $\widetilde f_{3}(r)$
 given in Eqs.~(\ref{eq_reg1pe}), (\ref{eq_reg2pe}), and (\ref{eq_regpia}), respectively, with $n=5$ and
 $R_\pi=1.0$.}
\label{fig_reg}
\end{figure}

The pion-exchange contributions up to N$^3$LO are all local
from the outset~\cite{Sah23}.
To preserve the local character, the pion-exchanges are multiplied with {\it local} regulator functions, which suppress the potential at short distances, at which the 2PE expressions diverge up to
$\sim 1/r^7$.  Commonly used local regulators are
\begin{eqnarray}
\widetilde f_{1}(r) & = & 1 - \exp \left[ - \left( \frac{r}{R_\pi}\right)^{2n} \right]  \,,
\label{eq_reg1pe} \\
\widetilde f_{2}(r) & = & \left[ 1 - \exp \left( - \frac{r^2}{R_\pi^2} \right) \right]^n \,,
\label{eq_reg2pe} \\
\widetilde f_{3} (r) & = & 1 - \frac{1}{ \left( {r}/{R_\pi} \right)^6 \, 
 \exp \left[{2(r-R_{\pi})}/{R_\pi} \right] \, + \, 1 } \,.
 \label{eq_regpia}
\end{eqnarray}
 Regulator $\widetilde f_{1}$  is used in Ref.~\cite{Gez14} with $n=2$.
Regulator $\widetilde f_{2}$ is applied in Refs.~\cite{EKM15a,Som23} with $n=6$.
Reference~\cite{Sah23} takes a more distinct approach and (using $n=5$) employs
$\widetilde f_{2}$ to the 2PE contributions, while utilizing $\widetilde f_{1}$
for 1PE, to preserve the beneficial nature of the 1PE also in the intermediate range.
Finally, $\widetilde f_{3}$ is chosen for the potentials by Piarulli {\it et al.}~\cite{Pia15,Pia16}. To provide an idea of the differences between the regulators,
we show in Fig.~\ref{fig_reg}  the shape of the different functions for $R_\pi=1.0$ fm and $n=5$.
In summary, in all configuration-space potentials, the regularized pion-exchange contributions are strictly local---and there is no problem achieving this.

In contrast, the locality of the contact terms is a more involved issue, which we will discuss now.
In principle, the most general set of contact terms at each order is provided by all combinations of 
spin, isospin, and momentum operators that are allowed by the usual symmetries~\cite{OM58}
at the given order.
Two momenta are available, namely, the final and initial nucleon momenta 
in the center-of-mass system, ${\vec p}\,'$ and $\vec p$.
This can be reformulated in terms of two alternative momenta, viz.,
the momentum transfer $\vec q = {\vec p}\,' - \vec p$ 
and the average momentum $\vec k = ({\vec p}\,' + \vec p)/2$.
Functions of $\vec q $\, lead to local interactions, that is, 
to functions of the relative distance
${\vec r}$ 
between the two nucleons
after Fourier transform.
On the other hand, functions of $\vec k$ lead to gradient operators
($-i\nabla$) and, thus, nonlocal interactions.

As discussed, chiral potentials are multiplied by
a regulator function that suppresses the large momenta (or, equivalently, the short distances).
Depending on the type of momenta used, the regulator can be local or nonlocal.

When chiral $NN$ potentials are constructed in momentum-space and regulated by nonlocal
cutoff functions, like Eq.~(\ref{eq_f}), then it is possible
to reduce the number of contact operators (by a factor of two) due to 
Fierz ambiguity~\cite{Hut17},
which is a consequence of the fact that nucleons are Fermions and obey the Pauli exclusion principle.
The momentum-space contact terms shown in Eqs.~(\ref{eq_ct0}),
(\ref{eq_ct2}), and (\ref{eq_ct4}) are all Fierz reduced.

However, when a local (regulator) function is applied to the contact terms, then
the Fierz rearrangement freedom is violated~\cite{Hut17,Sah23}.
For example, consider a contact operator of order zero ($\sim Q^0$, LO).
After a partial-wave decomposition and when multiplied by either no regulator or a nonlocal regulator, such operator produces no contributions for states
with orbital angular momentum $L>0$, i.e., $P$ and higher partial waves.
However, this property is violated when the operator is multiplied with a
local regulator function~\cite{Hut17,Sah23}. 
Attempts can be undertaken to partially restore Fierz reordering as tried in Ref.~\cite{Hut17}
by way of contributions of higher order. 
Based upon this argument, some local potentials assume Fierz reordering
for their contacts~\cite{Gez14,Pia15,Pia16,Som23}.

Alternatively, one may argue
 that it does not make sense to apply a symmetry that is invalid for the problem under consideration, and simply
use for the contacts all combinations of 
spin, isospin, angular momentum, and momentum $\vec q$  that are allowed by the usual symmetries---in each of the given orders.
This attitude was taken for the local potentials of Ref.~\cite{Sah23}.
On a historical note,
this is also the approach that was used for the very first chiral $NN$ potentials ever 
constructed~\cite{ORK96}.

Ignoring Fierz reordering, the LO or zeroth order charge-independent contact terms are given
in momentum-space by~\cite{Sah23}
\begin{eqnarray} 
{V}_{\rm ct}^{(0)}(q) \, &  = &
 \, \left(  C_c \, + C_\tau \, \bm{\tau}_1 \cdot \bm{\tau}_2 \, 
 + \, C_\sigma \, \vec\sigma_1 \cdot \vec \sigma_2 \,
 + \, C_{\sigma \tau} \, \vec\sigma_1 \cdot \vec \sigma_2 \; \bm{\tau}_1 \cdot \bm{\tau}_2 \right) \,
 f_{\rm ct}(q) \,,
\label{eq_ct0q}
\end{eqnarray}
while, with Fierz reordering, only two of these terms are retained, e.g., the ones shown in Eq.~(\ref{eq_ct0})~\cite{Gez14,Pia15,Pia16,Som23}.

To maintain the local nature, the contacts are multiplied with a {\it local} 
regulator function for which, in the case of most configuration-space potentials~\cite{Pia15,Pia16,Sah23,Som23},
a local Gaussian is chosen,
\begin{equation}
f_{\rm ct}(q) = e^{-(q/\Lambda)^2} \,,
\label{eq_regct}
\end {equation}
simply for practical reasons, because the Fourier transform of a Gaussian
is a Gaussian, that is
\begin{equation}
\widetilde f_{\rm ct} (r) = \frac{1}{\pi^{3/2} \, R_{\rm ct}^3} \, e^{-(r/R_{\rm ct})^2} \,.
\label{eq_cut0}
\end{equation}
The relationship between momentum-space cutoff $\Lambda$ and
position-space cutoff $R_{\rm ct}$ is then
\begin{equation}
 R_{\rm ct}=2/\Lambda \,.
 \end{equation}
Note that we use units such that $\hbar=c=1$.

Moving up to subleading orders, additional problems occur because
the Fourier transforms of some of the momentum-space contacts at second
order, Eq.~(\ref{eq_ct2}), and fourth order, Eq.~(\ref{eq_ct4}), are nonlocal.
However, at second order, an alternative set of seven linear independent local operators can be found~\cite{Gez14,Pia15,Pia16,Hut17,Sah23,Som23}, namely
\begin{eqnarray} 
{V}_{\rm ct}^{(2)}({\vec p}~', \vec p) \, &  = &  \Big\{
 \left(  \widetilde C_1 \, +  \widetilde C_2 \, \bm{\tau}_1 \cdot \bm{\tau}_2 \, 
 + \,  \widetilde C_3 \, \vec\sigma_1 \cdot \vec \sigma_2 \,
 + \,  \widetilde C_4 \, \vec\sigma_1 \cdot \vec \sigma_2 \; \bm{\tau}_1 \cdot \bm{\tau}_2 \right) \, 
 q^2 \nonumber \\
     &&  +
 \left(  \widetilde C_5 \,  +  \widetilde C_6 \, \bm{\tau}_1 \cdot \bm{\tau}_2 \, \right) \,
\widehat S_{12}(\vec q) \;\;   \nonumber \\
     &&   +
  \,\,\,  \widetilde C_7 \, \,
\left[-i \vec S \cdot (\vec q \times \vec k) \,\right]  \Big\}  \,  f_{\rm ct}(q)  \,,
\label{eq_ct2q}
\end{eqnarray}
where 
\begin{equation}
\widehat S_{12}(\vec q) = 3 \, \vec \sigma_1 \cdot \vec q \,\,\: \vec \sigma_2 \cdot \vec q - q^2 \, \vec \sigma_1 \cdot  \vec \sigma_2 
\end{equation}
is the spin-tensor operator in momentum-space.
Note that some models~\cite{Sah23} include an eighth term,
$\widetilde C_8 \, \bm{\tau}_1 \cdot \bm{\tau}_2  \,
\left[-i \vec S \cdot (\vec q \times \vec k) \,\right]  $, for convenience.

Fourier transform of Eq.~(\ref{eq_ct2q}) creates the second order {\it local} contact contribution in position space
\begin{eqnarray} 
\widetilde{V}_{\rm ct}^{(2)}(\vec r) \, &  = &
 \, \left(   \widetilde C_1 \, +  \widetilde C_2 \, \bm{\tau}_1 \cdot \bm{\tau}_2 \, 
 + \,  \widetilde C_3 \, \vec\sigma_1 \cdot \vec \sigma_2 \,
 + \,  \widetilde C_4 \, \vec\sigma_1 \cdot \vec \sigma_2 \; \bm{\tau}_1 \cdot \bm{\tau}_2 \right) \,\,
 ^{\rm ct}\widetilde{V}_C^{(2)}(r) \nonumber \\
     &&  +
 \left(  \widetilde C_5 \,  +  \widetilde C_6 \, \bm{\tau}_1 \cdot \bm{\tau}_2 \, \right) \,
S_{12}(\hat r) \;\;  ^{\rm ct}\widetilde{V}_T^{(2)}(r) \nonumber \\
     &&   +
  \,\,\, \widetilde C_7 \, \,\,
(\vec L \cdot \vec S) \;\;  ^{\rm ct}\widetilde{V}_{LS}^{(2)}(r) \,,
\label{eq_ct2r}
\end{eqnarray}
where
$\vec L$ denotes the operator of total angular momentum.
Furthermore,
\begin{eqnarray}
 ^{\rm ct}\widetilde{V}_C^{(2)}(r) &=& - \widetilde f_{\rm ct}^{(2)} (r)  - \frac{2}{r} \, \widetilde f_{\rm ct}^{(1)} (r) \,,
 \label{eq_cut2}
 \\
 ^{\rm ct}\widetilde{V}_T^{(2)}(r) &=&  - \widetilde f_{\rm ct}^{(2)} (r)  + \frac{1}{r} \, \widetilde f_{\rm ct}^{(1)} (r) \,,
 \\
^{\rm ct}\widetilde{V}_{LS}^{(2)}(r) &=&  - \frac{1}{r} \, \widetilde f_{\rm ct}^{(1)} (r) \,,
\end{eqnarray}
with
\begin{equation}
\widetilde f_{\rm ct}^{(n)} (r) = \frac{d^n  \widetilde f_{\rm ct}(r)}{dr^n} \,.
\end{equation}

Turning now to
the N$^3$LO or fourth order contact contributions,
we have, when Fierz reordering is ignore and only the local momentum $q$
taken into account~\cite{Sah23},
\begin{eqnarray} 
{V}_{\rm ct}^{(4)}({\vec p}~', \vec p)  \, &  = &   \Big\{
  \left(  \widetilde D_1 \, + \widetilde D_2 \, \bm{\tau}_1 \cdot \bm{\tau}_2 \, 
 + \, \widetilde D_3 \, \vec\sigma_1 \cdot \vec \sigma_2 \,
 + \, \widetilde D_4 \, \vec\sigma_1 \cdot \vec \sigma_2 \; \bm{\tau}_1 \cdot \bm{\tau}_2 \right) \,
q^4 \nonumber \\
     &&  +
 \left( \widetilde D_5 \,  + \widetilde D_6 \, \bm{\tau}_1 \cdot \bm{\tau}_2 \, \right) \, q^2 \,
\widehat S_{12}(\vec q) \;\;   \nonumber \\
     &&   +
      \left( \widetilde D_7 \,  + \widetilde D_8 \, \bm{\tau}_1 \cdot \bm{\tau}_2 \, \right) \, q^2 \,
 \left[-i \vec S \cdot (\vec q \times \vec k) \,\right]  \;\;   \nonumber \\
   &&   +
      \left( \widetilde D_9 \,  + \widetilde D_{10} \, \bm{\tau}_1 \cdot \bm{\tau}_2 \, \right) \,
 \left[-i \vec S \cdot (\vec q \times \vec k) \,\right]^2 
 \label{eq_LS2} \\
   &&  +
    \, \left(  \widetilde D_{11} \, + \widetilde D_{12} \, \bm{\tau}_1 \cdot \bm{\tau}_2 \, 
 + \, \widetilde D_{13} \, \vec\sigma_1 \cdot \vec \sigma_2 \,
 + \, \widetilde D_{14} \, \vec\sigma_1 \cdot \vec \sigma_2 \; \bm{\tau}_1 \cdot \bm{\tau}_2 \right) \,\,
\left[-i (\vec q \times \vec k) \,\right]^2 \Big\} f_{\rm ct}(q) \,,
\label{eq_L2}
\end{eqnarray}
which converts to
\begin{eqnarray} 
\widetilde{V}_{\rm ct}^{(4)}(\vec r) \, &  = &
 \, \left(  \widetilde D_1 \, + \widetilde D_2 \, \bm{\tau}_1 \cdot \bm{\tau}_2 \, 
 + \, \widetilde D_3 \, \vec\sigma_1 \cdot \vec \sigma_2 \,
 + \, \widetilde D_4 \, \vec\sigma_1 \cdot \vec \sigma_2 \; \bm{\tau}_1 \cdot \bm{\tau}_2 \right) \,\,
 ^{\rm ct}\widetilde{V}_C^{(4)}(r) \nonumber \\
     &&  +
 \left( \widetilde D_5 \,  + \widetilde D_6 \, \bm{\tau}_1 \cdot \bm{\tau}_2 \, \right) \,
S_{12}(\hat r) \;\;  ^{\rm ct}\widetilde{V}_T^{(4)}(r) \nonumber \\
     &&   +
      \left( \widetilde D_7 \,  + \widetilde D_8 \, \bm{\tau}_1 \cdot \bm{\tau}_2 \, \right) \,
(\vec L \cdot \vec S) \;\;  ^{\rm ct}\widetilde{V}_{LS}^{(4)}(r)  \nonumber \\
   &&   +
      \left( \widetilde D_9 \,  + \widetilde D_{10} \, \bm{\tau}_1 \cdot \bm{\tau}_2 \, \right) \,
(\vec L \cdot \vec S)^2 \;\;  ^{\rm ct}\widetilde{V}_{LS2}^{(4)}(r)  \nonumber \\
   &&  +
    \, \left(  \widetilde D_{11} \, + \widetilde D_{12} \, \bm{\tau}_1 \cdot \bm{\tau}_2 \, 
 + \, \widetilde D_{13} \, \vec\sigma_1 \cdot \vec \sigma_2 \,
 + \, \widetilde D_{14} \, \vec\sigma_1 \cdot \vec \sigma_2 \; \bm{\tau}_1 \cdot \bm{\tau}_2 \right) \,\,
{\vec L}^2 \, ^{\rm ct}\widetilde{V}_{LL}^{(4)}(r) \,,
\label{eq_ct4r}
\end{eqnarray}
with
\begin{eqnarray}
 ^{\rm ct}\widetilde{V}_C^{(4)}(r) &=&  \widetilde f_{\rm ct}^{(4)} (r)  + \frac{4}{r} \, \widetilde f_{\rm ct}^{(3)} (r) \,,
 \label{eq_cut4}
 \\
 ^{\rm ct}\widetilde{V}_T^{(4)}(r) &=&  \widetilde f_{\rm ct}^{(4)} (r)  + \frac{1}{r} \, \widetilde f_{\rm ct}^{(3)} (r) 
 - \frac{6}{r^2} \, \widetilde f_{\rm ct}^{(2)} (r) + \frac{6}{r^3} \, \widetilde f_{\rm ct}^{(1)} (r)  \,,
 \\
^{\rm ct}\widetilde{V}_{LS}^{(4)}(r) &=&  \frac{1}{r} \, \widetilde f_{\rm ct}^{(3)} (r) 
 + \frac{2}{r^2} \, \widetilde f_{\rm ct}^{(2)} (r) - \frac{2}{r^3} \, \widetilde f_{\rm ct}^{(1)} (r)   \,,
\\
^{\rm ct}\widetilde{V}_{LS2}^{(4)}(r) &=&   \frac{1}{r^2} \, \widetilde f_{\rm ct}^{(2)} (r) - \frac{1}{r^3} \, \widetilde f_{\rm ct}^{(1)} (r) \,,
\\
^{\rm ct}\widetilde{V}_{LL}^{(4)}(r) &=&   \frac{1}{r^2} \, \widetilde f_{\rm ct}^{(2)} (r) - \frac{1}{r^3} \, \widetilde f_{\rm ct}^{(1)} (r) \,,
\end{eqnarray}
where, from the Fourier transforms of
Eqs.~(\ref{eq_LS2}) and (\ref{eq_L2}),
only the local terms are retained~\cite{Pia15,Sah23}.

However, at fourth order, the full scenario is more complicated than this.
With Fierz reordering, there should be 15
linear independent operators, cf.\ Eq.~(\ref{eq_ct4}). But there exist only 11 linearly
independent local ones (in the above, the $\widetilde D_{10}$,
$\widetilde D_{12}$, and $\widetilde D_{14}$ terms are redundant
for strict Fierz reordering).
As a consequence, one finds two different philosophies in local N$^3$LO potential constructs: One is to insist the potential to be strictly local
and, thus, use only the 11 local ones~\cite{Pia16} (or all of the above 14 ones~\cite{Sah23}). 
In the alternative approach, one insists on using 
a complete set of 15 operators and, hence, includes four nonlocal ones.
Hoping that the nonlocal terms are small, their contributions can be included
in perturbation theory, thus, making QMC calculations possible.
This attitude is assumed in Refs.~\cite{Pia15,Som23}, which
explains the attributes ``minimally nonlocal'' or ``maximally local'' 
 used to characterize these potentials.

The operator set used in Eq.~(\ref{eq_ct4r}) is the same that 
 the phenomenological Argonne $v_{18}$ potential (AV18)~\cite{WSS95}
 is based upon. 
 Thus, for local chiral potentials applying this set,
 a detailed comparison with established phenomenology can be made.
 Such comparison, conducted in Ref.~\cite{Sah23},
 revealed substantial agreement between the AV18 and the chiral N$^3$LO potentials in the intermediate range
ruled by chiral symmetry, hence, providing a chiral underpinning for the
phenomenological AV18 potential.

\subsection{$NN$ potentials order by order}
\label{sec_pot2}

\begin{figure}[t]\centering
\vspace*{-0.5cm}
%\hspace*{-0.4cm}
\scalebox{0.4}{\includegraphics{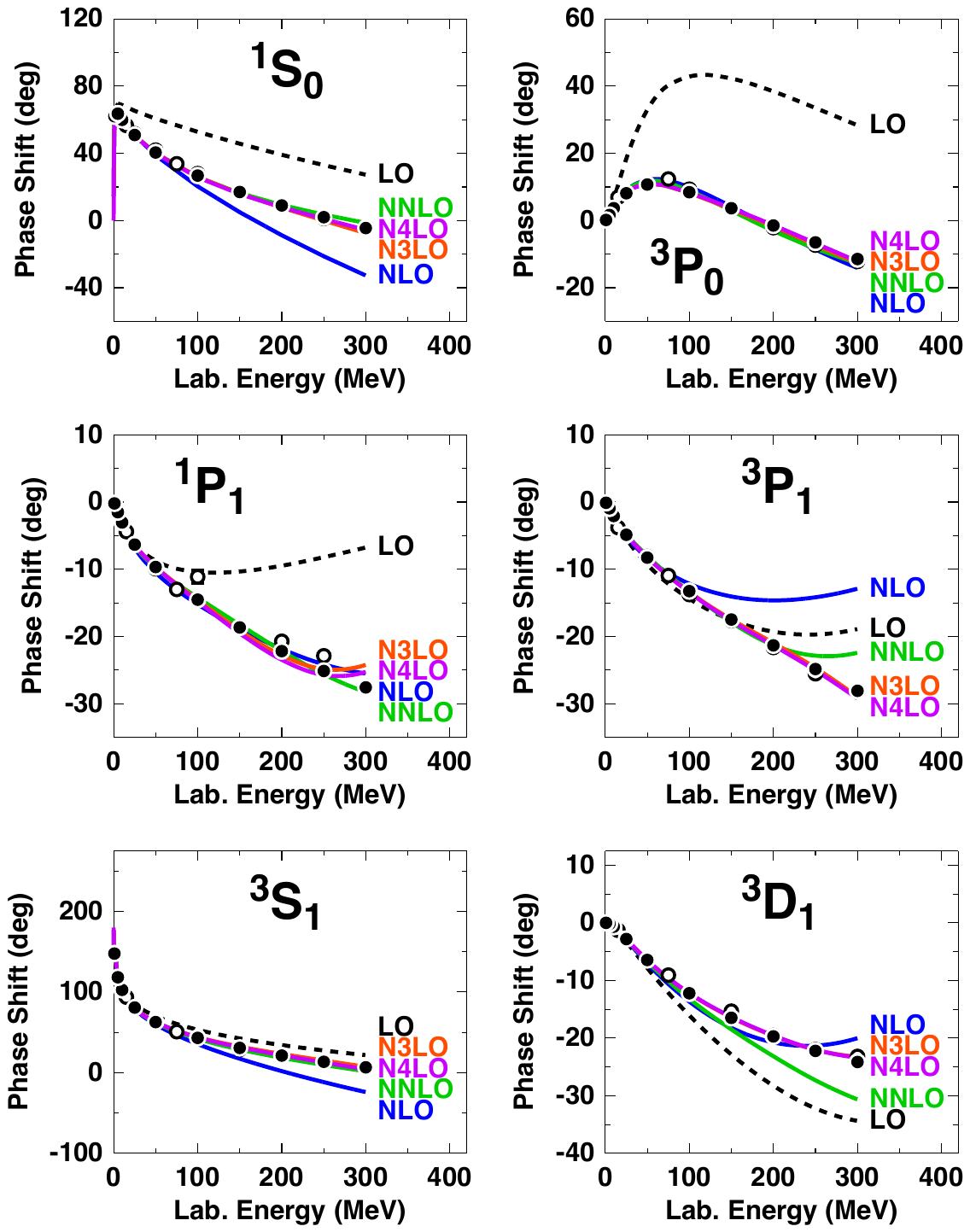}}
%\hspace*{-0.5cm}
\scalebox{0.4}{\includegraphics{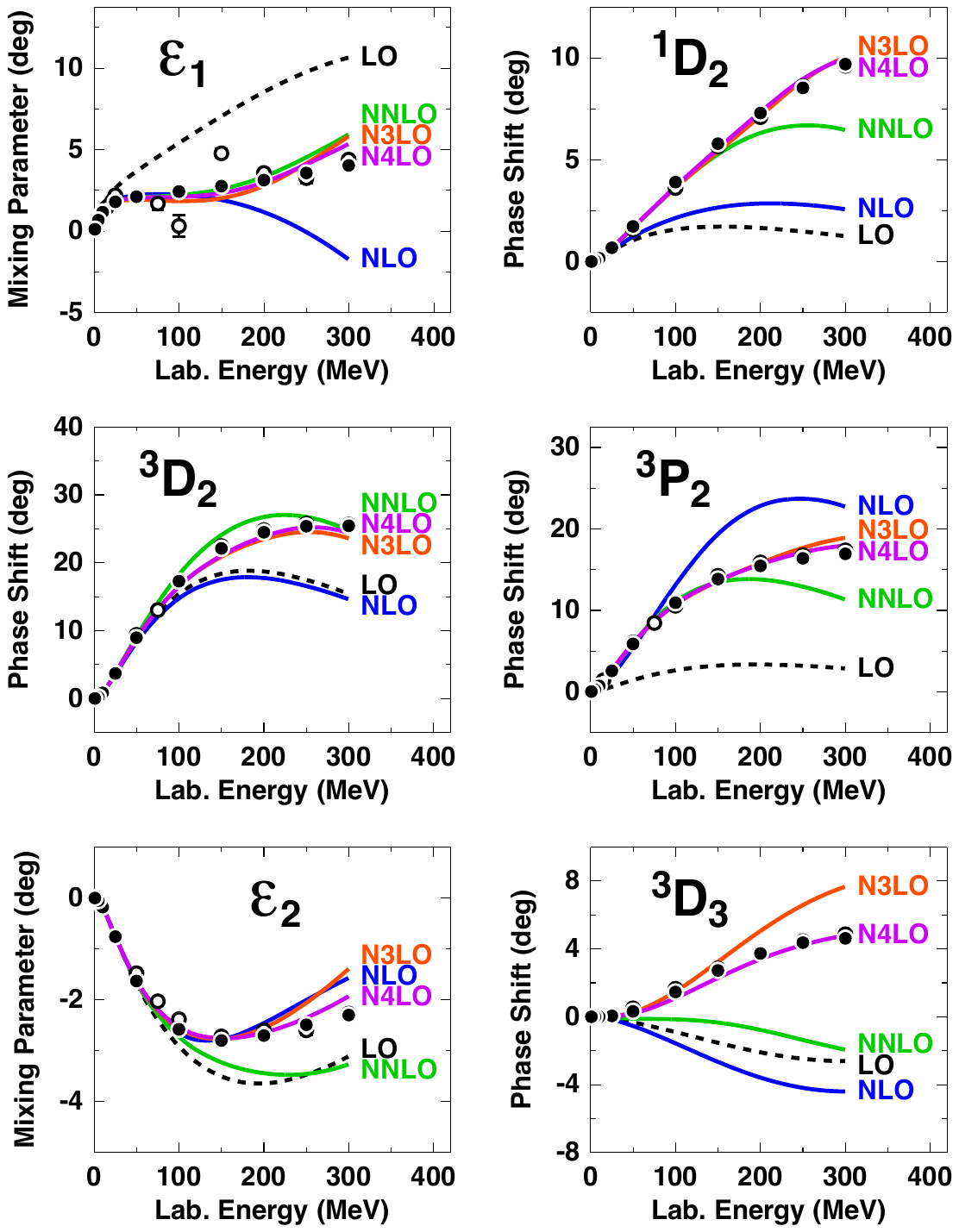}}
\vspace*{-0.5cm}
%\vspace*{-1cm}
%\scalebox{0.5}{\includegraphics{figph1a.pdf}}
%\vspace*{-1.0cm}
\caption{
Chiral expansion of neutron-proton scattering as represented by the phase shifts 
in $S$, $P$, and $D$ waves and mixing parameters $\epsilon_1$ and $\epsilon_2$.
 Five orders ranging from LO to N$^4$LO are shown as 
denoted ($\Lambda = 500$ MeV in all cases).
Filled and open circles as in Fig.~\ref{fig_ph6a}.
(From Ref.~\cite{EMN17})
\label{fig_ph1a}}
\end{figure}

\begin{table}[t]
\caption{$\chi^2/$datum for the fit of the 2016 $NN$ data base by $NN$ potentials at various orders of chiral EFT ($\Lambda = 500$ MeV in all cases). (From Ref.~\cite{EMN17})
\label{tab_chi}}
\smallskip
\begin{tabular*}{\textwidth}{@{\extracolsep{\fill}}ccccccc}
\hline 
\hline 
\noalign{\smallskip}
 $T_{\rm lab}$ bin (MeV) & No.\ of data & LO & NLO & NNLO & N$^3$LO & N$^4$LO \\
\hline
\noalign{\smallskip}
\multicolumn{7}{c}{\bf proton-proton} \\
0--100 & 795 & 520 & 18.9  & 2.28   &  1.18 & 1.09 \\
0--190 & 1206  & 430 & 43.6  &  4.64 & 1.69 & 1.12 \\
0--290 & 2132 & 360 & 70.8  & 7.60  &  2.09  & 1.21 \\
\hline
\noalign{\smallskip}
\multicolumn{7}{c}{\bf neutron-proton} \\
0--100 & 1180 & 114 & 7.2  &  1.38  & 0.93  & 0.94 \\
0--190 & 1697 &  96 & 23.1  &  2.29 &  1.10  & 1.06 \\
0--290 & 2721 &  94 &  36.7 & 5.28  &  1.27 & 1.10 \\
\hline
\noalign{\smallskip}
\multicolumn{7}{c}{\boldmath $pp$ plus $np$} \\
0--100 & 1975 & 283 &  11.9 &  1.74  & 1.03   & 1.00  \\
0--190 & 2903 & 235 &  31.6 &  3.27 & 1.35   &  1.08 \\
0--290 & 4853 & 206 &  51.5 & 6.30  & 1.63   &  1.15 \\
\hline
\hline
\noalign{\smallskip}
\end{tabular*}
\end{table}

$NN$ potentials depend on two different sets of parameters, the $\pi N$ and the $NN$ LECs.
The $\pi N$ LECs are the coefficients that appear in the $\pi N$ Langrangians. They are determined in $\pi N$ analysis~\cite{Hof16}, cf.\ Table~\ref{tab_lecs}.
The $NN$ LECs are the coefficients of the $NN$ contact terms.
They are fixed by an optimal fit to the $NN$ data below pion-production threshold, see
Ref.~\cite{EMN17} for details.

$NN$ potentials are then constructed order by order
and the precision and accuracy improve as the order increases.
How well the chiral expansion converges in 
important lower partial waves
is demonstrated in Fig.~\ref{fig_ph1a},
where we show phase parameters for potentials developed
through all orders from LO to N$^4$LO (of the $\Delta$-less theory). As a representative example, we
show the predictions generated in Ref.~\cite{EMN17}, but for other families
of recently constructed chiral $NN$ potentials~\cite{EKM15a,Pia15,Pia16,RKE18,Sah23},
the picture is very similar.
Fig.~\ref{fig_ph1a} clearly reveals
substantial improvements in the reproduction of the empirical
phase shifts with increasing order.

At this point, it is instructive to talk about the uncertainties of the phase shift predictions. As to be discussed in Sec.~\ref{sec_uncert} below, the truncation error creates the largest uncertainty, for which the simplest formula is given by 
Eq.~(\ref{eq_err}). Following this prescription, the error at a certain order is the difference between the given order and the next higher one. For example, the uncertainties of the NNLO phase shifts are given by the differences between the (green) NNLO curves and the (red) N3LO curves in Fig.~\ref{fig_ph1a}. For the
uncertainty at N4LO, Eq.~(\ref{eq_err2}) has to be invoked.
The factor $Q$ in this formula is, of course, energy dependent but, as a simple rule of thumb, one may assume  $Q \approx 1 / 3$.

The $\chi^2$/datum for the reproduction of the $NN$ data at various orders of chiral EFT 
are shown in Table~\ref{tab_chi} for different energy intervals below 290 MeV laboratory energy ($T_{\rm lab}$). 
The bottom line of Table~\ref{tab_chi} summarizes the essential results.
For the nearly 5000 $pp$ plus $np$ data below 290 MeV (pion-production threshold),
the $\chi^2$/datum 
is 51.4 at NLO and 6.3 at NNLO. Note that the number of $NN$ contact terms is the same for both orders. The improvement is entirely due to an improved description of the 2PE contribution, which is responsible for the crucial intermediate-range attraction of the nuclear force.
At NLO, only the uncorrelated 2PE is taken into account which is insufficient. From the classic meson-theory of nuclear forces~\cite{MHE87}, it is wellknown that $\pi$-$\pi$ correlations and nucleon resonances need to be taken into account for a realistic model of 2PE.
As discussed, in the chiral theory, these contributions are encoded in the 
subleading $\pi N$ vertexes. 
These enter at NNLO and are the reason for the substantial improvements we encounter at that order.

To continue on the bottom line of Table~\ref{tab_chi}, after NNLO,
the $\chi^2$/datum then further improves to 1.63 at N$^3$LO and, finally, reaches the almost perfect value of 1.15 at N$^4$LO---great convergence.

\begin{table}[t]
\small
\caption{Two- and three-nucleon bound-state properties as predicted by
  $NN$ potentials at various orders of chiral EFT ($\Lambda = 500$ MeV in all cases).
(Deuteron: Binding energy $B_d$, asymptotic $S$ state $A_S$,
asymptotic $D/S$ state $\eta$, structure radius $r_{\rm str}$,
quadrupole moment $Q$, $D$-state probability $P_D$; the predicted
$r_{\rm str}$ and $Q$ are without meson-exchange current contributions
and relativistic corrections. Triton: Binding energy $B_t$.)
$B_d$ is fitted, all other quantities are predictions. (From Ref.~\cite{EMN17})
\label{tab_deu}}
\smallskip
\begin{tabular*}{\textwidth}{@{\extracolsep{\fill}}lllllll}
\hline 
\hline 
\noalign{\smallskip}
 & LO & NLO & NNLO & N$^3$LO & N$^4$LO & Empirical$^a$ \\
\hline
\noalign{\smallskip}
{\bf Deuteron} \\
$B_d$ (MeV) &
 2.224575& 2.224575 &
 2.224575 & 2.224575 & 2.224575 & 2.224575(9) \\
$A_S$ (fm$^{-1/2}$) &
 0.8526& 0.8828 &
0.8844 & 0.8853 & 0.8852 & 0.8846(9)  \\
$\eta$         & 
 0.0302& 0.0262 &
0.0257& 0.0257 & 0.0258 & 0.0256(4) \\
$r_{\rm str}$ (fm)   & 1.911
      & 1.971 & 1.968
       & 1.970
       & 1.973 &
 1.97507(78) \\
$Q$ (fm$^2$) &
 0.310& 0.273&
 0.273 & 
 0.271 & 0.273 &
 0.2859(3)  \\
$P_D$ (\%)    & 
 7.29& 3.40&
4.49 & 4.15 & 4.10 & --- \\
\hline
\noalign{\smallskip}
{\bf Triton} \\
$B_t$ (MeV) & 11.09  & 8.31  & 8.21 & 8.09  & 8.08 & 8.48 \\
\hline
\hline
\noalign{\smallskip}
\end{tabular*}
\footnotesize
$^a$See Table XVIII of Ref.~\cite{Mac01} for references;
the empirical value for $r_{\rm str}$ is from Ref.~\cite{Jen11}.  \\
\end{table}

The evolution of the deuteron properties from LO to N$^4$LO  of chiral EFT are shown in Table~\ref{tab_deu}.
In all cases, the deuteron binding energy is fit to its empirical value of 2.224575 MeV
using the non-derivative $^3S_1$ contact. All other deuteron properties are predictions.
Already at NNLO, the deuteron has converged to its empirical properties and stays there
through the higher orders.

At the bottom of Table~\ref{tab_deu}, we also show the predictions for the triton binding
as obtained in 34-channel charge-dependent Faddeev calculations using only 2NFs. The results show smooth and steady convergence, order by order, towards a value around 8.1 MeV that is reached at the highest orders shown. This contribution from the 2NF will require only a moderate 3NF. 
The relatively low deuteron $D$-state probabilities ($\approx 4.1$\% at N$^3$LO and N$^4$LO) and the concomitant generous triton binding energy predictions are
a reflection of the fact that the $NN$ potentials are soft (which is, at least in part, due to the nonlocal character of these momentum-space potentials).

For a comparison with  $\chi^2$ values from the older generation of high-precision potentials, the reader is referred to Ref.~\cite{Ep+22}.

\section{Nuclear many-body forces \label{sec_manyNF}}

Two-nucleon forces derived from chiral EFT 
have been applied, often successfully, in the many-body system.                                  
On the other hand, over the past several years we have learnt that, for some few-nucleon
reactions and nuclear structure issues, 3NFs cannot be neglected.              
The most well-known cases are the so-called $A_y$ puzzle of $N$-$d$ scattering~\cite{EMW02},
the ground state of $^{10}$B~\cite{Cau02}, and the saturation of nuclear matter~\cite{Sam12,Cor14,Sam15,MS16,Sam18}.
As we observed previously, 
the EFT approach generates          
consistent two- and many-nucleon forces in a natural way 
(cf.\ the overview given in Fig.~\ref{fig_hi}).
We now shift our focus to chiral three- and four-nucleon forces.
For a recent review on this topic, see Ref.~\cite{Heb21}.

\subsection{Three-nucleon forces}
\label{sec_3nfs}

Weinberg~\cite{Wei92} was the first to discuss     
nuclear three-body forces in the context of ChPT. Not long after that, 
the first 3NF at NNLO was derived by van Kolck~\cite{Kol94}.

For a 3NF, we have $A=3$ and $C=1$ and, thus, Eq.~(\ref{eq_nu})
implies
\begin{equation}
\nu = 2 + 2L + 
\sum_i \Delta_i \,.
\label{eq_nu3nf}
\end{equation}
We will use this equation to analyze 3NF contributions
order by order.

\subsubsection{Next-to-leading order}

The lowest possible power is obviously $\nu=2$ (NLO), which
is obtained for no loops ($L=0$) and 
only leading vertices
($\sum_i \Delta_i = 0$). 
As discussed by Weinberg~\cite{Wei92} and van Kolck~\cite{Kol94}, 
the contributions from these diagrams
vanish at NLO. So, the bottom line is that there is no genuine 3NF contribution at NLO---in the $\Delta$-less theory that we will mainly focus on
when discussing many-body forces. For the modifications of the 3NFs when explicit $\Delta$ are introduced, see Fig.~\ref{fig_delta_3nf}.

\subsubsection{Next-to-next-to-leading order}

\begin{figure}[t]\centering
%\vspace{-10.0cm}
%\hspace*{-2.0cm}
\scalebox{0.65}{\includegraphics{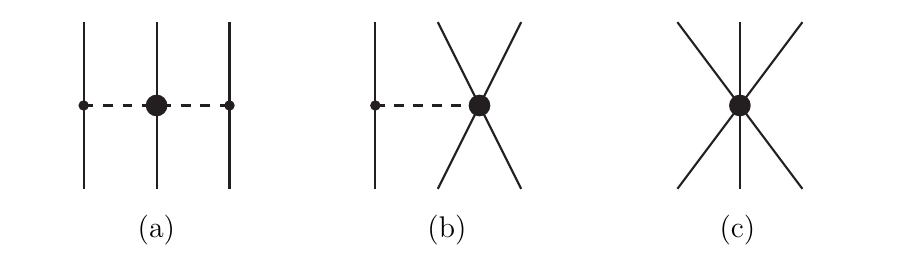}}
\vspace{-0.2cm}
\caption{The three-nucleon force at NNLO
with (a) 2PE, (b) 1PE, and (c) contact diagrams.
Notation as in Fig.~\ref{fig_hi}.}
\label{fig_3nf_nnlo}
\end{figure}

The power $\nu=3$ (NNLO) is obtained when
there are no loops ($L=0$) and 
$\sum_i \Delta_i = 1$, i.e., 
$\Delta_i=1$ for one vertex 
while $\Delta_i=0$ for all other vertices.
There are three topologies which fulfill this condition,
known as the 2PE, 1PE,
and contact graphs~\cite{Kol94,Epe02}
(Fig.~\ref{fig_3nf_nnlo}).

The 2PE 3N-potential is derived to be
\begin{equation}
V^{\rm 3NF}_{\rm 2PE} = 
\left( \frac{g_A}{2f_\pi} \right)^2
\frac12 
\sum_{i \neq j \neq k}
\frac{
( \vec \sigma_i \cdot \vec q_i ) 
( \vec \sigma_j \cdot \vec q_j ) }{
( q^2_i + m^2_\pi )
( q^2_j + m^2_\pi ) } \;
F^{ab}_{ijk} \;
\tau^a_i \tau^b_j
\label{eq_3nf_nnloa}
\end{equation}
with $\vec q_i \equiv \vec{p_i}' - \vec p_i$, 
where 
$\vec p_i$ and $\vec{p_i}'$ are the initial
and final momenta of nucleon $i$, respectively, and
\begin{equation}
F^{ab}_{ijk} = \delta^{ab}
\left[ - \frac{4c_1 m^2_\pi}{f^2_\pi}
+ \frac{2c_3}{f^2_\pi} \; \vec q_i \cdot \vec q_j \right]
+ 
\frac{c_4}{f^2_\pi}  
\sum_{c} 
\epsilon^{abc} \;
\tau^c_k \; \vec \sigma_k \cdot [ \vec q_i \times \vec q_j] \; .
\label{eq_3nf_nnlob}
\end{equation}  

It is interesting to observe that there are clear analogies between this force and earlier          
2PE 3NFs already proposed decades ago, particularly the Fujita-Miyazawa~\cite{FM57} and
the Tucson-Melbourne (TM)~\cite{Coo79} forces. 
In fact, based upon the chiral 3NF at NNLO, the TM force was corrected~\cite{FHK99}
leading to what became known as the TM' or TM99 force~\cite{CH01}.

The 2PE 3NF does not introduce additional fitting constants, 
since the LECs $c_{1,3,4}$ are fixed in $\pi N$ analysis 
(cf.\ Table~\ref{tab_lecs}) and are already present in the 2PE 2NF.

The other two 3NF contributions shown in Fig.~\ref{fig_3nf_nnlo}
are  the 1PE contribution 
\begin{equation}
V^{\rm 3NF}_{\rm 1PE} = 
- \frac{c_D \, g_A}{8f^4_\pi \, \Lambda_\chi} 
\sum_{i \neq j \neq k}
\frac{\vec \sigma_j \cdot \vec q_j}{
 q^2_j + m^2_\pi }
( \mbox{\boldmath $\tau$}_i \cdot \mbox{\boldmath $\tau$}_j ) 
( \vec \sigma_i \cdot \vec q_j ) 
\label{eq_3nf_nnloc}
\end{equation}
and the 3N contact potential 
\begin{equation}
V^{\rm 3NF}_{\rm ct} =  \frac{c_E}{2 f_\pi^4 \, \Lambda_\chi}
\sum_{i \neq j \neq k}
 \mbox{\boldmath $\tau$}_i \cdot \mbox{\boldmath $\tau$}_j  
\label{eq_3nf_nnlod}
\end{equation}
with $\Lambda_\chi=700$ MeV.
These 3NF potentials introduce 
two additional constants, $c_D$ and $c_E$, which can be constrained in             
 more than one way.                                    
One may use 
the triton binding energy and the $nd$ doublet scattering
length $^2a_{nd}$~\cite{Epe02}
or an optimal global fit of the properties of light nuclei~\cite{Nav07}.
 Alternative choices include 
the binding energies of $^3$H and $^4$He~\cite{Nog06} or
the binding energy of $^3$H and the point charge radius of $^4$He~\cite{Heb11}.
Another method makes use of
the triton binding energy and the Gamow-Teller matrix element of tritium $\beta$-decay~\cite{Mar12}.
When the values of $c_D$ and $c_E$ are fixeded, the results for other
observables involving three or more nucleons are true theoretical predictions.

A comprehensive study of all possible operator structures that enter the leading and subleading contact 3NF can be found in Ref.~\cite{Gir19}. In Ref.~\cite{PT20}, the NM EoS was calculated with different choices of the 3NF contact operator and found to be consistent with other {\it ab initio} determinations within uncertainties.

Applications of the leading 3NF include few-nucleon 
reactions~\cite{Epe02,NRQ10,Viv13}, structure of light- and medium-mass 
nuclei~\cite{Nav07a,Rot11,Rot12,Hag12a,Hag12b,BNV13,Her13,Hag14a,Bin14,HJP16,Sim16,Sim17,Mor18,Som20},
and infinite matter~\cite{Sam12,Cor14,Sam15,MS16,Sam18,Heb11,HS10,Hag14b,Cor13}.
One of the greatest successes of the NNLO 3NF is that it solves the
problem of the saturation of nuclear 
matter~\cite{Sam12,Cor14,Sam15}---a problem that has a long history.

Originally, it was hoped that nuclear matter saturation and the structure of finite nuclei
could be understood in terms of just the 
2NF~\cite{Neg70}---if one would only find the ``right'' 2NF.
However, in the course of the 1970's, when more reliable microscopic calculations became available, growing evidence accumulated that showed that it was impossible
to saturate nuclear matter at the right energy and density when applying only 2NFs~\cite{Coe70,Day83,Mac89}. Another problem was the triton binding energy, which was considerably underpredicted
with the 2NFs available at the time~\cite{BKT74,BSM77}.
These failures were interpreted as an indication for the need of nuclear many-body forces.

\begin{figure}[t]\centering
%\hspace*{-0.2cm}
\scalebox{0.4}{\includegraphics{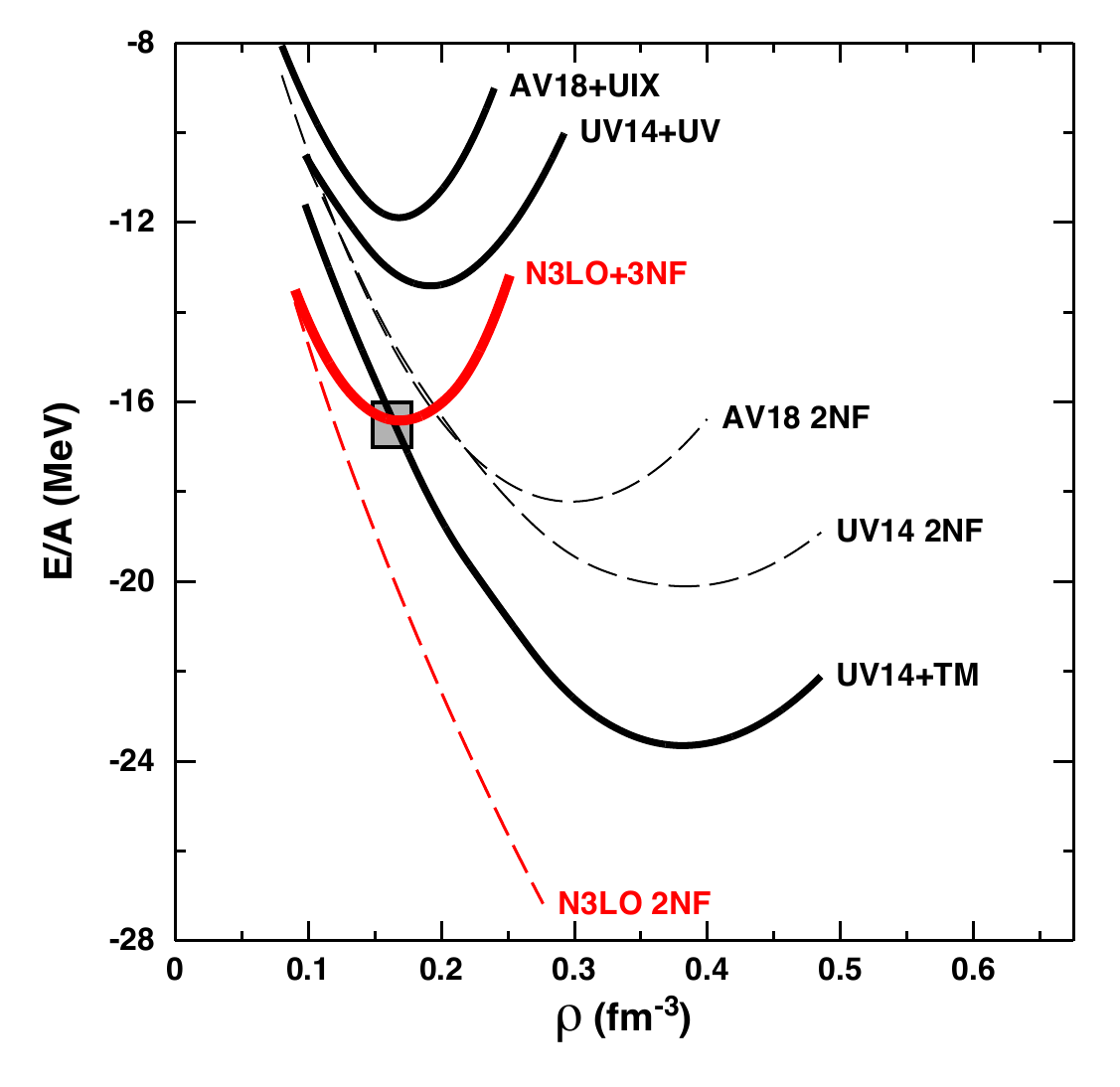}}
%\vspace*{-0.3cm}
\caption{Energy per nucleon, $E/A$, as a function of density, $\rho$, of symmetric nuclear matter (SNM). Black dashed curves represent predictions applying purely phenomenologial 2NFs with UV14 referring to the Urbana $v_{14}$ 2NF~\cite{LP81} and AV18 signifying the
Argonne $v_{18}$ 2NF~\cite{WSS95}. Solid black curves include phenomenological 3NFs with TM
 denoting the Tucson-Melbourne~\cite{Coo79}, UV the Urbana V~\cite{CPW83},
 and UIX the Urbana IX~\cite{Pud95} 3NF models.
 Red curves depict results from chiral EFT based forces~\cite{EMN17,Sam18}.
 The shaded box marks the area in which empirical nuclear matter saturation is presumed to occur.}
\label{fig_snm}
\end{figure}

Thus, in the late 1970's and early 1980's, various phenomenological
3NFs were constructed, like the TM~\cite{Coo79} and the Urbana~\cite{CPW83}
forces.
The attraction provided by the TM 2PE 3NFs has proven to be useful in explaining
the binding energies of light nuclei (particularly, $^3$H and $^4$He) which are, in general, 
underbound when only 2NFs are applied.
However, this added attraction leads to overbinding and too high a saturation density
in nuclear matter (cf.\ Fig.~\ref{fig_snm}, curve labeled UV14+TM). Therefore, some groups added
a repulsive short-range 3NF which ameliorates the problem,
but does not solve it~\cite{Day83,CPW83} (Fig.~\ref{fig_snm}, curve UV14+UV).
In the work by the Urbana group, 
many versions of such 3NF were developed with Urbana IX (UIX) 
being the most popular one (Fig.~\ref{fig_snm}, curve AV18+UIX).
In later work~\cite{Pie01}, the Urbana group extended their model for the 3NF
by including the $2\pi$-exchange $S$-wave contribution plus three-pion exchange ring diagrams
with one $\Delta$ excitation. The peculiar spin and isospin dependencies of $\Delta$-ring diagrams were found to be helpful in the explanation of spectra of light nuclei. This has become known as the Illinois 3NFs~\cite{Pie01}, which so far have
evolved up to Illinois-7 (IL7)~\cite{Pie08}.

 The 3NFs of the Urbana type, adjusted to the ground
state and the spectra of light nuclei, do not saturate nuclear matter
properly~\cite{CPW83,APR98}
(Fig.~\ref{fig_snm}, curves UV14+UV and  AV18+UIX)
 and severely underbind intermediate-mass nuclei~\cite{Lon17}.
The AV18 2NF plus IL7 3NF yield a pathological equation of state
of pure NM~\cite{Mar13a}.
In addition, the so-called $A_y$ puzzle of nucleon-deuteron scattering~\cite{EMW02}
is not resolved by any of the phenomenological 3NFs~\cite{Kie10}.

In view of this long and desperate history of failed attempts to solve the 
nuclear matter problem,
the success of the 
the chiral NNLO 3NF is outstanding,
see red `N3LO+3NF'
curve in Fig.~\ref{fig_snm}.

\begin{figure}[t] 
\centering
\includegraphics[width=7cm]{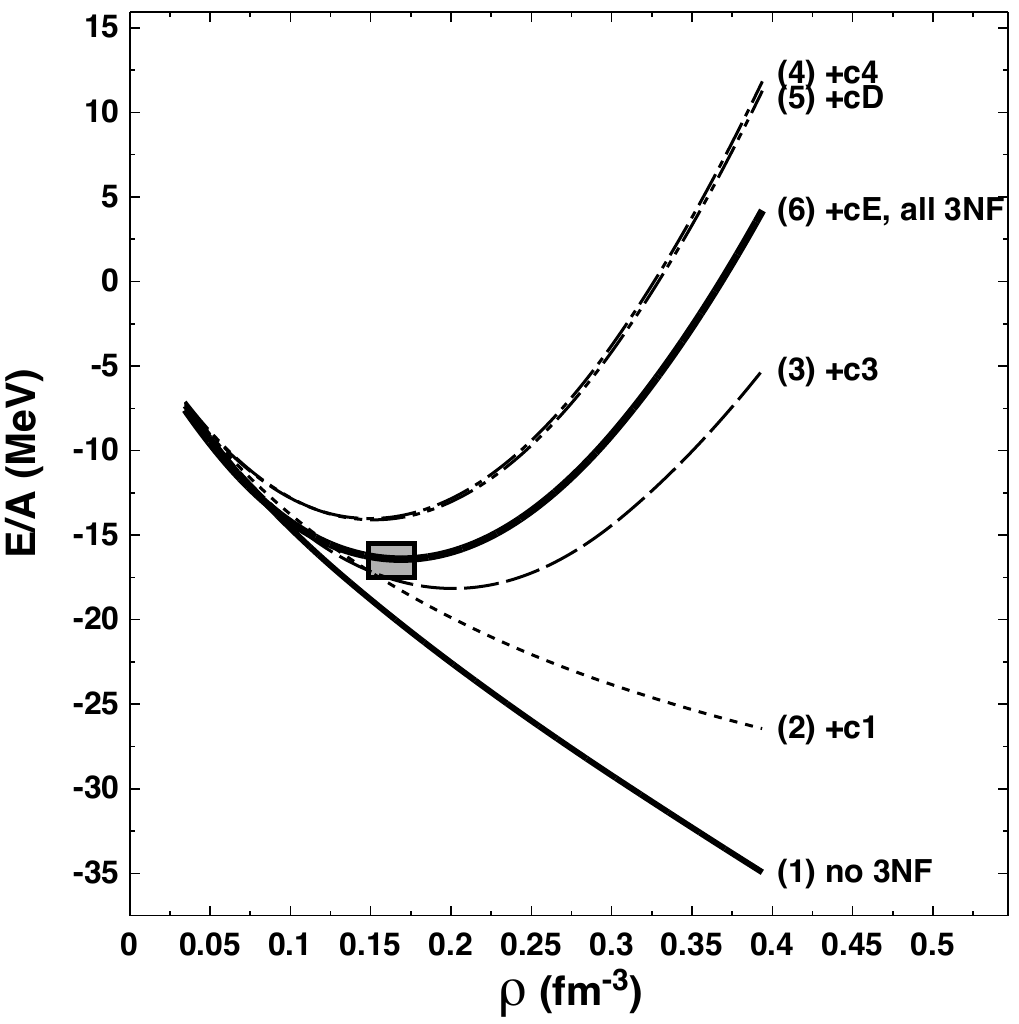}
\caption{Energy per nucleon, $E/A$, as a function of density, $\rho$, of SNM from the chiral EFT based interaction. Curve (1) shows the result 
when only the 2NF is used.
Curve (6) displays the final result after all (NNLO) 3NF contributions have been added. The individual 3NF contributions are added up successively in the
order given by the parenthetical numbers.
The LECs $c_{1,3,4}$ parametrize the 2PE 3NF, while $c_D$ and $c_E$ are the coefficients of the 1PE and contact 3NF diagrams, respectively.
}
\label{fig_snm2}
\end{figure}

It it interesting to dissect how the chiral 3NF achieves saturation. For this purpose,
we display in Fig.~\ref{fig_snm2} the effects of the individual chiral NNLO
3NF terms.
Curves (2) to (4) show the impact of the 
2PE 3NF [Fig.~\ref{fig_3nf_nnlo}(a)], while curves (5) and (6) exhibit the contributions
from the 3NF diagrams Fig.~\ref{fig_3nf_nnlo}(b) and (c), respectively.
Obviously, the 2PE 3NF component does the job, namely, provides a strongly density-dependent
repulsive force that generates saturation at the right energy and density.

In summary, the leading (NNLO) 3NF of ChPT is a remarkable contribution. 
It gives validation to, 
and provides a better framework for, 
3NFs which were proposed already five decades ago; it alleviates existing problems            
in few-nucleon reactions and the spectra of
light nuclei and, particularly, solves the nuclear matter problem.

Some problems, though, remain unresolved at this order (NNLO), such as 
the well-known `$A_y$ puzzle' in nucleon-deuteron                
scattering~\cite{EMW02,Epe02} which, however, will be solved
by the 3NF at N$^4$LO, see below.
                 
As discussed earlier, for the 2NF, it turned out to be necessary 
to go to order four or even five for convergence and high-precision predictions. 
Thus, the 3NF at N$^3$LO and N$^4$LO must be considered simply as a matter of consistency with the 2NF sector. 
At the same time, one hopes that its 
inclusion may result in improvements with the still unresolved problems.

\subsubsection{Next-to-next-to-next-to-leading order}
\label{sec_3nfn3lo}

\begin{figure}[t]\centering
\vspace*{-0.5cm}
%\hspace*{-1.5cm}
\scalebox{0.85}{\includegraphics{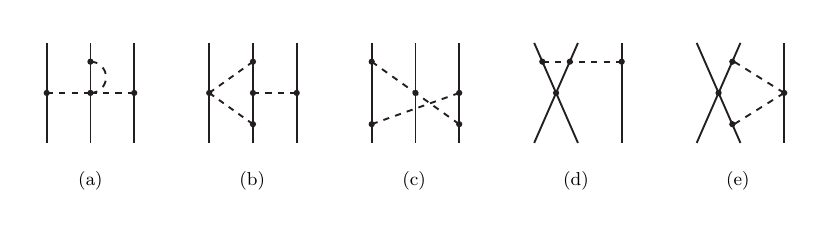}}
\vspace*{-0.75cm}
\caption{Leading one-loop 3NF diagrams at N$^3$LO.
We show one representative example for each of five topologies,
which are: (a) 2PE, (b) 1PE-2PE, (c) ring, (d) contact-1PE, (e) contact-2PE.
Notation as in Fig.~\ref{fig_hi}.}
\label{fig_3nf_n3lo}
\end{figure}

At N$^3$LO, there are loop and tree diagrams.
For the loops (Fig.~\ref{fig_3nf_n3lo}), we have
$L=1$ and, therefore, all $\Delta_i$ have to be zero
to ensure $\nu=4$. 
Thus, these one-loop 3NF diagrams can include
only leading order vertices, the parameters of which
are fixed from $\pi N$ and $NN$ analysis.
The diagrams
have been evaluated by the Bochum-Bonn group~\cite{Ber08,Ber11}.
The long-range part of the chiral N$^3$LO 3NF has been
tested in the triton and in three-nucleon scattering~\cite{Gol14}
leaving the $N-d$ $A_y$ puzzle unresolved.
 The long- and short-range parts of this
force have been applied in calculations of both SNM and NM~\cite{Kru13,Dri16,Heb15,DHS19,SM20}
as well as
in the structure of medium-mass nuclei~\cite{Hop19,Hut20} with, partially, great success.

\subsubsection{The 3NF at N$^4$LO}
\label{sec_3nfn4lo}

\begin{figure}[t]\centering
\vspace*{-0.5cm}
\scalebox{0.85}{\includegraphics{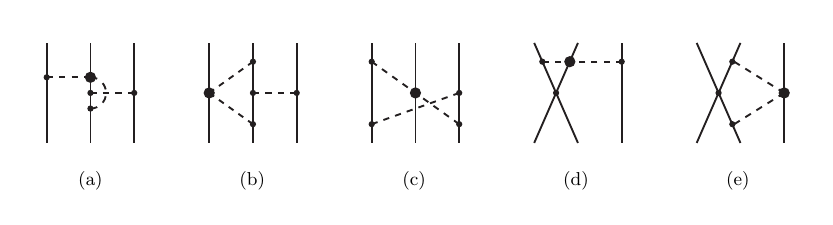}}
\vspace*{-0.75cm}
\caption{Sub-leading one-loop 3NF diagrams which appear at N$^4$LO
with topologies similar to Fig.~\ref{fig_3nf_n3lo}.
Notation as in Fig.~\ref{fig_hi}.}
\label{fig_3nf_n4loloops}
\end{figure}

\begin{figure}[t]\centering
\vspace*{-0.5cm}
\scalebox{0.9}{\includegraphics{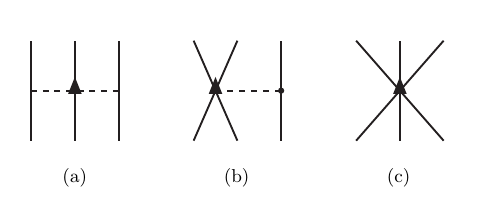}}
\vspace*{-0.3cm}
\caption{3NF tree graphs at N$^4$LO ($\nu=5$) denoted by: (a) 2PE, (b) 1PE-contact, and (c) contact. Notation as in Fig.~\ref{fig_hi}.}
\label{fig_3nf_n4lotrees}
\end{figure}

\begin{figure*}[t]\centering
\hspace*{-0.2cm}
\scalebox{1.2}{\includegraphics{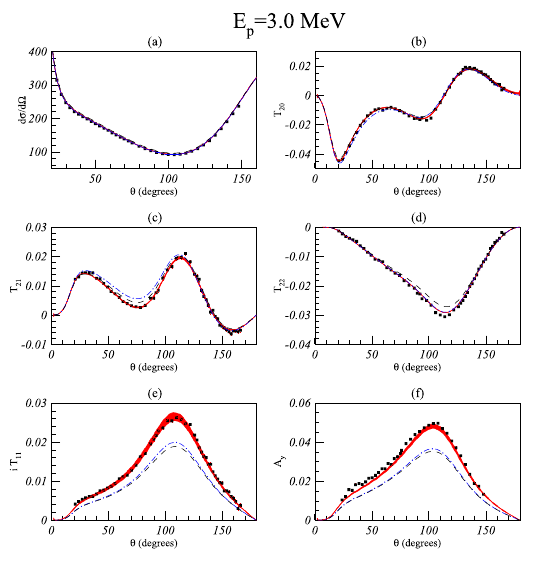}}
%\vspace*{-0.3cm}
\caption{Cross section and polarization observables of $p$-$d$ scattering at 3 MeV
proton energy as predicted by the AV18 2NF~\cite{WSS95} (black dashed lines),
 the AV18 2NF + UIX 3NF~\cite{Pud95}
(blue dash-dotted lines),
and calculations that include chiral N$^4$LO 3NF contact terms~\cite{Gir19} (red bands). 
The width of the red bands reflect the uncertainty when the cutoff varies between $\Lambda$=200 MeV and 500 MeV.
Data from Ref.~\cite{Shi95}. (Figure reproduced 
from Ref.~\cite{Gir19} with permission.)}
\label{fig_ay}
\end{figure*}

In regard to some unresolved issues, one may go ahead and look
at the next order of 3NFs, which is N$^4$LO or $\nu=5$.
The loop contributions that occur at this order
are obtained by replacing in the N$^3$LO loops
one vertex by a $\Delta_i=1$ vertex (with LEC $c_i$), Fig.~\ref{fig_3nf_n4loloops},
which is why these loops may be more sizable than the N$^3$LO loops.
The 2PE, 1PE-2PE, and ring topologies have been evaluated~\cite{KGE12,KGE13} so far.
In addition, we have three `tree' topologies (Fig.~\ref{fig_3nf_n4lotrees}), which include
a new set of 3N contact interactions that has been derived
by the Pisa group~\cite{GKV11}.
{\it The N$^4$LO 3NF contacts have been applied with success in calculations of few-body reactions at low energy 
solving the $p$-$d$ $A_y$ puzzle~\cite{Gir19}}, Fig.~\ref{fig_ay}.
This is an outstanding success in view of the  $A_y$ puzzle's long history~\cite{EMW02}.

\subsection{Four-nucleon forces}

\begin{figure}[t]\centering
%\vspace*{-1.5cm}
\scalebox{0.68}{\includegraphics{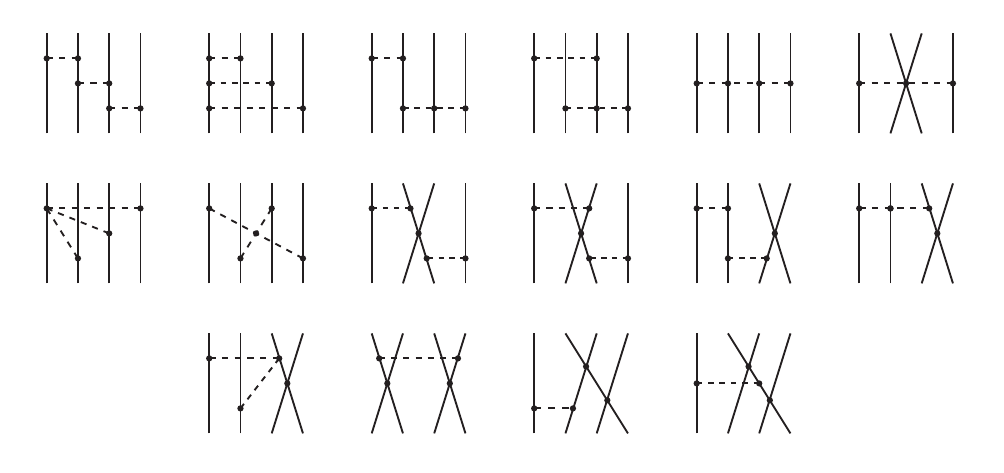}}
\vspace*{-0.3cm}
\caption{Leading four-nucleon force at N$^3$LO.}
\label{fig_4nf_n3lo}
\end{figure}

For connected ($C=1$) $A=4$ diagrams, Eq.~(\ref{eq_nu}) yields
\begin{equation}
\nu = 4 + 2L + 
\sum_i \Delta_i \,.
\label{eq_nu4nf}
\end{equation}
%HERE8
We then see that the first (connected) non-vanishing 4NF is generated at $\nu = 4$ (N$^3$LO), with                   
all vertices of leading type, Fig.~\ref{fig_4nf_n3lo}. 
This 4NF has no loops and introduces no novel parameters~\cite{Epe07}.

For a reasonably convergent series, terms                      
of order $(p/\Lambda_b)^4$ should be small and, therefore,              
chiral 4NF contributions are expected to be very weak.    
This has been confirmed in calculations of the energy of              
$^4$He~\cite{Roz06} as well as
NM and SNM~\cite{Kru13}.

The effects of the leading chiral 4NF in SNM and pure NM have been
worked out by Kaiser {\it et al.}~\cite{Kai12,KM16}.

\section{Uncertainty quantification}
\label{sec_uncert}

When applying chiral two- and many-body forces in {\it ab initio} calculations producing predictions for observables of nuclear structure and reactions, 
major sources of uncertainties are~\cite{FPW15}:
\begin{enumerate}
\item
Experimental errors of the input $NN$ data that the 2NFs are based upon and the input
few-nucleon data to which the 3NFs are adjusted.
\item
Uncertainties in the Hamiltonian due to 
       \begin{enumerate}
       \item
       uncertainties in the determination of the $NN$ and $3N$ contact LECs,
       \item
       uncertainties in the $\pi N$ LECs, 
       \item
       regulator dependence, 
       \item
       EFT truncation error.
       \end{enumerate}
\item
Uncertainties associated with the few- and many-body methods applied.
\end{enumerate}

\begin{enumerate}
\item
Experimental errors of the input $NN$ data that the 2NFs are based upon and the input
few-nucleon data to which the 3NFs are adjusted. \\
In Ref.~\cite{Nav+14}, the authors systematically investigated this error propagation by constructing a family of
smooth local potentials the parameters of which carry the uncertainties implied by the errors in the $NN$ data.
With 205 Monte Carlo samples of these potentials, they found an uncertainty of 15 keV for the triton binding
energy. In a more recent study~\cite{Nav+15} that used 33 Monte Carlo samples, 
 the uncertainty of 15 keV for the triton binding energy was reproduced, and the uncertainty for
the $^4$He binding energy was detrmined to be 55 keV. Our assessment is that statistical error propagation from the $NN$ input data
 is negligible as compared to uncertainties from other sources discussed below.\\
Regarding the propagation of
experimental errors from the few-nucleon data used to fit the 3NF contact terms, this uncertainty can be made very small
by adjusting the 3NFs to data with very small experimental errors; for instance, the measured
binding energy of the triton is 8.481795 $\pm$ 0.000002 MeV, which will lead to negligible propagation.
\item
Uncertainties in the Hamiltonian due to 
       \begin{enumerate}
       \item uncertainties in the determination of the $NN$ and $3N$ contact LECs.\\
We have fitted the $NN$ contact LECs to the $NN$ data below 100 MeV at LO and NLO, below 190 MeV at NNLO, and
below 290 MeV at N3LO and N4LO. Based on our experience~\cite{Mar13}, we do not expect significant errors from variations of the fit intervals at the low energies appropriate for chiral EFT. \\
As for the contact 3NF LECs, they can be fixed through a number of different procedures, that is, fitting different observables. For observables that were not part of the fitting protocol, predictions obtained with LECs from different methods will be different, and such differences can be taken as a measure of the uncertainty from this source.
       \item
       Uncertainties in the $\pi N$ LECs. \\
 Since the high-precision determination of the $\pi N$ LECs from Roy-Steiner analyses~\cite{Hof15, Hof16}, this uncertainty can essentially be ignored.
       \item
       Regulator dependence, see discussion below.
       \item
       EFT truncation error, see discussion below.
       \end{enumerate}
\item
Uncertainties associated with the few- and many-body methods applied.\\ 
This is unrelated to chiral EFT. While few-body systems can be solved exactly, predictions for heavier nuclei and nuclear matter will carry this uncertainty regardless the chosen method. Many-body theory practitioners may probe the size of this uncertainty by comparing their predicted observable with predictions from different many-body methods and otherwise identical input, if available.
\end{enumerate}

The choice of the regulator function and its cutoff parameter creates uncertainty.  
Originally, cutoff variations were perceived as a demonstration of the uncertainty at a given order
(equivalent to the truncation error).
However, in various investigations~\cite{Sam15,EKM15a} it has been shown that this is not correct and that cutoff variations,
in general, underestimate this uncertainty. 
Therefore, the truncation error is better determined by sticking literally to what
 `truncation error' means, namely, the error due to
 ignoring contributions from orders beyond the given order $\nu$. 
 The largest such contribution is the one of order $(\nu + 1)$,
 which one may, therefore, consider as representative for the magnitude of what is left out.

 This suggests that the truncation error at order $\nu$ can reasonably be estimated as
\begin{equation}
\Delta X_\nu (p)= |X_\nu (p) - X_{\nu+1} (p) | \,, 
\label{eq_err}
\end{equation}
where $X_\nu (p)$ denotes the prediction for observable $X$ at order $\nu$ and momentum $p$. If $X_{\nu+1}$ is not available, then one may use, 
\begin{equation}
\Delta X_\nu (p) = |X_{\nu-1} (p) - X_\nu (p) | Q \,,
\label{eq_err2}
\end{equation}
 with the expansion parameter $Q$ chosen as
  \begin{equation}
  Q = \max \left\{ \frac{m_\pi}{\Lambda_b}, \; \frac{p}{\Lambda_b} \right\} \,,
  \end{equation}
  where $p$ is the characteristic center-of-mass (CMS) momentum scale and $\Lambda_b$
  the breakdown scale.
 
Alternatively, one may also apply the more elaborate scheme suggested in Ref.~\cite{EKM15a}
where  the truncation error at, e.g.,  N$^3$LO is  calculated in the following way:
   \begin{eqnarray}
  \Delta X_{\rm N^3LO}(p) &=& \max \left\{ Q^5 \times \left| X_{\rm LO}(p) \right|, \;\;
  Q^3 \times \left| X_{\rm LO}(p) - X_{\rm NLO}(p) \right|, \;\;
  Q^2 \times \left| X_{\rm NLO}(p) - X_{\rm NNLO}(p) \right|, \;\;
        \right.      \\  
        && \left.
   Q \times \left| X_{\rm NNLO}(p) - X_{\rm N^3LO}(p) \right|
   \right\} \,,
   \label{eq_error}
  \end{eqnarray}
  with $X_{\rm N^3LO}(p)$ denoting the N$^3$LO prediction for observable $X(p)$, etc..
  This more sophisticated formula is especially useful, whenever---by accident---the difference between the highest two orders is uncharacteristically small. ``Extrapolating'' the errors of the lower orders may then provide a more realistic estimate.

Note that one should not add up (in quadrature) the uncertainties due to regulator dependence and the truncation error, because they are not independent. In fact, it is appropriate to 
leave out the uncertainty due to regulator dependence entirely and just focus on the truncation error~\cite{EKM15a}. The latter should be estimated using the same cutoff 
in all orders considered.
In summary, the truncation error is the dominant source of (systematic) error that can be reliably estimated in the EFT approach. 

The above is not a comprehensive statistical analysis.
Statistically robust  Bayesian methods have been developed and applied, see, for instance, Ref.~\cite{Sve+23}. The BUQEYE collaboration (Bayesian Uncertainty Quantification: Errors in Your EFT) works with projection-based, reduced-order emulators for applications in low-energy nuclear physics~\cite{Dri22}.

\section{Applications in the nuclear many-body problem}
\label{sec_manybod}

\begin{table}%\centering
\caption{Microscopic calculations of finite nuclei and nucleonic matter
applying chiral forces  published during the past decade (see Table ~\ref{tab_acro} for abbreviations and acronyms used).}
\smallskip
%\begin{tabular}{cccccccccc}
%\begin{tabular*}{\textwidth}{@{\extracolsep{\fill}}llllllllll}
%\footnotesize
\scriptsize
%\tiny
\begin{tabular}{llllllll}
\hline
\hline
\noalign{\smallskip}
 Year & Authors/collab.\  &  2NF & 3NF & $\Delta$s$^c$ & Nuclear systems(s) & Many-body method(s) & Ref(s).   \\
    &.  & Order (cutoff$^a$) locality$^b$ & Order (cutoff$^a$) locality$^b$ & &  &  \\
\hline
\hline
\noalign{\smallskip}
\multicolumn{4}{l}{\underline{Set 1: Momentum space, nucleonic matter:}} \\
2013 & Tews {\it et al.} & N$^3$LO (450, 500) nl & N$^3$LO (450, 500) nl + 4NF & No & NM & MBPT & \cite{Tew13} \\
2013 & Coraggio {\it et al.} & N$^3$LO (414-500) nl & NNLO (414, 500) nl  & No & NM & MBPT & \cite{Cor13} \\
2014 & Coraggio {\it et al.} & N$^3$LO (414-500) nl & NNLO (414, 500) nl  & No & SNM & MBPT & \cite{Cor14} \\
2014 & Hagen {\it et al.} & NNLO (500) nl & NNLO (400, 500) loc, nl  & No & SNM, NM & BHF & \cite{Hag14b} \\
2015 & Sammarruca {\it et al.} & NNLO, N$^3$LO (450-600) nl & NNLO (450-600) nl  & No & SNM, NM & BHF & \cite{Sam15} \\
2019 & Drischler {\it et al.} & NNLO, N$^3$LO (450, 500) nl & NNLO, N$^3$LO  (450, 500) nl  & No & SNM, NM & MBPT & \cite{DHS19} \\
2020 & Drischler {\it et al.} & LO-N$^3$LO (450, 500) nl & NNLO, N$^3$LO  (450, 500) nl  & No & NM & MBPT & \cite{Dri20} \\
2021 & Sammarruca {\it et al.} & LO-N$^3$LO (450) nl & NNLO-N$^3$LO (450) nl  & No & NM & BHF & \cite{SM21a} \\
2021 & Sammarruca {\it et al.} & LO-N$^3$LO (450) nl & NNLO-N$^3$LO (450) nl  & No & SNM & BHF & \cite{SM21b} \\
2021 & Keller {\it et al.} & NNLO, N$^3$LO (450, 500) nl & NNLO, N$^3$LO (450, 500) nl & No & NM & MBPT & \cite{Kel21} \\
2023 & Keller {\it et al.} & NNLO, N$^3$LO (450, 500) nl & NNLO, N$^3$LO (450, 500) nl & No & ASNM & MBPT & \cite{Kel23} \\
\noalign{\smallskip}
\multicolumn{4}{l}{\underline{Set 2: Momentum space, finite nuclei (and nucleonic matter):}} \\
2013 & Ekstr\"om {\it et al.} & NNLO (500) nl & NNLO (500) nl  & No & 
$^{10}$Be - $^{56}$Ca, NM & NCSM, CC & \cite{Eks13} \\
2013 & Hergert {\it et al.} & N$^3$LO (500) nl & NNLO (400) loc  & No & $^{4}$He - $^{56}$Ni & IM-SRG & \cite{Her13} \\
2014 & Binder {\it et al.} & N$^3$LO (500) nl & NNLO (350, 400) loc & No  & $^{16}$O - $^{132}$Sn & SRG, CC & \cite{Bin14} \\
2015 & Ekstr\"om {\it et al.} & NNLO (450) nl & NNLO (450) nl  & No & 
$^{4}$He - $^{40}$Ca, SNM & CC & \cite{Eks15} \\
2017 & Stroberg {\it et al.} & N$^3$LO (500) nl & NNLO (400) loc  & No & C - Ni & IM-SRG & \cite{Str17} \\
2017 & Simonis {\it et al.} & N$^3$LO (1.8, RG evolved) nl & NNLO (2.0) nl & No & $^{4}$He - $^{78}$Ni & IM-SRG & \cite{Sim17} \\
2018 & Morris {\it et al.} & N$^3$LO (1.8, RG evolved) nl & NNLO (2.0) nl & No & $^{100-111}$Sn, $^{105}$Te & IM-SRG & \cite{Mor18} \\
2018 & Ekstr\"om {\it et al.} & NNLO (450, 500) nl & NNLO (450, 500) nl  
& Yes & 
$^{4}$He - $^{48}$Ca, SNM, NM & CC & \cite{Eks18} \\
2019 & Hoppe {\it et al.} & NLO-N$^3$LO (420-500) nl & NNLO-N$^3$LO (420-500) nl & No & $^{4}$He - $^{68}$Ni & IM-SRG & \cite{Hop19} \\
2020 & Huether {\it et al.} & NLO-N$^3$LO (450-550) nl & NNLO-N$^3$LO (450-550) nl  & No & $^{16}$O - $^{78}$Ni & IM-SRG & \cite{Hut20} \\
2020 & Soma {\it et al.} & N$^3$LO (500) nl & NNLO (500/650) lnl & No & Li - Ni & NCSM, ADC(3) & \cite{Som20} \\
2020 & Jiang {\it et al.} & NNLO (394, 450) nl & NNLO (394, 450) nl  
& Yes & $^{4}$He - $^{132}$Sn, SNM, NM & CC & \cite{Jia20} \\
2022 & Miyagi {\it et al.} & N$^3$LO (1.8, RG evolved) nl & NNLO (2.0) nl & No & $^{132}$Sn, $^{127}$Cd & HF-MBPT, IM-SRG & \cite{Miy22} \\
2022 & LENPIC & LO-N$^4$LO$^+$ (400-550) nl & NNLO (400-550) nl  
& No & $^{3}$H - $^{48}$Ca & NCCI, IM-SRG & \cite{Mar22} \\
\noalign{\smallskip}
\multicolumn{4}{l}{\underline{Set 3: Configuration space calculations:}} \\
2014 & Gezerlis {\it et al.} & LO-NNLO (1.0-1.2) loc & None
& No & NM & AFDMC, MBPT & \cite{Gez14} \\
2016 & Tews {\it et al.} & LO-NNLO (1.0-1.2) loc & NNLO (1.0-1.2) loc
& No & NM & AFDMC & \cite{Tew16} \\
2016 & Lynn {\it et al.} & NLO-NNLO (1.0-1.2) loc & NNLO (1.0-1.2) loc
& No & $^3$H-$^4$He, NM& GFMC & \cite{Lyn16,Lyn17} \\
2016 & Piarulli {\it et al.} & LO-N(3)LO$^d$ (0.8-1.2) loc & None
& Yes & $^3$H-$^6$Li & HH, GFMC & \cite{Pia16} \\
2018 & Piarulli {\it et al.} & N(3)LO$^d$ (1.0-1.2) loc & NNLO (1.0-1.2) loc
& Yes & $^3$H-$^{12}$C & HH, GFMC & \cite{Pia18} \\
2018 & Lonardoni {\it et al.} & LO-NNLO (1.0-1.2) loc & NNLO (1.0-1.2) loc
& No & $^3$H-$^{16}$O & AFDMC, GFMC & \cite{Lon18} \\
2018 & Tews {\it et al.} & LO-NNLO (1.0) loc & NNLO (1.0) loc
& No & NM & AFDMC & \cite{Tew18} \\
2020 & Lonardoni {\it et al.} & NNLO (1.0) loc & NNLO (1.0) loc
& No & SNM, NM & AFDMC & \cite{Lon20} \\
2022 & Lovato {\it et al.} & N(3)LO$^d$ (1.0, 1.2) loc & NNLO (1.0, 1.2) loc
& Yes & NM & BHF, FHNC, AFDMC & \cite{Lov22} \\
 \hline
\hline
\noalign{\smallskip}
%\end{tabular*}
\end{tabular}

\footnotesize
$^a$ Cutoff numbers above 100 are in units of MeV, below 100 in units of fm.
\\
$^b$ loc = local, nl = nonlocal, lnl = local/nonlocal.
\\
$^c$ $\Delta(1232)$-isobar excitations included, yes or no?
\\
$^d$ 2PE at NNLO, contacts at N$^3$LO.
\label{tab_calcs}
\end{table}

\subsection{The {\it ab initio} approach to finite nuclei}
\label{sec_nuc}

The tenet of microscopic nuclear theory is that atomic nuclei
can be accurately described as collections of point-like nucleons interacting via
two- and many-body forces obeying nonrelativistic quantum 
mechanics---the forces being fixed in free-space scattering.
The microscopic or {\it ab initio} approach to nuclear structure
and reactions is then defined as calculating the properties of nuclei
in accordance with the tenet.
Before we will report on the latest calculations of this kind,
we will briefly review the history of this approach, because some
mistakes of history are being repeated today.
 
\subsubsection{Historical perspective}

The microscopic approach to nuclear structure is almost as old as 
nuclear physics itself. Brueckner and co-workers introduced Brueckner theory as early as 1954~\cite{BLM54} and performed the first semi-realistic microscopic nuclear matter 
calculation in 1958~\cite{BG58}. 
Already that same year, Brueckner discussed  finite
nuclei proposing  the local
density approximation~\cite{BGW58}.

In the second half of the 1960's, 
one of the hottest topics in nuclear structure physics was calculating
the properties of
finite nuclei without recourse through nuclear matter using 
Brueckner-Hartree-Fock (BHF) theory.
The Oak Ridge National
Laboratory (ORNL) with its computer power played a leading role in this effort that was guided by Thomas Davies and Michel Baranger~\cite{Bar69,Dav69}.
BHF (and coupled cluster) calculations of finite nuclei continued into the early 1970s with work by the Bochum~\cite{KZ73} and the Bonn-J\"ulich groups~\cite{MMF75}.

In parallel to the above developments, research on the microscopic derivation of the shell-model effective interaction was conducted (again, applying Brueckner theory) that had been kicked off by Kuo and Brown in 1966~\cite{KB66}.

Applying the nucleon-nucleon ($NN$) potentials available at the time, 
the BHF approach reproduced about one half of the binding energies of closed-shell nuclei which, in the early phase, was seen
as a great success~\cite{Bar69}, but in the long run did not satisfy demands for
more quantitative predictions.
Therefore, a departure from the microscopic approach happened around 1973
as reflected most notably in a lead-talk by Michel Baranger at the
 International Conference on Nuclear Physics in Munich in 1973~\cite{Bar73}.
 
 The shell-model effective interaction suffered a similar fate at the
International Conference on Effective Interactions 
and Operators in Nuclei in Tucson, Arizona, in 1975, organized
by Bruce Barrett~\cite{Bar75}.

And so it happened that in the early 1970s, the microscopic approach was abandoned and replaced by phenomenological
effective interactions (also know as mean-field models):
the Skyme interaction~\cite{Sky59} as revived by Vautherin and co-workers~\cite{VB72,Vau73}, the Gogny force~\cite{Gog73,DG80}, and the relativistic  
mean-field model of Walecka~\cite{Wal74,SW86}.

Ironically, the calculations with those effective interactions continued to be called ``microscopic'', for which John Negele had provided
the (debatable) justification in his Ph.D.\  thesis of 1970~\cite{Neg70}. Before calculating finite nuclei in the local density approximation, Negele had adjusted the insufficient binding of nuclear matter provided by the Reid soft-core
potential~\cite{Rei68} (11 MeV per nucleon) by hand to the presumed empirical value
of 15.68 MeV making ``the assumption that when higher-order corrections have been evaluated carefully, nuclear-matter theory will indeed produce the
correct binding''~\cite{Neg70}.
Negele had many followers~\cite{CS72,FN73,MHN75}.

However, the true ``deeper reason'' for those effective interactions was 
much simpler:
``To get better results!''~\cite{ano70}.
Clearly, the trends
that won popularity in the early 1970s were a setback for the fundamental
research in nuclear structure.

Nuclear structure theory at its basic level is not about fitting data
to get ``good'' results.
Fundamental nuclear structure theory is about answering the question:

\begin{quote}
{\it Do the same nuclear forces that explain free-space scattering experiments
also explain the properties of finite nuclei and nuclear matter when
applied in nuclear many-body theory?}
\end{quote}

One can think of many reasons why the basic tenet should be wrong.
According to the EMC effect, nucleons swell when inserted into nuclei which
might affect the force between nucleons~\cite{Ban92}. Meson exchange in the nuclear medium may be different than in free-space
for various reasons~\cite{KMS76,Wil79,BKR91}. The excitation of resonances, e.~g.\
$\Delta(1232)$ isobars, within the nucleon-nucleon interaction process is subject to changes when 
happening in a nuclear medium~\cite{GN75,Gre76,DC76,HM77}.
And many more ideas have been advanced, like e.~g., Brown-Rho scaling~\cite{BR91}.
In fact, in the 1970s, a popular belief was that medium effects on the $NN$
interaction may be the solution to the problem of lacking saturation~\cite{Mac89}.

Thus, it is a good question to ask whether medium modifications 
of nuclear forces show up in a noticeable way and/or are even needed for quantitative nuclear structure predictions.
But when we re-adjust the free-space forces arbitrarily to get ``good'' results, then we will never find out.
Note also that at some (high) energy and high density, the picture of
point-like nucleons is bound to break down~\cite{Ben23}. 
So, the issue behind the nuclear theory tenet is:
 Are the energies typically involved in conventional nuclear structure physics low enough to treat nucleons as structure-less objects?

To come back to history: the renunciation of the truly microscopic approach
lasted about two decades (essentially the 1970s and 80s).
Then, in the early 1990s, the microscopic
theory was revived by the Argonne-Urbana group~\cite{Wir93,Pud95}.
The crucial element in those new microscopic calculations was the
inclusion of a three-nucleon force (3NF).
The idea of a nuclear 3NF was not new. In fact, it is almost as old as 
meson theory itself~\cite{PH39}. But for years it had been considered
just an academic topic, too difficult to incorporate into actual
calculations, anyhow. But the persistent failure to saturate nuclear matter
at reasonable energies and densities, as well as the the underbinding of
nuclei, finally compelled nuclear structure physicists
to take a serious look at the 3NF issue, as explained in the
exemplary Comment by Ben Day~\cite{Day83} based upon
first test calculations by the Urbana group~\cite{CPW83}.
The 3NF definitely improved nuclear saturation and the properties
of light nuclei, even though nothing was perfect~\cite{Pud95} (cf.\ Fig.~\ref{fig_snm}).

\subsubsection{Recent advances}
\label{sec_recad}

After the year of 2000, two changes occurred.
First, the term `microscopic' was increasingly
replaced by the term {\it `ab initio'}~\cite{NVB00}---for reasons
nobody knows (but nothing to worry about because both mean the same).
Second and more importantly, nuclear forces based upon chiral effective field theory (EFT) entered the picture~\cite{EGM00,EM03}. This development was of great advantage.
Note that for a microscopic approach to be truly microscopic, the free-space forces need to be accurate.
But with phenomenological or meson-theoretic forces it was difficult to define what sufficiently accurate means, since the errors in those theories are unknown. 
However, in the framework of an EFT, the theoretical uncertainty can
be determined (cf.\ Sec.~\ref{sec_uncert}) and, thus, related with the accuracy of the predictions.
Hence, in the framework of an EFT:

\begin{quote}
{\it  Accurate free-space forces are forces that
predict experiment within the theoretical uncertainty of the EFT at the 
given order.}
\end{quote}

After 2000, it also became well established that predictive nuclear structure must include 
3NFs, besides the usual two-nucleon force (2NF) contribution.
Another 
advantage of chiral EFT is then that it generates 2NFs and multi-nucleon forces simultaneously and 
on an equal footing. In the $\Delta$-less theory~\cite{ME11,EHM09}, 3NFs occur for the first time at next-to-next-to-leading order
(NNLO) and continue to have additional contributions in higher orders.
 If an explicit
$\Delta$-isobar is included in chiral EFT ($\Delta$-full theory~\cite{ORK96,KGW98,KEM07,EKM08}), then 3NF contributions start already at next-to-leading order (NLO). 

In the initial phase, 
the 3NFs were typically adjusted in
 $A=3$ and/or the $A=4$ systems and the
 {\it ab initio} calculations were driven up to the oxygen region~\cite{BNV13}.
 It turned out that for $A \lea 16$ the ground-state energies and radii are predicted about right, no matter what type of chiral or phenomenological potentials were applied (local, nonlocal, soft, hard, etc.)
 and what the details of the 3NF adjustments to few-body systems were~\cite{BNV13,Rot11,Pia18,Lon18,Mar21,Mar22}.

 However, around the year of 2015, the picture changed, when the many-body practitioners were able to move up to medium-mass nuclei (e.~g., the calcium or even the tin regions). 
 Large variations of the predictions now occurred depending on what forces were used, and cases
 of severe underbinding~\cite{Lon17} as well as of substantial overbinding~\cite{Bin14} were observed. Ever since, the nuclear structure community understands that 
 accurate {\it ab initio} explanations of intermediate and heavy nuclei is an outstanding problem.
 
  There have been several attempts to 
  predict the properties of
  medium-mass nuclei with more accuracy.
Of the various efforts, we will now list four cases, which are representative
for the status, and will denote each case with a short label for
ease of communication. We restrict ourselves to cases, where 
the properties of medium-mass nuclei {\it and}\/ nuclear matter have been calculated, because the simultaneous description of both systems is part of the 
problem.\footnote{Other interesting cases are the models by
Soma {\it et al.}~\cite{Som20} and Maris {\it et al.}~\cite{Mar22}
for which, however, presently no nuclear matter results are available.}

 \begin{figure}[t]\centering
%\vspace*{-4.4cm}
\scalebox{0.60}{\includegraphics{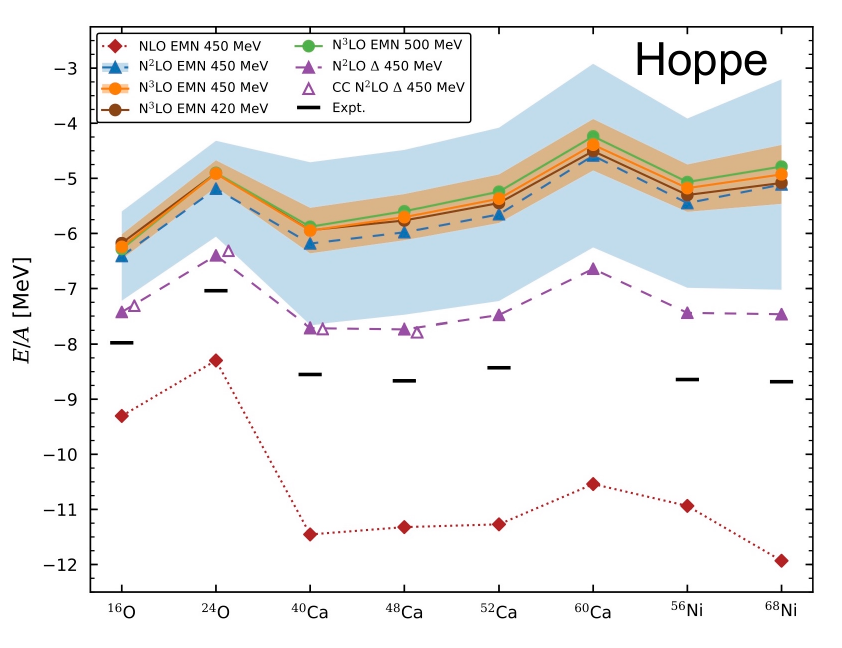}}

\scalebox{0.60}{\includegraphics{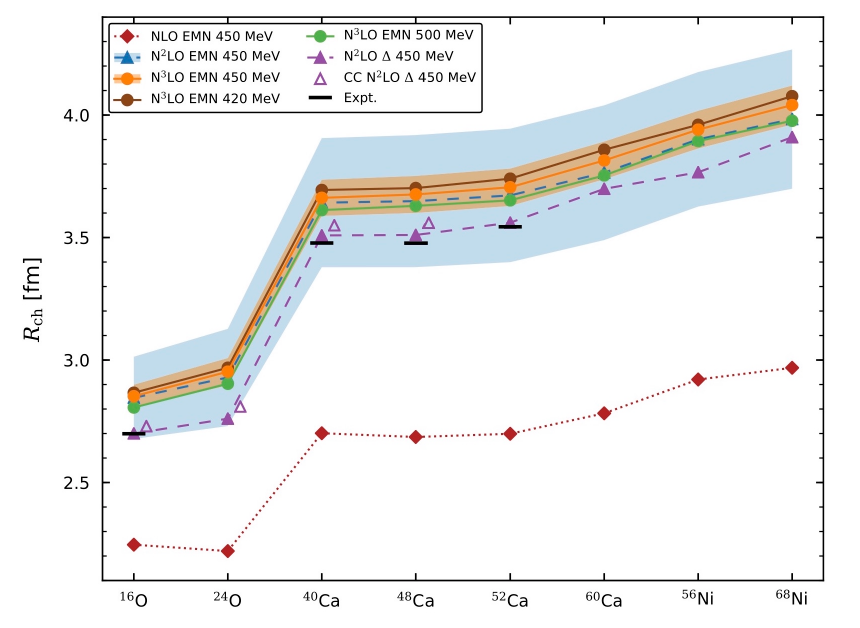}}

%\vspace*{-0.25cm}
\caption{Upper panel:
Ground-state energies per nucleon, $E/A$, of selected closed-shell oxygen, calcium, and nickel isotopes 
as obtained in the 
 ``Hoppe'' case~\cite{Hop19}.
Results are shown for various chiral interactions as denoted.
 The blue and orange bands give the NNLO and N$^3$LO uncertainty 
 estimates, respectively. 
 $\Lambda=450$ MeV in all cases except the green curve.
 Black bars indicate experimental data.
   Lower panel: 
 Same as upper panel, but for charge radii.
 (Reproduced from Ref.~\cite{Hop19} with permission.)
  }
 \label{fig_hop}
\end{figure}

 \begin{figure}[t]\centering
%\vspace*{-4.4cm}
\scalebox{1.2}{\includegraphics{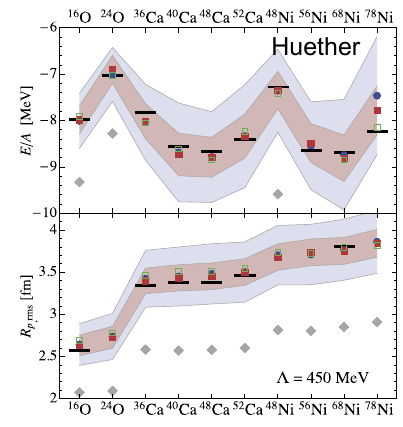}}
%\vspace*{-0.25cm}
\caption{
 Ground-state energies per nucleon (top panel)
 and point-proton rms radii (bottom panel)
 for selected medium-mass isotopes as obtained in the 
 ``H\"uther'' case~\cite{Hut20}.
The light blue and pink bands represent 
the theoretical uncertainties at NNLO and N$^3$LO, respectively.
$\Lambda=450$ MeV.
Black bars indicate the experimental data. (Figure courtesy of R. Roth)}
 \label{fig_roth}
\end{figure}

\begin{figure}[t]\centering
%\vspace*{-4.4cm}
\scalebox{0.7}{\includegraphics{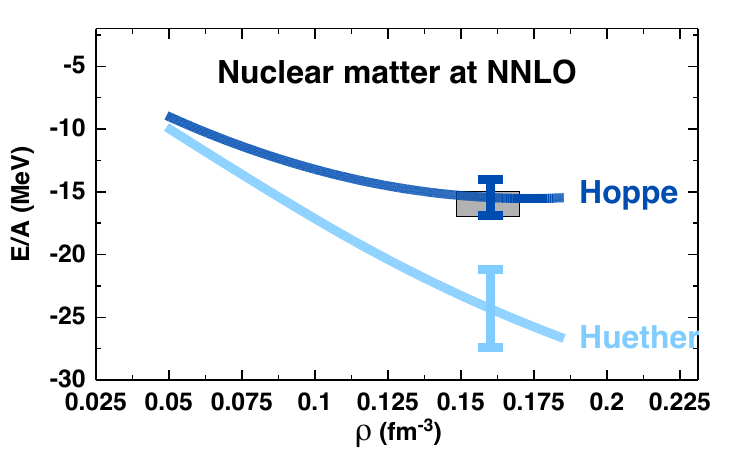}}
%\vspace*{-0.25cm}
\caption{
 Energy per nucleon, $E/A$, as a function of density, $\rho$, of SNM as obtained in
calculations with the 2NFs and 3NFs consistently at NNLO~\cite{SM20}. 
In the two cases shown, the 2NF is the same, while the 3NFs are the ones used
in the calculations of finite nuclei in the ``Hoppe'' and ``Huether'' cases as denoted.
$\Lambda=450$ MeV in both cases.
The error bars show the theoretical uncertainties around saturation,
which is expected to occur in the area of the gray box. }
 \label{fig_fsd}
\end{figure}

 \begin{itemize}
 \item {\bf ``Magic''}~\cite{Heb11,Heb21}:
 A seemingly successful interaction for the intermediate mass region 
 commonly denoted by ``1.8/2.0(EM)'' (sometimes
 dubbed ``the  Magic force''). It is a
 similarity renormalization group
   (SRG) evolved version of the N$^3$LO 2NF of Ref.~\cite{EM03} complemented by a NNLO 3NF adjusted to the triton binding energy and the 
 point charge radius  of $^4$He. With this force, the ground-state energies all the way up to the
 tin isotopes are reproduced perfectly---but with charge radii being on the smaller side~\cite{Sim17,Mor18}.
 Nuclear matter saturation is also reproduced reasonably well, but at a slightly too high saturation density~\cite{Heb11}.
 
 \item {\bf ``GO''}~\cite{Eks18,Jia20}:
 A family of $\Delta$-full 
NNLO potentials constructed by the G\"oteborg/Oak Ridge (GO) group.
The authors claim to obtain
 ``accurate binding energies and radii for a range of nuclei from $A=16$ to $A=132$,
 and provide accurate equations of state for nuclear matter''~\cite{Jia20}.
 
\item {\bf ``Hoppe''}~\cite{DHS19,Hop19}:
 Recently developed soft chiral 2NFs~\cite{EMN17} at NNLO and N$^3$LO
 complemented with 3NFs at NNLO and N$^3$LO, respectively, to fit the triton binding energy and nuclear matter saturation. These forces applied in
 in-medium similarity renormalization group (IM-SRG~\cite{Her16})
 calculations of finite nuclei up to $^{68}$Ni predict underbinding
and slightly too large radii~\cite{Hop19}, see Fig.~\ref{fig_hop}.

\item {\bf ``H\"uther''}~\cite{Hut20}: The same 2NFs used in ``Hoppe'', but with the 3NFs adjusted
 to the triton and $^{16}$O ground-state energies.  The interactions so obtained reproduce
 accurately experimental energies and point-proton radii of nuclei up to $^{78}$Ni~\cite{Hut20}, see Fig.~\ref{fig_roth}.
 However, when the 2NF plus 3NF combinations of ``H\"uther'' are utilized in nuclear matter, then overbinding and no saturation
 at realistic densities is obtained~\cite{SM20}, see Fig.~\ref{fig_fsd}.
\end{itemize}

\begin{figure}[t]\centering
%\vspace*{-2cm}
\scalebox{1.0}{\includegraphics{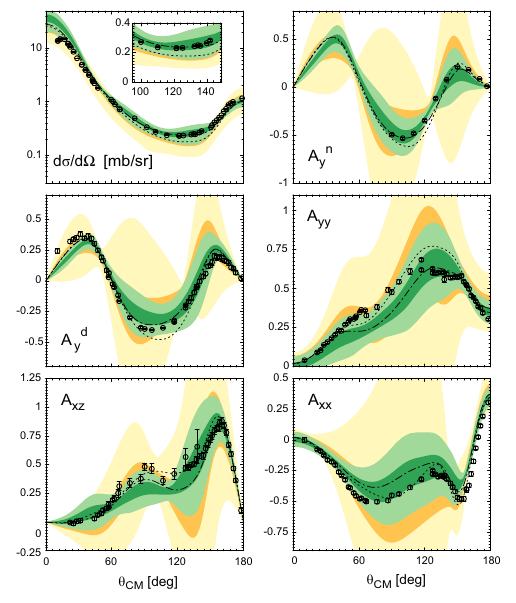}}
%\vspace*{-0.25cm}
\caption{
Predictions for the differential cross section, nucleon and
deuteron analyzing powers $A^n_y$ and $A^d_y$
 as well as deuteron tensor analyzing
powers $A_{yy}$, $A_{xz}$, and $A_{xx}$ in elastic nucleon–deuteron scattering
at a laboratory energy of 135 MeV at NLO (yellow bands) and
NNLO (green bands). 
 The light- (dark-) shaded bands indicate 95\% (68\%)
 confidence levels.
 The dotted (dashed)
lines show the results based on the CD-Bonn $NN$ potential~\cite{Mac01} 
 (CD-Bonn $NN$ potential in combination with the Tucson-Melbourne 3NF~\cite{CH01}).
 Black symbols represent the data together with their experimental errors.
(Reproduced from Ref.~\cite{Epe20}.)}
\label{fig_3N}
\end{figure}

  Obviously, in some cases, there appears to be a problem with achieving simultaneously accurate results
 for nuclear matter and medium-mass nuclei: In the ``Hoppe'' case, nuclear matter is saturated correctly, but nuclei are underbound; while in the
 ``H\"uther'' case, nuclei are bound accurately, but nuclear matter is overbound.
 Other cases seem to have solved this problem.
 But are they all truly {\it ab initio}?
 Our assessment:

 \begin{itemize}
 \item {\bf ``Magic'':}
 The construction of this force includes some inconsistencies.
The 2NF is SRG evolved, while the 3NF is not.
 Moreover, the SRG evolved 2NF
 is used like an original force with the induced 3NFs omitted. 
 Note that {\it ab inito} also implies that the forces are based upon
 some sort of theory in a consistent way. This is here not true
 and, thus, this case is not {\it ab initio}.
 
  \item {\bf ``GO'':}
  In Ref.~\cite{NEM21} it has been shown that the predictions by
  the $\Delta$-full $NN$ potentials at NNLO constructed by the 
G\H{o}teborg-Oak Ridge (GO) group~\cite{Jia20} are up to 40 times outside the 
theoretical error of  chiral EFT at NNLO.
So, they fail on accuracy.
 The reason for their favorable reproduction of the energies (and radii) of intermediate-mass nuclei,
can be traced to incorrect $P$-wave and $\epsilon_1$ mixing parameters~\cite{NEM21}. 
Thus, this case is especially far from being {\it ab initio}.
It is just a repetition of the mistakes of the early 1970s.

\item {\bf ``Hoppe'':}
In this case, the 2NF and 3NF forces are consistently chiral EFT based.
Moreover, the 2NFs are accurate. 
However, there is another accuracy aspect that is, in general, quietly 
ignored~\cite{KVR01,Mar09}: Are the 3NFs accurate?
The accuracy of the chiral 3NF at NNLO was thoroughly investigated in Ref.~\cite{Epe20} 
for a variety of cutoffs ranging from 400-550 MeV 
and large variations of the NNLO 3NF parameters, $c_D$ and $c_E$.
A typical result is shown in Fig.~\ref{fig_3N}.
It is seen that the 3$N$ data are reproduced within the truncation errors
at NNLO (green bands). On the other hand, it is also clearly seen that the theoretical
uncertainties are very large.
Moreover, it was found in Ref.~\cite{Epe20} that the cutoff dependence is weak and that the variations of the 3NF LECs $c_D$ and $c_E$ make only small differences relative to the
large uncertainties.
Thus, we can assume that the NNLO 3NFs used
in ``Hoppe'' will yield results
that lie within the NNLO uncertainties
shown in Fig.~\ref{fig_3N} by the green bands
and, consequently,
the ``Hoppe'' 3NF is accurate.
Hence, ``Hoppe'' passes on all accounts and is, therefore, truly {\it ab initio}.
\item {\bf ``H\"uther'':}
An assessment similar to ``Hoppe'' applies. Thus,  this case is
also truly {\it ab initio}.
 \end{itemize}
 
One may conclude that not all calculations, which have been published in the 
literature under the label of {\it ab initio}, are really {\it ab initio}.
Indeed, of the cases we considered here, only 50\% pass the test.
But we need to point out that even in the two cases we declared {\it ab initio},
there are concerns. The NNLO predictions by Hoppe and H\"uther
for finite nuclei barely overlap within their theoretical uncertainties and, for nuclear matter, they do not overlap at all.
Obviously, there are problems with the error estimates
and the uncertainties are much larger than the shown ones.
The true NNLO truncation errors of the Hoppe and H\"uther calculations
are probably as large as the differences between the two predictions.
 In this way, the two predictions are actually consistent with
each other, in spite of their seeming discrepancy. Chiral EFT is a
model-independent theory and, thus, different calculations at the same order
should agree within truncation errors.

At N$^3$LO the predictions differ even more. However, 
for current N$^3$LO calculations, a strong caveat is in place.
As pointed out in Ref.~\cite{EKR20}, 
there is a problem with the regularized 3NF at N$^3$LO (and higher orders)
in all present nuclear structure calculations. 
The N$^3$LO 3NFs currently in use are all regularized
by a multiplicative regulator applied to the 3NF expressions that are
derived from dimensional regularization.
This approach leads to a violation of chiral symmetry at N$^3$LO
and destroys the consistency between two- and three-nucleon forces~\cite{EKR20, Ep+22}. Consequently, all current calculations that
include a N$^3$LO 3NF contain an uncontrolled error and are, therefore,
unreliable. When a consistent regularization scheme has been found, the calculations have to be repeated.
At the present time, reliable predictions exist only at
NNLO, NLO, and LO.

 \subsubsection{The future: {\it ab initio} plus precision}

 \begin{figure}[t]\centering
%\vspace*{-2cm}
\scalebox{1.0}{\includegraphics{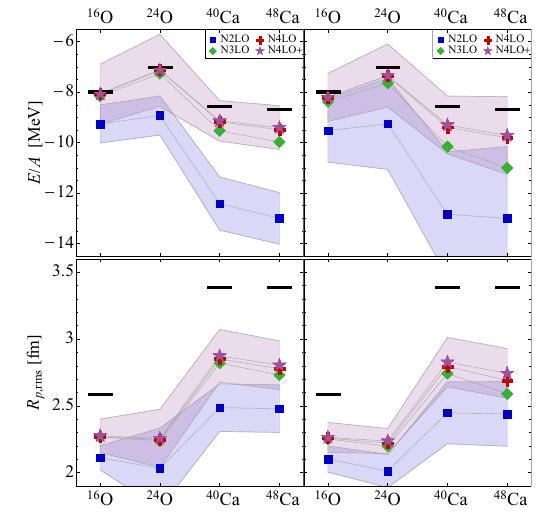}}
%\vspace*{-0.25cm}
\caption{Latest {\it ab initio} predictions by the LENPIC collaboration~\cite{Mar22}:
Ground-state energies and point-proton radii for doubly magic oxygen and
calcium isotopes obtained from the $NN$ potential of Ref.~\cite{RKE18}
complemented by NNLO 3NFs using a cutoff of 450 MeV (left-hand panel)
and of 500 MeV (right-hand panel). The blue squares
represent the predictions by complete NNLO calculations with the 
blue error bands showing the chiral NNLO truncation
uncertainties at the 95\% confidence level.
The green and purple points and pink error bands are based upon
incomplete calculations and are to be ignored.
Black bars indicate the experimental data.
(Reproduced from Ref.~\cite{Mar22} with permission.)
}
\label{fig_LENPIC}
\end{figure}

It is encouraging to know that at least a few correct {\it ab initio} 
calculations do exist. But these cases show that the precision at NNLO 
is very poor. The same
is true for the latest LENPIC calculations~\cite{Mar22}, see Fig.~\ref{fig_LENPIC} (which we did not include in our case study,
because nuclear matter results are lacking). 
At N$^3$LO (if one day correct such calculations become available)
the precision will most likely not be substantially better.

The purpose of the {\it ab initio} approach
is to test if the tenet of nuclear theory is correct or not.
Within huge errors as, e.~g. in Fig.~\ref{fig_3N}, any approach may
come out right. So, that is not a good basis for a reliable test.
We need more precision!
This is in particular true for the 3NF and the reproduction of the 3$N$ data,
which has been thoroughly investigated in Refs~\cite{Epe20,WGS22}
with the conclusion that, at N$^4$LO, there is a chance to achieve
the desirable precision---for several reasons.
The long- and intermediate-range
topologies of the 3NF at N$^4$LO are expected to be much larger than the corresponding ones at N$^3$LO because, at N$^4$LO, the subleading $\pi NN$ seagull vertex is involved
with LECs $c_i$, which are large~\cite{KGE12,KGE13}. 
This will provide the 3NF at N$^4$LO with more leverage as compared to  N$^3$LO.
Moreover, at N$^4$LO,
13 new 3$N$ contact terms occur~\cite{GKV11} with essentially free parameters introducing considerable flexibility~\cite{Gir19,WGS22} (see also Ref.~\cite{Gir23}).
Worth mentioning is also that,
 at N$^4$LO, the 3NF includes all 20 operators of the most general 3NF~\cite{Epe15}.
 Furthermore, the plentiful N$^4$LO 3NF terms may also provide what is needed
 to improve the status of the medium-mass nuclei and nuclear matter.
 
 Thus, the future of truly microscopic nuclear structure is to go for
 complete N$^4$LO calculations---a gigantic task.

\subsection{Infinite matter: A path to reach out to multiple systems, from nuclei to neutron stars}

As discussed in the previous sections, during the past two decades there has been remarkable progress in understanding nuclear forces at a fundamental level, through the concept of EFT~\cite{Wei90, Wei92}. Meson theoretic or entirely phenomenological models of the $NN$ interaction, augmented with phenomenological few-nucleon forces, are the typical paradigm used in the past.  Within the framework of chiral EFT, we discussed applications to medium-mass nuclei.

We now move to infinite matter. Although an idealized system, its equation of state (EoS) is a powerful tool for exploring nuclear interactions in the medium. Isospin-asymmetric nuclear matter (IANM) is characterized by the degree of neutron excess, all the way to pure NM. 
Because neutrons do not form a bound state, the presence of excess neutrons in a nucleus reduces the binding energy, namely, it is a necessary but destabilizing effect which gives rise to the symmetry energy. As a consequence, neutron-rich structures have common features, which explain the formation of a neutron skin (in isospin-asymmetric nuclei) or the outward pressure in dense systems with high neutron concentration, which supports neutron stars against gravitational collapse. Studies of nuclear interactions in systems with high or extreme neutron to proton ratio are crucial towards understanding of the neutron driplines, the location of which is not well known. The new Facility for Rare Isotope Beams (FRIB), operational since May 2022, is expected to increase the number of known rare isotopes from 3000 to about 6000~\cite{frib}.

After a brief discussion of SNM in Sec.~\ref{sec_snm}, we focus on the NM EoS (Sec.~\ref{sec_nm}) and the symmetry energy (Sec.~\ref{sec_nm_2}). A focal point of this section is a comparison of {\it ab initio} predictions with phenomenological and empirical findings. Since many years, several groups have sought constraints on the density dependence
of the symmetry energy. Intense experimental effort has been and continues to be devoted to this question using various measurements, which are typically analyzed with the help of correlations obtained through different parametrizations 
of  phenomenological models. Popular examples are the Skyrme forces (Refs.~\cite{skyrm1,skyrm2,skyrm3} are only a few of the many review articles on the Skyrme model), and relativistic mean-field models (RMF)~\cite{rmf1,rmf2,rmf3}. Reference~\cite{Bro00} and Ref.~\cite{PF19} are representative applications of Skyrme forces or RMF models, respectively, to the issues confronted here.  
Generally, the extracted constraints vary considerably depending on the methods employed.

We follow with a discussion on the neutron skin, see Sec.~\ref{sec_ns_skin}, and its sensitivity to the density slope of the symmetry energy at saturation. We underline the sensitivity of the discussed predictions to free-space $NN$ scattering phases. In Sec.~\ref{sec_ns_skin}, we also present an extended analysis of current issues related to recent extractions of the neutron skins in $^{208}$Pb and $^{48}$Ca from weak electron scattering measurements.

The density dependence of the symmetry energy is also a major component in the physics of neutron stars, in particular the radius of a medium-mass neutron star. Of course, a very large range of densities can be found in a neutron star, from the density
of iron in the outer crust up to several times normal nuclear density in the core, and thus no theory of hadrons can be applicable over the entire range. With that in mind,
state-of-the-art {\it ab initio} theories of SNM and NM can be taken as the foundation for any extension method, which will unavoidably
involve some degree of phenomenology. Note, though, that the radius of a 1.4 M$_{\odot}$ neutron star is sensitive to the pressure in neutron matter around saturation, and thus it is useful to constrain microscopic theories of the EoS at those densities where the theories are reliable~\cite{Universe}. In fact, the radius of light to medium neutron stars has been found to correlate with the density slope of the symmetry energy at saturation, with correlation coefficient of 0.87 for $ M = 1.0 M_{\odot}$ and 0.75 for  $ M = 1.4 M_{\odot}$~\cite{LR1}.

To advance our understanding of intriguing systems such as neutron skins
and neutron stars, it is important to build on the enormous progress in nuclear theory
since the days of one-boson-exchange or phenomenological $NN$
potentials and phenomenological 3NFs, selected with no clear scheme or guidance.  
Predictions from state-of-the-art nuclear theory favor a softer density dependence of the symmetry energy -- on the low-to-medium end of the spectrum cited in the literature -- and, naturally, smaller neutron skins and radii of canonical mass neutron stars.

\begin{figure*}[!t] 
\centering
\hspace*{-1cm}
\includegraphics[width=11.5cm]{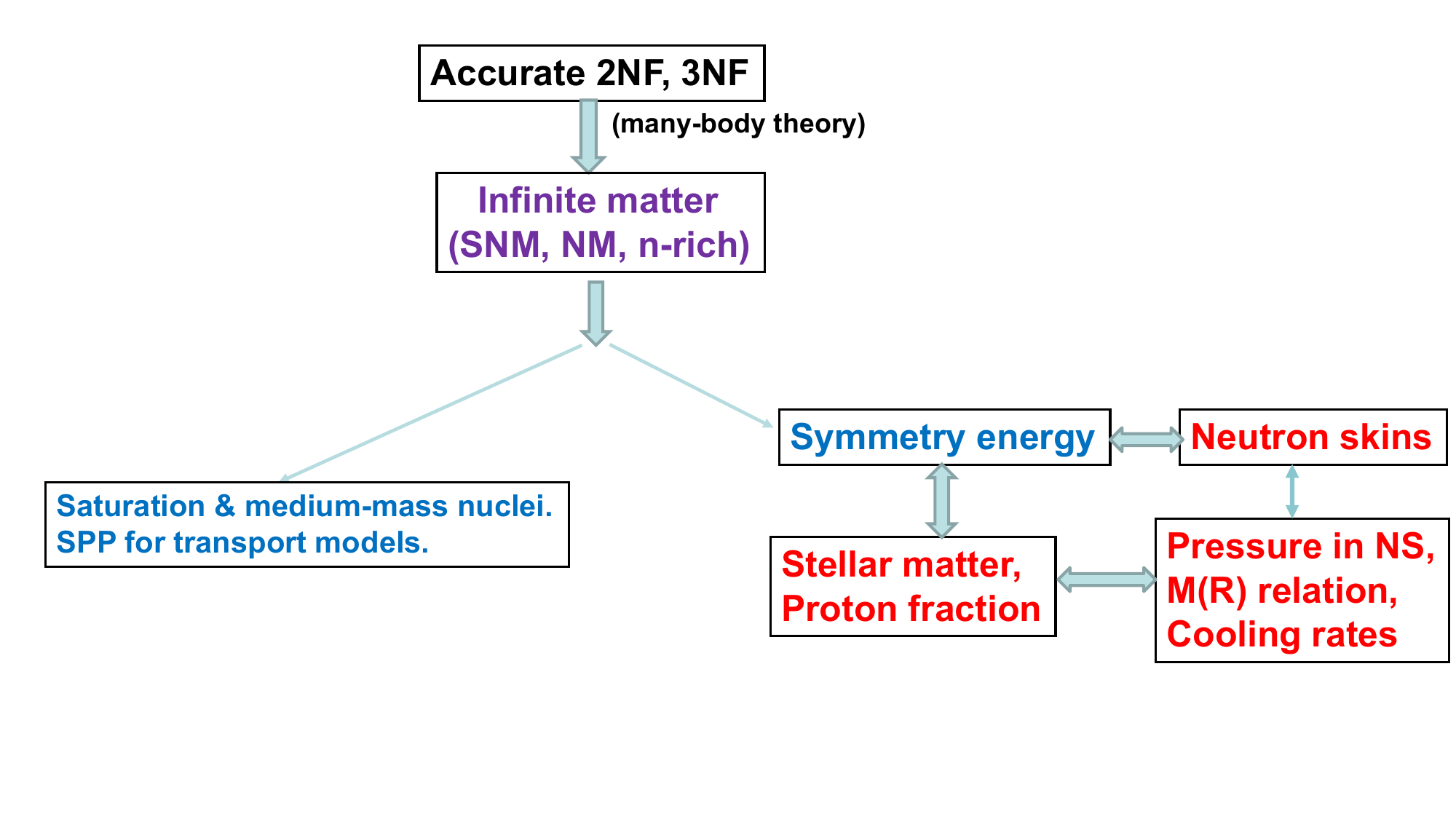}\hspace{0.01in} 
\vspace*{-0.5cm}
 \caption{ Schematic diagram showing the connections among the topics addressed in the remaining parts of this article. ``SPP "stands for single-particle potential.
}
\label{ddd}
\end{figure*}

The purpose and overarching theme of this section are to perform an analysis of existing literature addressing the nuclear symmetry energy, with the objective to identify and discuss current gaps or problems, and provide recommendations for future research. The diagram in Fig.~\ref{ddd} is a "flow chart" showing how the various applications we discuss are linked to one another and originate from a common source, state-of-the-art nuclear forces.

%Cutoff variations have sometimes been used to estimate contributions beyond truncation. However, they do not allow to estimate the impact of neglected long-range contributions. Also, due to the intrinsic limitations of the EFT, a meaningful %cutoff range is hard to estimate precisely, and often very limited. The method of Eq.~(\ref{eq_error}) allows to determine truncation errors from predictions at all lower orders, without the need to use cutoff variations. Comments concerning %regulator variations and error quantification were also presented in Sec.~\ref{sec_uncert}, below Eq.~\ref{eq_error}.

\subsubsection{Symmetric nuclear matter}
\label{sec_snm}

Constructing the equation of state (EoS) of infinite nuclear matter microscopically from state-of-the-art few-body interactions remains a very 
challenging problem in nuclear theory. The EoS gives fundamental
insight into nuclear forces in the medium, and is a crucial input in a variety of fields, ranging from heavy-ion (HI) reactions to astrophysical processes.

In Sec.~\ref{sec_nuc}, we discussed the current problems with reconciling SNM saturation and bulk properties of medium-mass nuclei, as well as recently proposed unsuccessful solutions. In this section, we wish to discuss several aspects related to the EoS of SNM, from bulk to single-particle properties~\cite{SM21b}. First, we will show order-by-order predictions for the EoS with errors quantified as described in Sec.~\ref{sec_uncert}.
 
After addressing bulk properties, we will study the impact of 3NFs on the single-particle potential. Single-particle energies, often parametrized in terms of effective masses, provide insight into both density and momentum dependence of the in-medium interaction, and  are an important part of the input for transport calculations. 

\paragraph{Theoretical tools.}
\label{tools}
In what follows, our highest-order predictions are obtained using $NN$ forces and 3NF consistently at N$^3$LO. The chiral $NN$ potential is the one from Ref.~\cite{EMN17}, with cutoff equal to 450 MeV. 
To calculate the energy per nucleon in nuclear (or neutron) matter, we use the nonperturbative particle-particle ladder approximation, which generates the leading-order terms in the hole-line expansion of the energy per particle. We compute the
single-particle spectrum self-consistently.
The Bethe-Goldstone (or Brueckner) equation for two particles having center-of-mass momentum $\vec{P}$, initial relative momentum $\vec{q_0}$, and starting energy $\epsilon_0$ in nuclear matter with Fermi momentum $k_F$ is 
\begin{equation}
G_{P,e_0,k_F}(q,q_0) = V(q,q_0, k_F) + \int_0^{\infty} dk k^2 \frac{V(q,k,k_F)Q(k,P,k_F)G(k,q_0,k_F)}
{\epsilon_0 - (\epsilon(\vec{P} + \vec{k}) + \epsilon(\vec{P} - \vec{k}) ) + i\delta }  \; ,
\label{BG}
\end{equation} 
where the Pauli operator, $Q$, has been angle-averaged, and 
 $V$ is the $NN$ potential augmented with effective 3NFs as density dependent potentials.
When calculating the single-particle potential, $U$, one must recall that, in order to avoid double-counting, 
a factor of $1/2$  must be applied to the density-dependent part of $V$ at the Hartree-Fock level~\cite{HF}:
\begin{equation}
V = V_{NN} + (1/2)V_{DD} \; ,
\label{HF}
\end{equation} 
where, again, the first term on the right-hand side is the actual $NN$ potential and the second one is a density-dependent effective 3NF. For the latter, we employ the density dependent
$NN$ interactions derived in Refs.~\cite{HKW10, KN18, KS19,Kai20} from the leading and subleading
chiral three-body force. This effective interaction is
obtained by summing one particle line over the occupied
states in the Fermi sea, such that 
the resulting $NN$ interactions can be expressed in analytical or semi-analytical
form with operator structures identical to those of free-space
$NN$ interactions. A comprehensive review on implementation and applications of 3NFs~\cite{Heb21} examines various approximations. We agree with the author of Ref.~\cite{Heb21} that the analytical or semi-analytical approaches, while 
having the advantage of not requiring large input files of 3NF matrix elements, can quickly become cumbersome. 
As for the quantitative impact from this approximation, in Ref.~\cite{SM21a} we examined different subleading contributions to the 3NF potential in neutron matter~\cite{Kai20_nn, Tre20} and found excellent agreement with the 3NF potentials from Ref.~\cite{Tew13}, where 3NFs are implemented differently, which we find encouraging.

Using partial wave decomposition, Eq.(\ref{BG}) can be solved for each partial wave using standard matrix inversion techniques. Note that the single-particle energy, $\epsilon$, contains the single-particle potential, $U$,  yet to be determined. Because the G-matrix depends on $U$ and $U$ depends on $G$, an iteration scheme is applied to obtain a self-consistent solution for $U$ and $G$, which are related as:
\begin{equation}
\label{uuu} 
U(k_{1},k_F) = \int  \ \langle \vec{q_{o}} | G_{P,e_0,k_F} | \vec{q_{o}} \rangle d^{3} k_{2}(\vec{q_{o}}, \vec{P})  \; ,
\end{equation}
where $k_{1,2}$ are sincle-particle momenta. Starting from some initial values and a suitable parametrization of the single-particle potential, a first solution is obtained for the G-matrix, which is then used in Eq.(\ref{uuu}). The procedure continues until convergence to the desired accuracy.

The energy per nucleon is then evaluated as:
\begin{equation}
\frac{E}{N} = \langle T (k_{1}) \rangle_{(k_{F})} + \langle U (k_{1},k_F) \rangle_{(k_{F})} \; ,
\end{equation}
where the averages of the single-nucleon kinetic ($T(k_{1})$) and potential ($U(k_{1},k_F)$) energies are taken over the Fermi sea. 

Some comments are in place regarding the choice of the many-body method we employ, namely the nonperturbative particle-particle ladder approximation.
In the traditional hole-line expansion, the particle-particle ladder diagrams comprise
the leading-order contributions. The next set of diagrams contains
 the three hole-line contributions, which
includes the third-order particle-hole (ph) diagram considered
in Ref.~\cite{Cor14}. The third-order hole-hole (hh) diagram (fourth
order in the hole-line expansion) was found to give a negligible
contribution to the energy per particle at normal density
regardless the cutoff (see Tables II and III of Ref.~\cite{Cor14}). The
ph diagram is relatively much larger, bringing in an uncertainty
of about $\pm$ 1 MeV on the potential energy per particle at normal
density.

\paragraph{Order by order predictions of the EoS.}
\label{sec_snm_1}

\begin{figure*}[!t] 
\centering
\hspace*{-1cm}
\includegraphics[width=7.5cm]{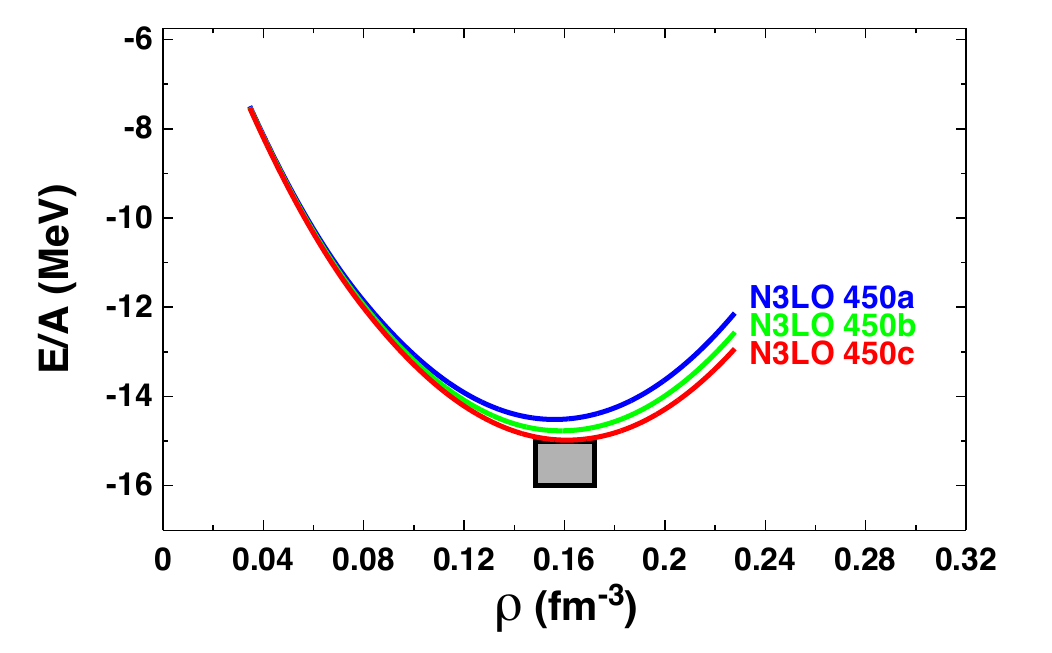}\hspace{0.01in} 
\vspace*{0.05cm}
 \caption{ Energy per particle as a function of density at N$^3$LO and cutoff equal to 450 MeV. The labels a, b, and c refer to the different sets of $c_D, c_E$ values given in Table~\ref{tab1}.
}
\label{eos_n3lo}
\end{figure*}

We begin with the study displayed in Fig.~\ref{eos_n3lo}. The curves are obtained with $\Lambda$ = 450 MeV  and the different sets of $c_D, c_E$ LECs displayed in Table~\ref{tab1}, of which set (c) produces the best saturating behavior. The method used to fit the $c_D, c_E$ LECs is described in the Supplemental Material of Ref.~\cite{DHS19}. The first step is to impose that the chosen $NN$ and 3NF combination reproduces the $^3$H binding energy. The result is a trajectory in the $c_D, c_E$ plane. The best pairs of  couplings are then selected looking at the SNM saturation point, see Fig.~\ref{eos_n3lo}.

\begin{figure*}[!t] 
\centering
\hspace*{-1cm}
\includegraphics[width=6.5cm]{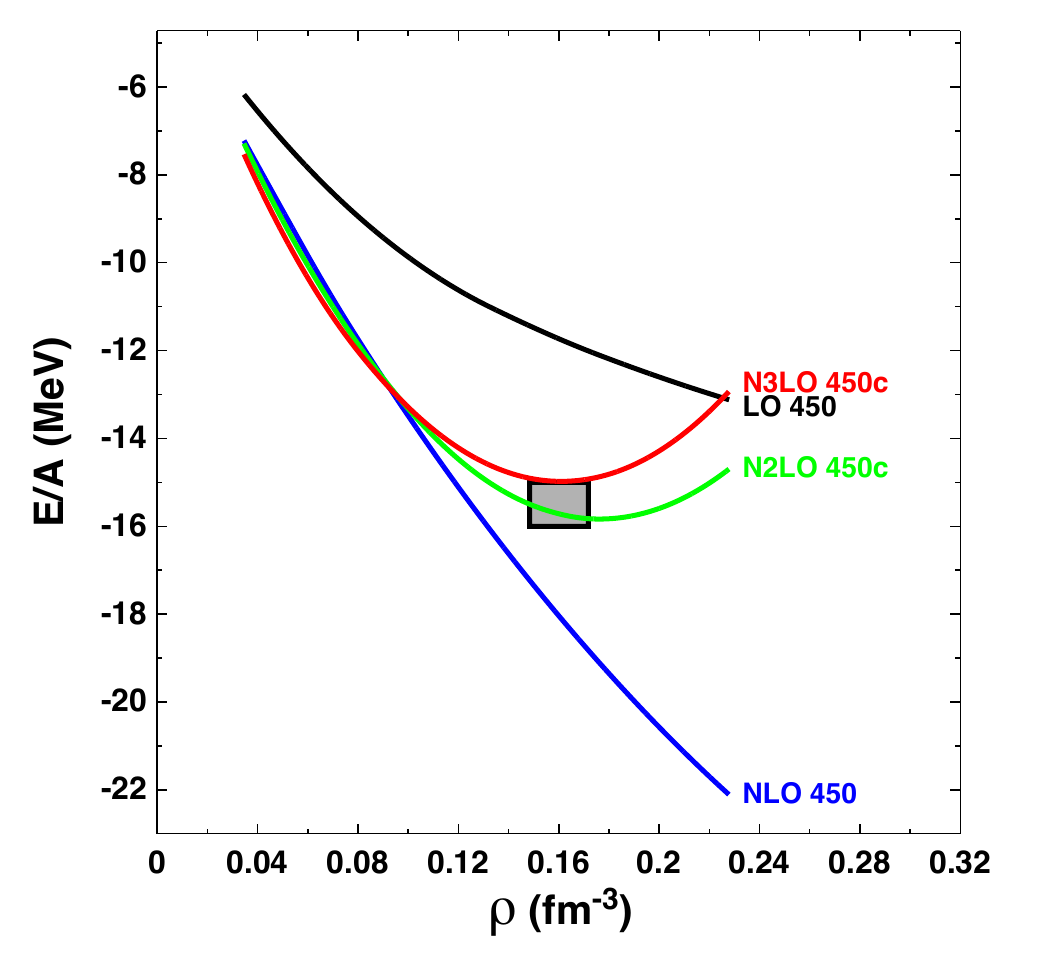}\hspace{0.01in}
 \includegraphics[width=8.5cm]{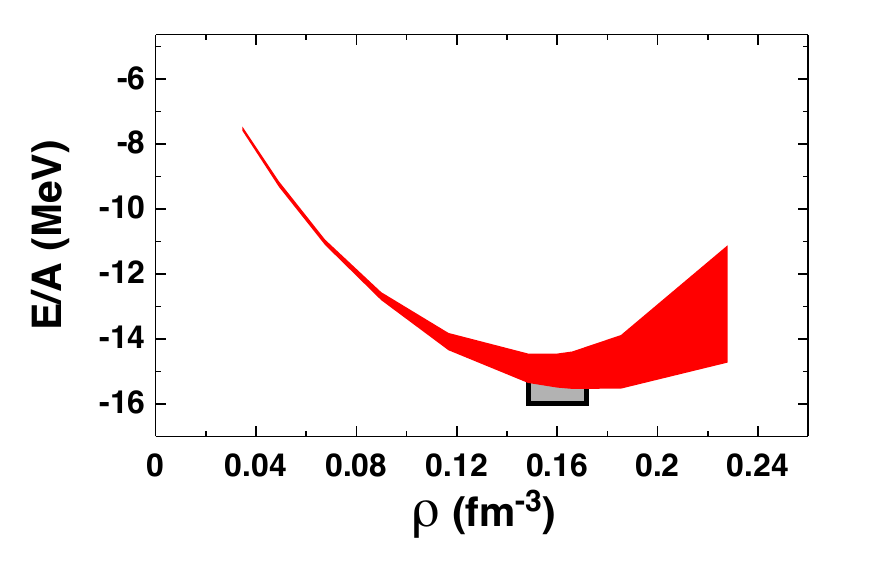}\hspace{0.01in} 
\vspace*{0.05cm}
 \caption{ Left: Energy per particle as a function of density from leading to fourth order of the chiral expansion. Right: Energy per particle as a function of density at fourth order of the chiral expansion. The band shows the uncertainty calculated from Eq.~\ref{eq_error}.
The cutoff is fixed at 450 MeV.}
\label{obo_c}
\end{figure*}

   On the left side of Fig.~\ref{obo_c}, we show the energy per nucleon from leading to fourth order.  While  the EoS has already a realistic behavior at the first order where 3NFs appear (N$^2$LO), there is a definite improvement when moving to N$^3$LO, for both saturation density and energy. This is an important validation of the predictive power of the chiral EFT -- of course, $NN$ data and the three-nucleon system must be described accurately for any subsequent many-body predictions to be meaningful. On the right side, we displays the predictions at N$^3$LO with the uncertainty band calculated from Eq.~\ref{eq_error}.
We note that our N$^3$LO(450) results for the energy per particle at saturation agree with those from Ref.~\cite{DHS19} within uncertainties.

\begin{table*}[t]
\caption{
Values of the LECs $c_{1,3,4}$, $c_D$, $c_E$, and $C_S$ and $C_T$
for different orders in the chiral EFT expansion. The momentum-space
cutoff $\Lambda$ is equal to 450 MeV.
The LECs $c_{1,3,4}$ are given in units of GeV$^{-1}$, 
$C_S$ and $C_T$ in units of $10^4$ GeV$^{-2}$,
and $c_D$ and $c_E$
 dimensionless. $c_D$, $c_E$ are taken from Ref.~\cite{DHS19}.}
\label{tab1}
\begin{tabular*}{\textwidth}{@{\extracolsep{\fill}}ccccccccc}
\hline
\hline
  & $\Lambda$ (MeV) & $c_1$ & $c_3$ & $c_4$ &  $c_D$ & $c_E$  & $C_S$  & $C_T$ \\
\hline    
\hline
N$^2$LO & 450 &  --0.74 & --3.61 & 2.44  &  (a) 2.25 &   0.07 & --- & --- \\
     &     &              &      &                              & (b)  2.50 &   0.1 & &  \\
     &     &              &      &                              & (c)   2.75 &   0.13 & &  \\    
\hline 
N$^3$LO & 450 & --1.07 & --5.32 & 3.56  &  (a)  0.00 &   -1.32  & -0.011828 &  -0.000010 \\
      &     &              &      &                              &(b)   0.25  &   -1.28 & &  \\
     &     &              &      &                              &  (c)  0.50 &   -1.25 &  & \\
\hline
\hline
\end{tabular*}
\end{table*}

\paragraph{Single-particle properties.}
\label{sec_snm_2}
While bulk properties of nuclear matter are very insightful for testing theories as well as providing a connection with bulk properties of nuclei, momentum- and density-dependent single-particle potentials (SPP) in nuclear matter provide complementary and more detailed information which is needed for heavy ion (HI) transport simulations.

Together with the SPP in neutron matter (NM), one can construct the momentum and density dependent SPP in IANM -- and thus the so-called symmetry potential -- to be used, for instance, in Boltzmann-Uehling-Uhlenbeck calculations of collective nuclear dynamics. A number of HI collision observables have been found to be sensitive to the symmetry potential, such as the neutron/proton ratio of pre-equilibrium nucleon emission, neutron-proton differential flow, and the proton elliptic flow at high transverse momenta. 

\begin{figure*}[!t] 
\centering
\hspace*{-1cm}
\includegraphics[width=7.0cm]{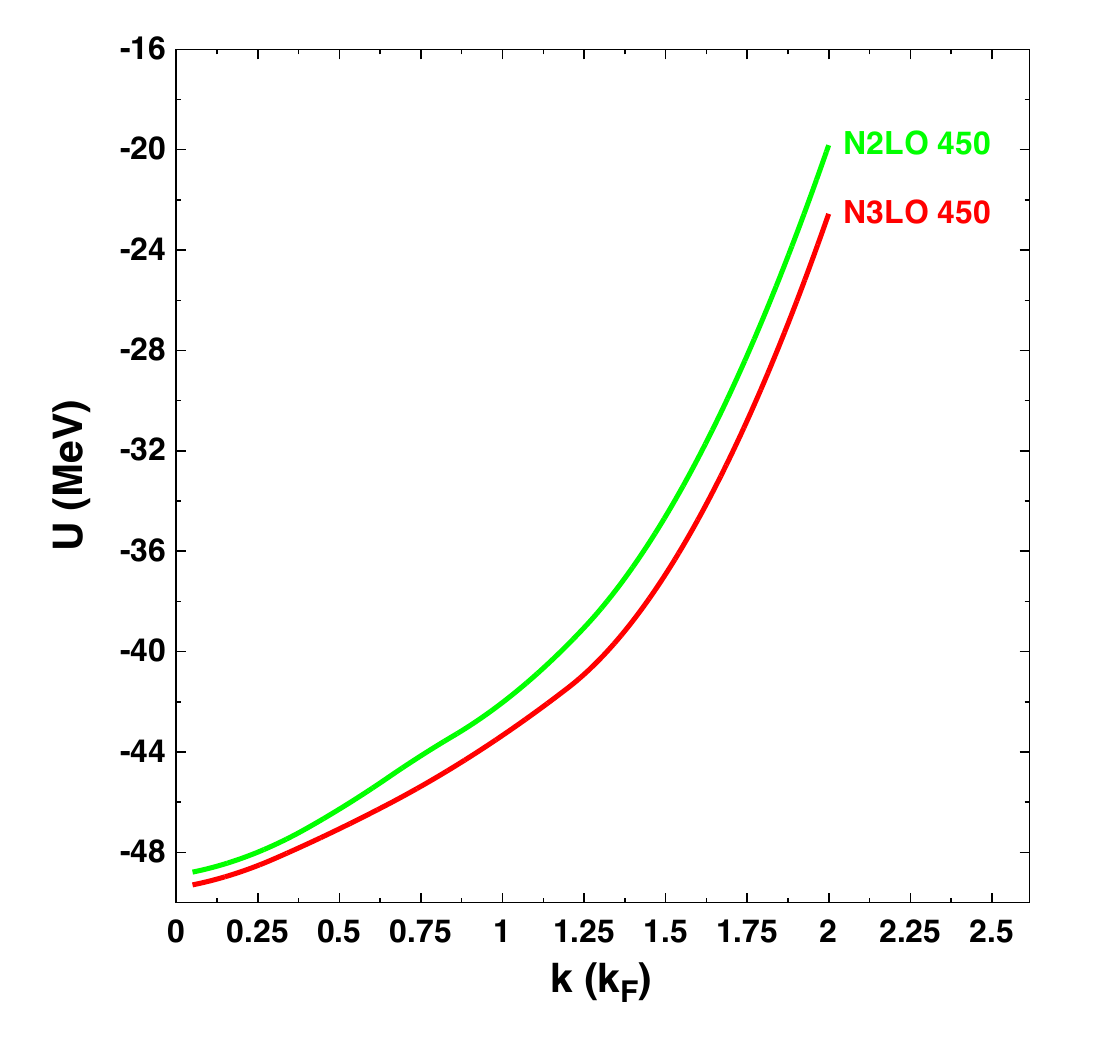}\hspace{0.01in} 
\includegraphics[width=7.0cm]{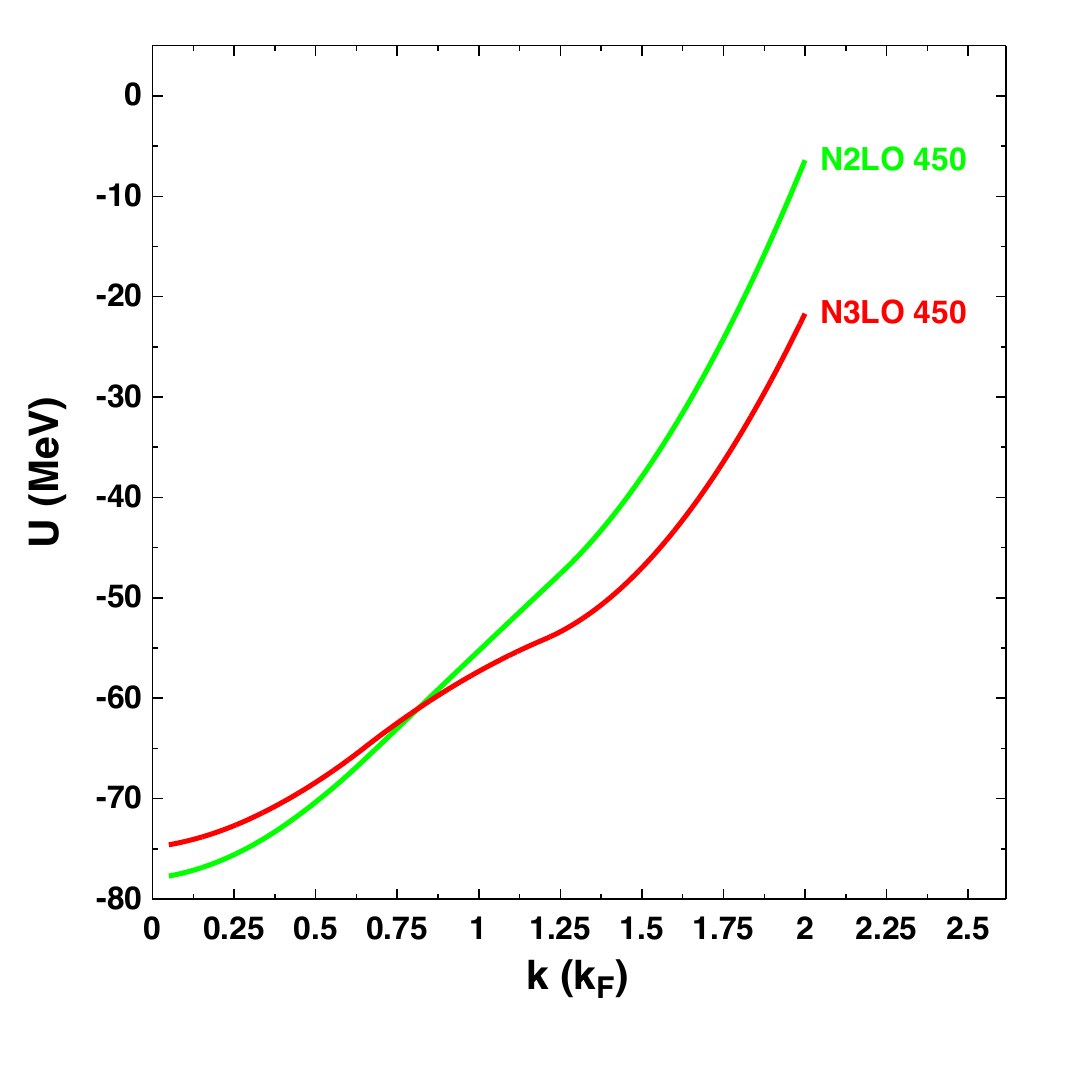}\hspace{0.01in} 
\vspace*{0.05cm}
\caption{Predictions for the SPP at N$^2$LO and N$^3$LO in SNM. The cutoff is fixed at 450 MeV. For the left (right) frame, the Fermi momentum is equal to $k_F$ = 1.0 (1.333) fm$^{-1}$.
}
\label{u_ord}
\end{figure*}

Next, we will take a look at the underlying SPP, derived self-consistently from the
Brueckner $G$-matrix and, thus, the EoS, to observe its momentum dependence and its changes with density and chiral order.
First, for two selected densities (saturation density and about 2/3 of it, corresponding approximately to $k_F$ = 1.0 fm$^{-3}$), we show the single-particle potential at third and fourth order, Fig.~\ref{u_ord}. 

\begin{figure*}[!t] 
\centering
\hspace*{-1cm}
\includegraphics[width=7.0cm]{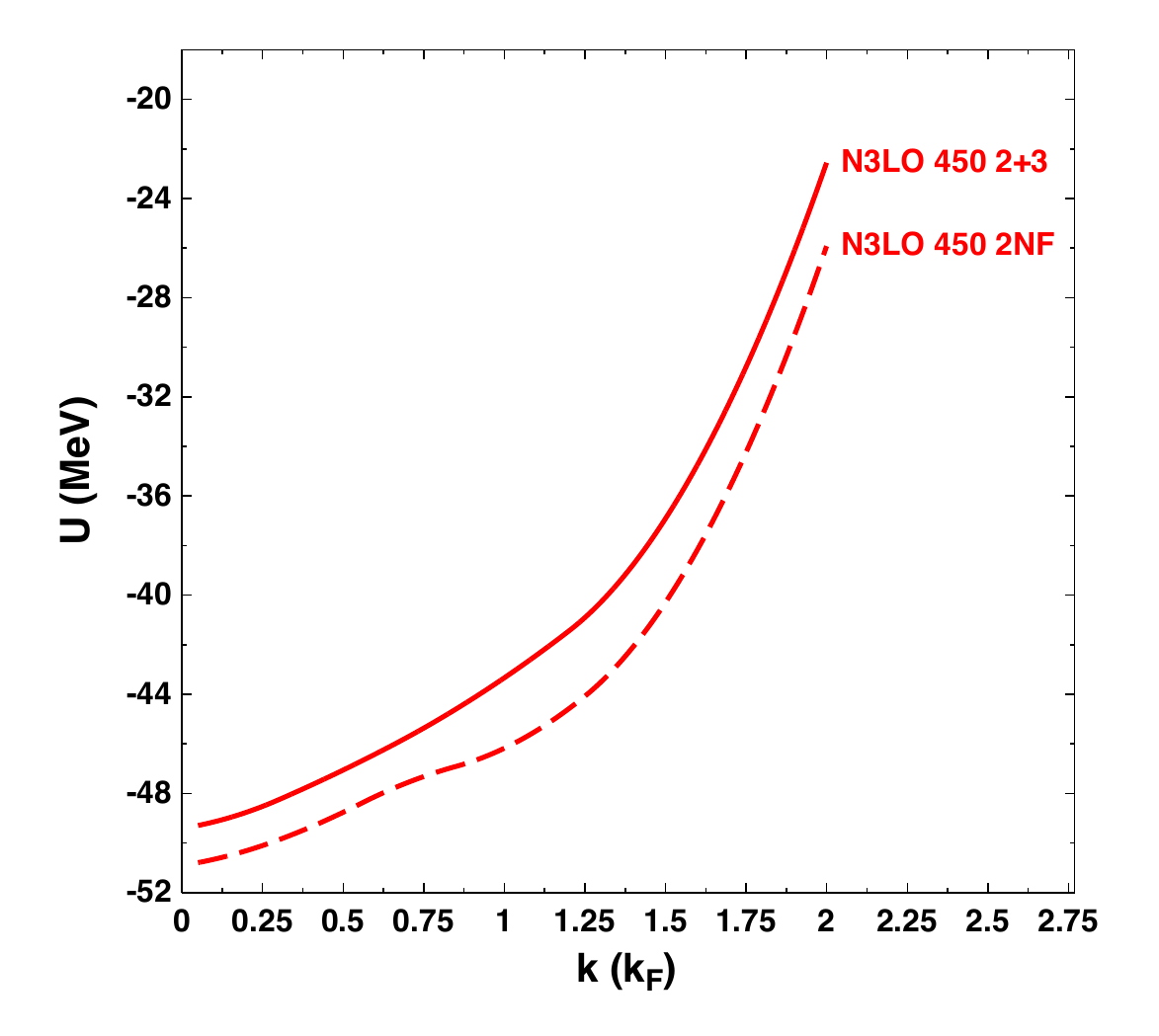}\hspace{0.01in} 
\includegraphics[width=7.0cm]{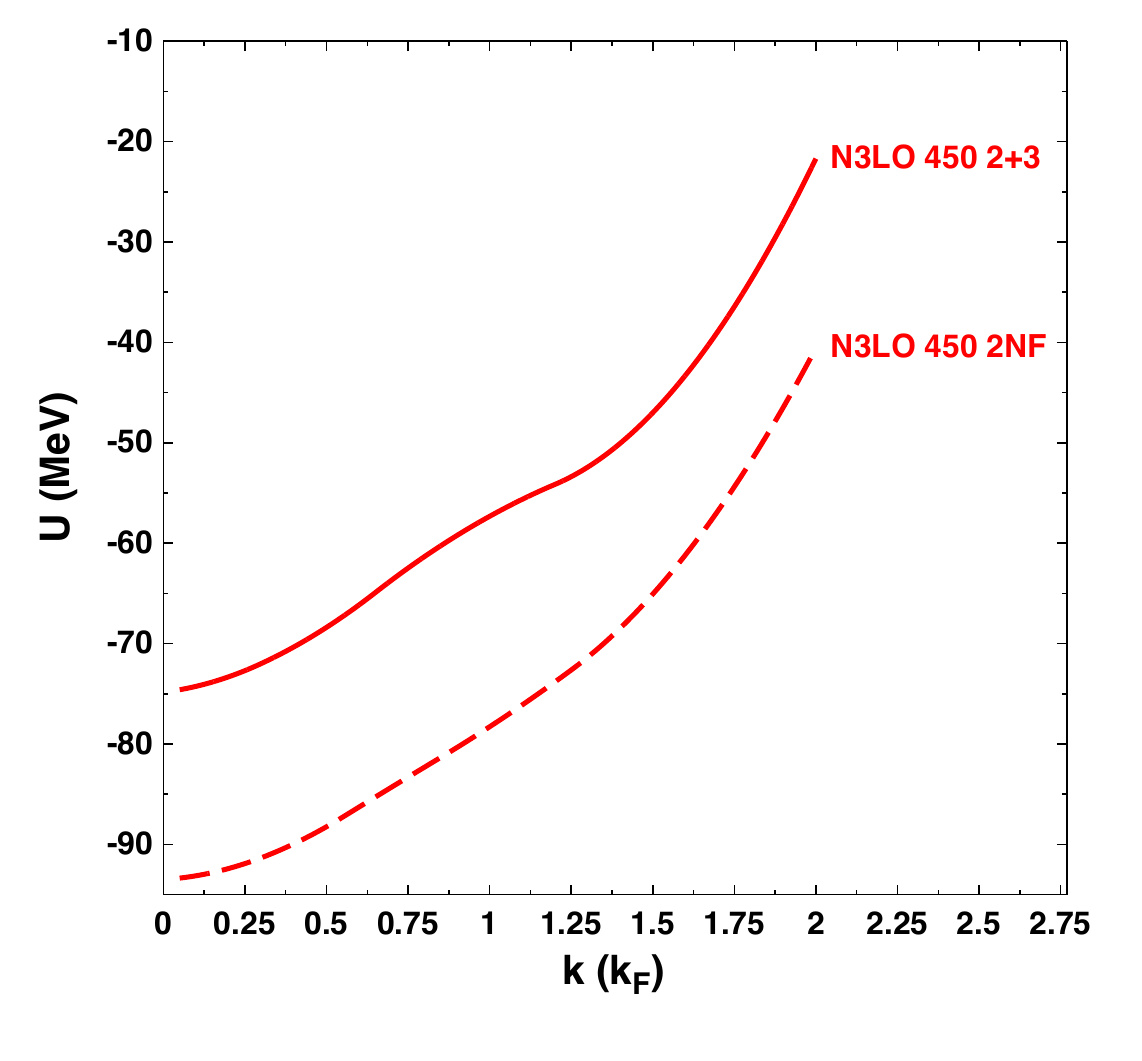}\hspace{0.01in} 
\vspace*{0.05cm}
\caption{Impact of the 3NF at N$^3$LO (solid curves) on the SPP at two different densities. Left: $k_F$ = 1.0 fm$^{-1}$; right: $k_F$ = 1.333 fm$^{-1}$. The cutoff is fixed at 450 MeV.
}
\label{u_ord_23}
\end{figure*}

Single-particle potentials derived from chiral interactions are generally deep and grow monotonically from the bottom of the Fermi sea.
  The impact of moving to fourth order is much larger at the higher density.

The impact of including the complete 3NF up to N$^3$LO is demonstrated in Fig.~\ref{u_ord_23}. The effect is to decrease the depth of the potential, and is strongly density dependent. This is a precursor of the repulsive and density dependent effect of the full 3NF on the average energy per nucleon.

Analyses of HI collision measurements are used to extract empirical constraints for the EoS. For instance, the elliptic flow in midperipheral to peripheral collisions was found to be particularly sensitive to the momentum dependence of the nucleon mean field~\cite{Dan00}. 
 More specifically, the 
  connection between the quantities displayed in Fig.~\ref{u_ord}-\ref{u_ord_23} and reaction data is through the nucleon isoscalar and isovector potentials, defined, respectively, as
\begin{equation}
U_0 (k,\rho) =\frac{ U_n + U_p}{2} \;
\end{equation}
and
\begin{equation}
U_{sym}(k,\rho, \alpha) =\frac{ U_n - U_p}{2\alpha} \; ,
\end{equation}
where $\alpha$ = $\frac{\rho_n - \rho_p}{\rho_n + \rho_p}$.
Naturally, the isovector potential, also known as the symmetry potential, is relevant for reactions with neutron-rich nuclei. Of particular interest are rare isotopes, which can now be studied with radioactive isotope beams.

\subsubsection{Neutron matter}
\label{sec_nm}

The physics of NM spans a broad range of densities. At low density, it approaches universal
behavior as a consequence of the large neutron-neutron scattering length in the spin-singlet channel. Around normal
nuclear density, it is a convenient laboratory to study neutron-rich nuclei, and, at even higher densities, it strongly
constrains the physics of neutron stars and, thus, the EoS of neutron-rich matter is the forefront of nuclear astrophysics.
 Although an idealized system, neutron matter also provides unique opportunities
 to test nuclear forces, because all low-energy couplings appearing in the three-neutron forces are predicted at
the two-body level.

 Theoretical predictions of neutron-rich matter are important  and particularly timely, as they complement on-going and planned experimental efforts. Following PREX-II~\cite{Ree21} and CREX~\cite{RRN22}, other experiments, such as MREX at MESA~\cite{Bec18}, seek to place constraints on the neutron radii and neutron skins of $^{48}$Ca and $^{208}$Pb, and to further elucidate the relation between the neutron skin and the energy and pressure in neutron-rich matter.

 Neutron stars are important natural laboratories for constraining theories of the EoS in neutron-rich matter, to which the mass-radius relationship of these stellar objects is sensitive. The radius of the average-mass neutron star is especially sensitive to the pressure gradient in neutron matter around normal densities~\cite{SM19}. Interest in these compact stars has increased considerably as we have entered the ``multi-messenger era" of astrophysical observation. The GW170817 neutron star merger event has yielded new and independent constraints on the radius  of a neutron star~\cite{Abb17a, Abb17b}. More precisely, with the constraint that  a mass of at least 2 M$_{\odot}$~\cite{Ant13} is supported by the EoS, the radii of the two neutron stars are reported as $R_1$ = (11.9 $\pm$ 1.4) km and $R_2$ = (11.9 $\pm$ 1.4) km,  with the respective mass ranges constrained at (1.18 - 1.36) M$_{\odot}$ (for M$_1$) and (1.36 - 1.58) M$_{\odot}$ (for M$_2$)~\cite{Abb19}.

\paragraph{Energy per neutron in neutron matter.}
\label{sec_nm_1}

\begin{figure*}[!t] 
\centering
\hspace*{-3cm}
\includegraphics[width=7.5cm]{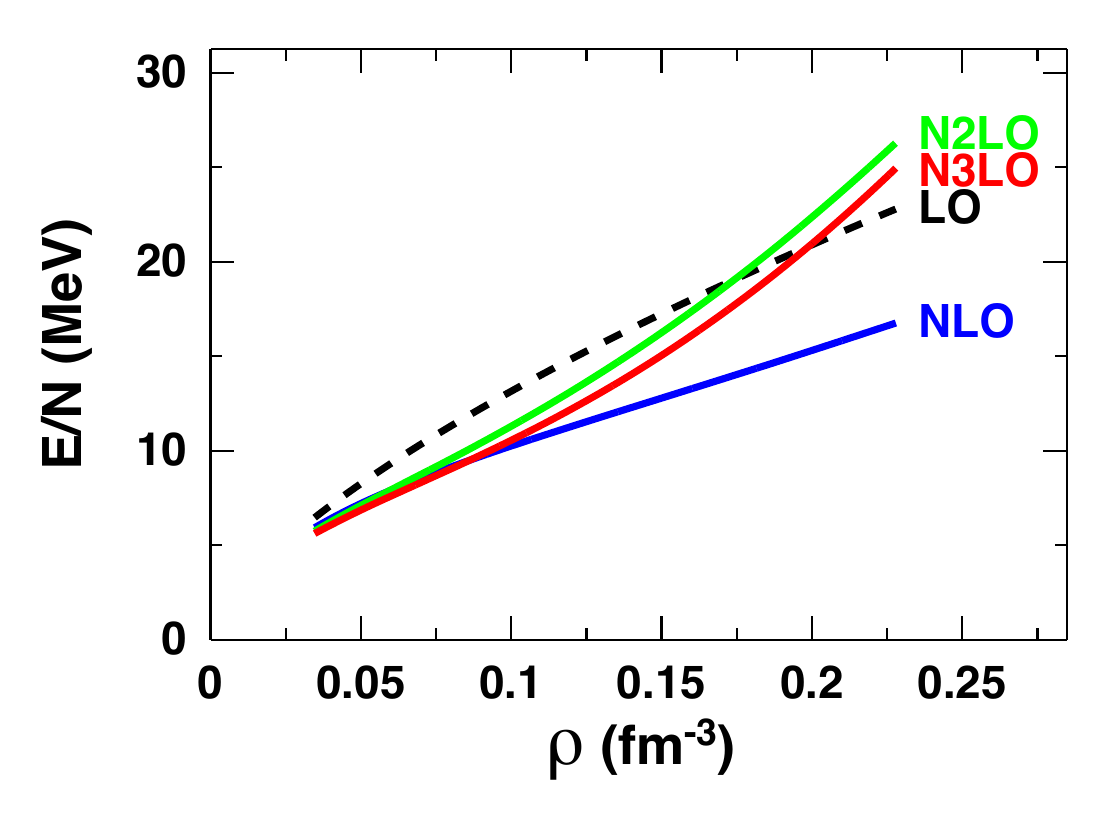}\hspace{0.01in} 
\includegraphics[width=8.7cm]{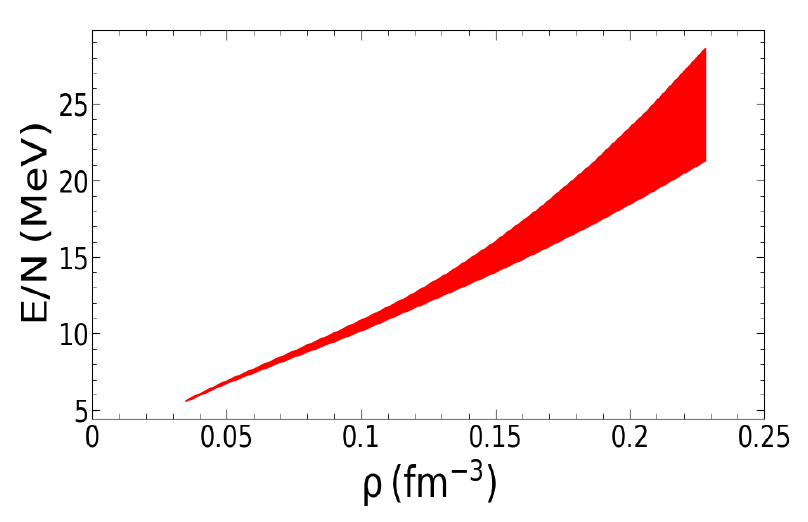}\hspace{0.01in} 
\vspace*{-0.5cm}
 \caption{(Color online) Left: Energy per neutron in NM as a function of density, from leading order (black dash) to fourth order (solid red). Right: Energy per particle as a function of density at fourth order of the chiral expansion. The band shows the uncertainty calculated from Eq.~\ref{eq_error}. 
The cutoff is fixed at 450 MeV.
}
\label{nm_eos}
\end{figure*}

In Fig.~\ref{nm_eos}, we show our results for the energy per neutron in neutron matter (NM) as a function of density over four orders, from LO to N$^3$LO. We see
large variations from leading order to the next, as to be expected at the lowest orders. It is interesting to notice the very large differences between NLO and N$^2$LO, mostly due to the first appearance of 3NFs. The predictions at N$^3$LO are slightly more attractive than those at N$^2$LO, in agreement with other calculations~\cite{Dri16}.

\begin{figure*}[!t] 
\centering
\hspace*{-3cm}
\includegraphics[width=7.7cm]{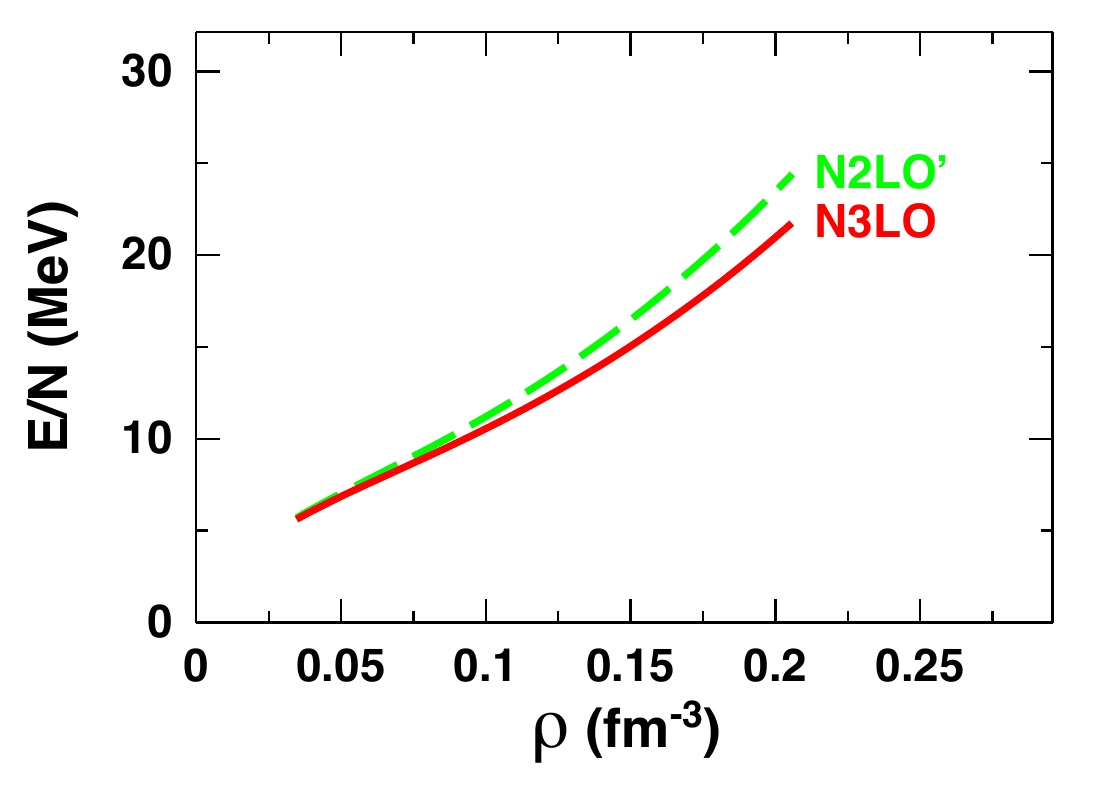}\hspace{0.01in} 
\vspace*{-0.5cm}
 \caption{(Color online) Energy per neutron in NM as a function of density. The solid (red) curve is the same as in Fig.~\ref{nm_eos}, while in the dashed green line (denoted by N$^2$LO'), 
the N$^3$LO 3NF contribution has been left out.}
\label{nm23}
\end{figure*}

Figure~\ref{nm23} shows the impact of the subleading 3NF contribution only, comparing the result of the complete calculation at N$^3$LO with the one obtained with the 2NF at N$^3$LO and the leading 3NF only,  model (N$^2$LO$'$). The effect is mildly attractive, about 2 MeV at the highest density, mostly due to the subleading 2PE 3NF. 

The first four-nucleon forces (4NFs) appear at N$^3$LO, but were found to be negligible~\cite{Tew13,Kai12, KM16}. Therefore, we omit them.

 Our highest-order results for the energy per neutron at different densities are shown in the second column of Table~\ref{par0} with their chiral uncertainty calculated as in Eq.~(\ref{eq_error}). 
Our NM EoS is rather soft within the large spectrum of theoretical predictions in the literature, which is generally true for predictions based on chiral EFT. The degree of softness is best discussed in the context of density dependence of the symmetry energy, as we do next.

\subsubsection{The symmetry energy}
\label{sec_nm_2}

We begin with a pedagogical introduction to establish the main concepts and definitions.

The simplest picture of the nucleus goes back to the semi-empirical mass formula (SEMF), also known as the liquid-droplet model. Its ability to capture basic bulk features of nuclei is remarkable in view of its simplicity. The binding energy per nucleon is written in terms of a handful of terms inspired by the dependence of nuclear radii on the cubic root of the mass number A:
\begin{equation}
\label{BoverA}
\frac{B(Z,A)}{A} = a_V - a_{sym}\frac{(A - 2Z)^2}{A^2} - \frac{a_s}{A^{1/3}} - \frac{a_C Z(Z-1)}{A^{4/3}} -\frac{\Delta}{A}  \; ,
\end{equation}
where the last term stands for additional, typically smaller, contributions.
Note that the second term on the RHS depends on the relative neutron-proton asymmetry, or isospin asymmetry, $\frac{N - Z}{A}$, and represents the loss in binding energy experienced by a nucleus due to the destabilizing presence of asymmetry in neutron/proton concentrations. Of course, Eq.~(\ref{BoverA}) is the simplest picture of a nucleus, but can be improved by replacing the number of nucleons $A$ and the number of protons $Z$ with the respective density profiles. 

To that end, we introduce the energy per nucleon, $e(\rho,\alpha)$, in an infinite system of nucleons at density $\rho$ and isospin asymmetry $\alpha = \frac{\rho_n - \rho_p}{\rho}$ -- namely, the EoS of neutron-rich matter -- and expand this quantity with respect to the isospin asymmetry parameter:
\begin{equation}
e(\rho,\alpha) = e(\rho, \alpha=0) + \frac{1}{2} \Big ( \frac{\partial^2 e(\rho,\alpha)}{\partial \alpha^2} \Big )_{(\alpha = 0)}  \alpha^2 + \mathcal{O} (\alpha^{4})  \; .
\label{e_exp}
\end{equation} 
 Neglecting terms of order $\mathcal{O} (\alpha^{4})$, Eq.~(\ref{e_exp}) takes the well-known parabolic form:
\begin{equation}
e(\rho,\alpha) \approx e_0(\rho) + e_{sym}(\rho) \ \alpha^2  \; , 
\label{asym_e}
\end{equation}
where  $e_{sym}$ = 
  $\frac{1}{2} \Big ( \frac{\partial^2 e(\rho,\alpha)}{\partial \alpha^2} \Big )_{\alpha = 0}$ and $e_0(\rho) =  e(\rho, \alpha=0)$, the EoS of SNM.

 With the SEMF as a guideline, one can write the main contributions to the total energy of a given nucleus (Z,A) with proton and neutron density profiles $\rho_p(r)$ and $\rho_n (r)$, respectively, as:
\begin{align}
\begin{split}
\label{eee}
E(Z,A)= \int d^3 r \ \rho(r) \ e(\rho,\alpha) + f_0 \int d^3 r  |\nabla \rho|^2  + \frac{e^2}{4 \pi \epsilon_o} (4\pi)^2 \int^{\infty}_{0} dr' \big[ r' \rho_{p}(r') \int^{r'}_{0} dr \ r^2 \rho_{p}(r) \big]  
\end{split} \; ,
\end{align}
where $f_0$ is a constant typically fitted to $\beta$-stable nuclei. Note that the second term on the RHS , dependent on the gradient of the density function, is a finite-size contribution -- the surface term proportional to the coefficient $a_s$ in Eq.~(\ref{BoverA}). The last term on the RHS stands for the Coulomb interaction among protons. The
 link to Eq.~(\ref {BoverA}) is apparent. In particular, the first integral on the RHS of Eq.~(\ref {eee}) comprises the first two terms on the RHS of Eq.~(\ref {BoverA}).

Within the parabolic approximation (Eq.~(\ref {asym_e})), the symmetry energy becomes the difference between the energy per neutron in NM and the energy per nucleon in SNM:
\begin{equation}
e_{sym} (\rho) = e_n (\rho) - e_0 (\rho) \; ,
\label{xxx}
\end{equation}
where $e_n(\rho) =  e(\rho, \alpha=1)$, the energy per neutron in pure neutron matter.

The minimum of $e_0(\rho)$ at a density approximately equal to the average central density of nuclei, $\rho_0$, is a reflection of the saturating nature of the nuclear force. Next, we
expand the symmetry energy about the saturation point:
\begin{equation}
\label{yyy}
e_{sym} (\rho) \ \approx \ e_{sym} (\rho_{0}) + L \ \frac{\rho -\rho_{0}}{3 \rho_0} + \frac{K_{sym}}{2} \frac{(\rho - \rho_{0})^2}{(3\rho_0)^2}  \; ,
\end{equation}
 which helps identifying several useful parameters. 
 $L$ is known as the slope parameter, as it is a measure of the slope of the symmetry energy at saturation:
\begin{equation}
\label{L}
L=3\rho_{0} \Big( \frac{\partial e_{sym}(\rho)}{\partial \rho} \Big)_{\rho_{0}}  \; .
\end{equation}
Furthermore, it is clear from Eqs. (\ref{xxx}) and (\ref{L}), recalling that the SNM EoS has
 vanishing derivative at that point, that $L$ measures the degree of ``stiffness of the NM EoS at saturation density.

 The parameter $K_{sym}$  characterizes the curvature of the symmetry energy at saturation density:
\begin{equation}
\label{ksym}
K_{sym}=9 \ \rho_{0}^2 \Big( \frac{\partial^2 e_{sym}(\rho)}{\partial \rho^2} \Big)_{\rho_{0}}  \; .
\end{equation}

Using the standard thermodynamic relation,
\begin{equation}
P(\rho) = \rho^2 \frac{\partial e}{\partial \rho} \; ,
\end{equation}
where $P$ is the pressure and $e$ is the energy per particle, 
 we define the symmetry pressure as:
\begin{equation}
\label{sym_pres}
P_{sym}(\rho) = \rho^2 \frac{\partial (e_{n} - e_0)}{\partial \rho} = P_{NM}(\rho) - P_{SNM}(\rho) \; .
\end{equation}
If the derivative is evaluated at or very near $\rho_0$, the symmetry pressure is essentially the pressure in NM because the pressure in SNM vanishes at saturation. Then:
\begin{equation}
\label{ppp}
P_{NM}(\rho_0) = \Big (\rho^2 \frac{\partial e_n(\rho)}{\partial \rho} \Big )_{\rho_0} \; .
\end{equation}
From  Eq.~(\ref{L}) and Eq.~(\ref{ppp}) it is clear that the slope parameter $L$ is a measure of the pressure in NM around saturation density:
\begin{equation}
\label{L_P}
P_{NM}(\rho_0) = \rho_0 \frac{L}{3}  \; ,
\end{equation}
showing that the pressure in NM is proportional to the slope of the symmetry energy at normal density. The value of $L$ is then a measure of 
the pressure gradient acting on excess neutrons and pushing them outward from the neutron-enriched core of the nucleus to the outer region, thus determining the formation and size of the neutron skin.

\begin{figure*}[!t] 
\centering
\hspace*{-3cm}
\includegraphics[width=7.2cm]{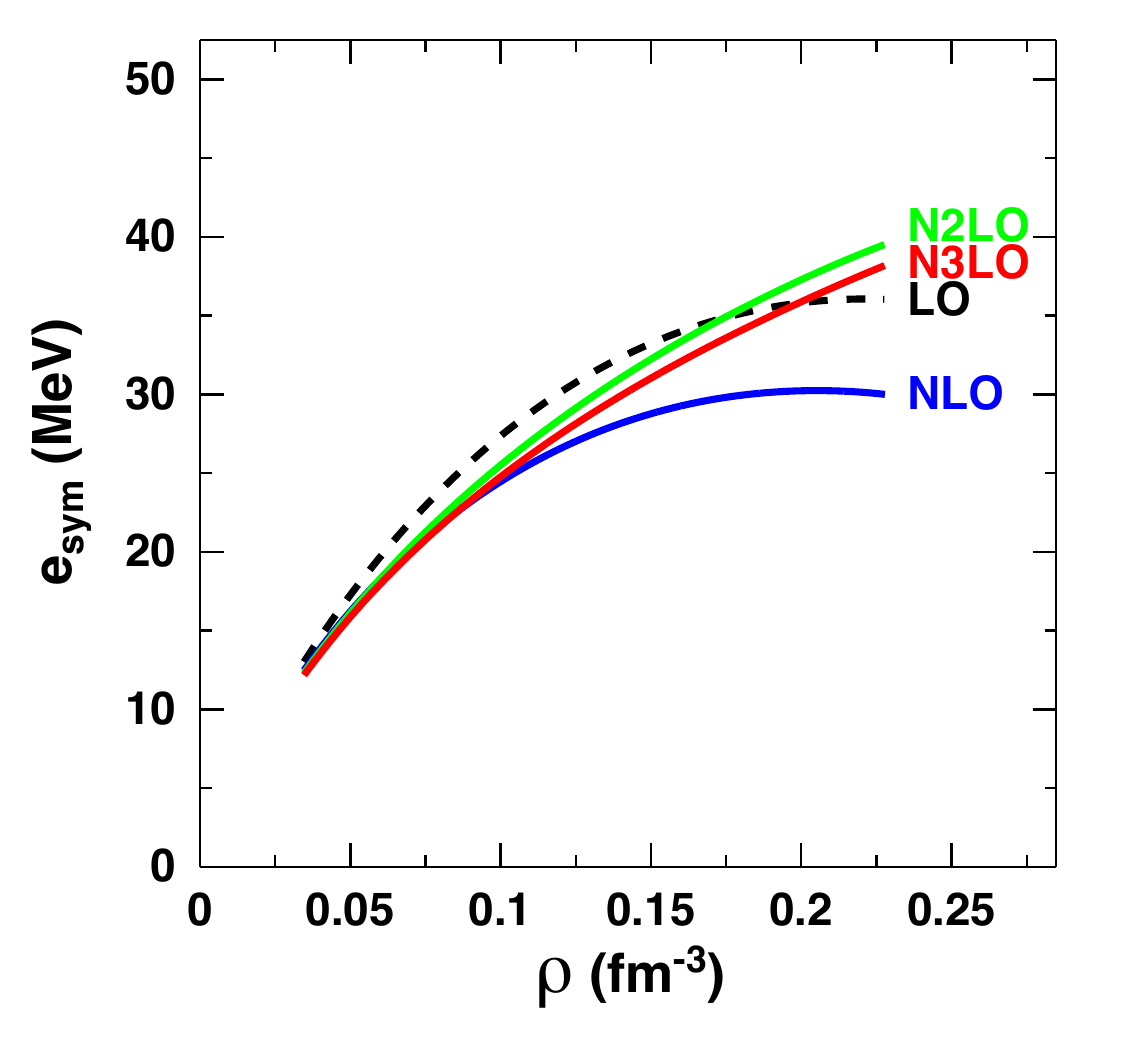}\hspace{0.01in} 
\includegraphics[width=8.2cm]{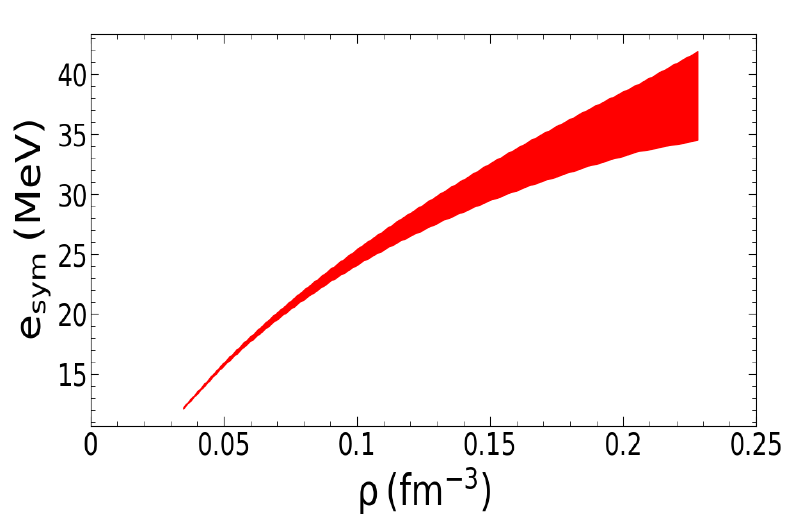}\hspace{0.01in} 
\vspace*{-0.5cm}
 \caption{(Color online) Left: The symmmetry energy as a function of density, from leading order (black dash) to fourth order (solid red). Right: The symmetry energy as a function of density at fourth order of the chiral expansion. The band shows the uncertainty calculated from Eq.~\ref{eq_error}. The cutoff is fixed at 450 MeV.
}
\label{esym}
\end{figure*}

Our order-by-order predictions for the symmetry energy are displayed in Fig.~\ref{esym}, and our results at N$^3$LO together with their truncation errors are given in Table~\ref{par0}.
To understand the full implications of these predictions, several comments are in place.
To study correlations, phenomenological models are parameterized so as to ensure that the empirical saturation properties are well described, while the models can differ wildly in the isovector properties. Earlier investigations with a family of Skyrme interactions concluded that there is a linear correlation between the slope parameter and the neutron skin thickness of $^{208}$Pb \cite{Bro00}. This inherent connection between the symmetry energy density derivative and the neutron skin of neutron-rich nuclei is of great interest, because accurate measurements of the skin should then allow to set stringent constraints on the density dependence of the symmetry energy around saturation.

Relativistic mean-field models predict a very wide range of $L$ values, for example IU-FSU \cite{Fat10} gives 47.2 MeV for $L$, while NL3 \cite{LKR97} yields a value of 118.2 MeV.  Naturally, these models also produce a large range of neutron skin values.  For neutron skin predictions from RMF models, see also Ref.~\cite{Roc11}, where the authors utilize a large set of RMF models constrained by accurate fits of the nuclear binding energies and charge radii.

 Constrains on $L$ vary considerably depending on the methods employed~ \cite{Agr13, Vin14}.  Analyses that employ laboratory data to extract constraints on the density dependence of the symmetry energy can be found in Refs.~\cite{Tsang+09, Tsang+12, LL13, Kort+10, DL14, Roca+15, Tam+11, Brown13, Rus11, Rus16}. Constraints on the symmetry energy and its density dependence have also been sought through techniques, such as:  data on nuclear masses across the periodic table~\cite{Mond+2015}, giant dipole resonance energies~\cite{Trip+2008}, electric dipole polarizability~\cite{Birk+2017}, measurements of directed and elliptic flows in heavy ion (HI) collisions~\cite{Rus16}, isobaric analog states~\cite{DL14}, isospin diffusion in HI collisions~\cite{Tsang+2004}, neutron and proton transverse emission ratio measurements~\cite{Fam+2006},
HI collisions at intermediate energies~\cite{Yong+2020}.   
As for the symmetry energy curvature, Eq.~(\ref{ksym}), 
 constraints carry a much larger uncertainty~\cite{Vid09, Duc11, San14}.

In addition to the energy per neutron at saturation density, we show in Table~\ref{par0} the symmetry energy at saturation, the slope parameter as defined in Eq.~(\ref{L}), and the pressure in neutron matter. As mentioned earlier, 
a softer nature is typical of chiral predictions, see, for instance, Ref.~\cite{DSS14}, where SRG-evolved interactions based on the potentials from Ref.~\cite{EM03} and the leading 3NF are employed. A comparison with phenomenological interactions of the past, such as Argonne V18 and the UIX 3NF~\cite{APR98}, is given in Ref.~\cite{DSS14}.
For a more recent analysis, see Ref.~\cite{Dri20}, where the reported values for $e_{sym}(\rho_o)$ and $L$  are (31.7 $\pm$ 1.1) MeV and (59.8 $\pm$ 4.1) MeV, respectively.

On the other hand, values such as those shown in the first row of Table~\ref{par0} -- approximately $L$=(50 $\pm$ 10) MeV, and pressure at $\rho_o$ between 2 and 3 MeV/fm$^3$--   are nowhere near those extracted from the PREX-II experiment~\cite{Ree21}, which are: 
$e_{sym}$ =  (38.29 $\pm$ 4.66) MeV 
and $L$  =  (109.56 $\pm$ 36.41) MeV.
The corresponding value of the pressure at $\rho_o$ is then, approximately, between 3.66 MeV/fm$^3$ and 7.30 MeV/fm$^3$. Furthermore, such stiff symmetry energy would allow rapid cooling through direct Urca processes to proceed at unusually low values of the neutron star mass and central density~\cite{Ree21}, which seems unlikely~\cite{Pag09}.  The various particle fractions in $\beta$-stable matter that we obtain with our N$^3$LO predictions are shown in Fig.~\ref{frac}. The proton fraction is close to 6\% at $\rho \approx$ 0.2 fm$^{-3}$, still far from the direct Urca threshold of approximately 11\%.

Back to Table~\ref{par0}, we also show the predictions at some specific densities below $\rho_o$. These are the  densities identified in Ref.~\cite{LT22} as ``sensitive" densities from the slope of the correlation in the plane of $e_{sym}(\rho_o)$  {\it vs.} $L$ obtained from the measurements of a specific observable. In fact, a particular slope reflects a specific density at which that observable is especially sensitive to the symmetry energy. Our ab initio predictions and the values taken from Ref.~\cite{LT22} shown in Table~\ref{par0}, compare favorably within uncertainties.
 We recall that, at $\rho$ = (2/3)$\rho_o \approx$ 0.1 fm$^{-3}$ (an average between central and surface densities in nuclei),  the symmetry energy is well constrained by the binding energy of heavy, neutron-rich nuclei~\cite{Ree21} -- hence, the relevance of this density region for the purpose of correlations between $L$ and the neutron skin of $^{208}$Pb.

Finally, in Table~\ref{kappa} we show our predictions for $K_{sym}$, defined in Eq.~(\ref{ksym}),
 in comparison with recent constraints~\cite{LT22}. Given the strong sensitivity of $K_{sym}$ to the details of the symmetry energy curvature, this parameter is very model dependent and difficult to constrain, with reported values ranging from large and negative to large and positive. As seen from Table~\ref{kappa}, our predictions are in reasonable agreement with the constraints discussed in Ref.~\cite{LT22}.

 It is our understanding that, focusing on the sensitive density for a given observable, better consistency among different analyses can be found~\cite{LT22}.
Perhaps these considerations may help with the interpretation of the large values from PREX-II. When the current constraint from PREX-II is included in the fits, its impact is weak due to the large experimental uncertainty~\cite{LT22}.  For the slope of the symmetry energy at $\rho$ = (2/3)$\rho_o \approx$ 0.1fm$^{-3}$, the value obtained from PREX-II data~\cite{Ree21} is  $L$ =(73.69 $\pm$ 22.28) MeV.

\begin{table*}
\caption{The energy per neutron, the symmetry energy, the slope parameter, and the pressure at N$^3$LO at various densities, $\rho$, in units of $\rho_o$=0.155 fm$^{-3}$.
$L$ is defined as in Eq.~(\ref{L}) at the specified density. The values in parentheses are taken from Ref.~\cite{LT22}. The constraint for $L$ in the third row ($\rho =0.67 \rho_o$) is given for  $\rho$=0.1 fm$^{-3}$. The constraint at $\rho =0.31 \rho_o$ is from Ref.~\cite{ZL15}. }
\label{par0}
\centering
\begin{tabular*}{\textwidth}{@{\extracolsep{\fill}}ccccc}
\hline
\hline
 $\rho$ $ (\rho_o) $ & $ \frac{E}{N}(\rho)$ (MeV) & $e_{sym}(\rho)$ (MeV)  & $L (\rho)$(MeV)  &  $P_{NM}(\rho)$ (MeV/fm$^3$)  \\
\hline
\hline
 1 & 15.67  $\pm$ 1.13 & 31.67 $\pm$ 1.64 (33.3 $\pm$ 1.3)  & 49.84 $\pm$ 8.71 (59.6 $\pm$ 22.1) & 2.52  $\pm$ 0.45 (3.2 $\pm$ 1.2)  \\  
0.72 (0.72 $\pm$ 0.01) & 11.55  $\pm$ 0.41 & 26.45 $\pm$ 0.82 (25.4 $\pm$ 1.1) & 45.59 $\pm$ 3.49 & 1.05  $\pm$ 0.13  \\  
 0.67 (0.66 $\pm$ 0.04) & 10.87  $\pm$ 0.35 & 25.30 $\pm$ 0.69 (25.5 $\pm$ 1.1) & 44.41 $\pm$ 2.67 (53.1 $\pm$ 6.1) & 0.86  $\pm$ 0.09  \\  
0.63 (0.63 $\pm$ 0.03) & 10.42  $\pm$ 0.32 & 24.47  $\pm$ 0.61 (24.7 $\pm$ 0.8) & 43.79 $\pm$ 2.28 & 0.75  $\pm$ 0.07  \\  
0.31 (0.31 $\pm$ 0.03) &   6.72 $\pm$ 0.08  &15.43 $\pm$ 0.12  (15.9 $\pm$ 1.0) &  32.44 $\pm$ 0.51    &     0.18 $\pm$ 0.01                            \\ 
0.21 (0.22 $\pm$ 0.07) & 5.44  $\pm$  0.04 & 11.63 $\pm$ 0.04 (10.1 $\pm$ 1.0) & 26.09 $\pm$ 0.11 & 0.096  $\pm$ 0.001  \\  
\hline
\hline
\end{tabular*}
\end{table*} 
 \begin{table*}
\caption{Predictions  for the parameter $K_{sym}$  at N$^3$LO and constraints
for $K_{sym}$  from
Ref.~\cite{LT22} 
 at two densities. }
\label{kappa}
\centering
\begin{tabular*}{\textwidth}{@{\extracolsep{\fill}}ccc}
\hline
\hline
 $\rho (\rho_o)$  & $ K_{sym}(\rho)$ (MeV) & $K_{sym}(\rho)$ (MeV) from Ref.~\cite{LT22}  \\
\hline
\hline
 1 & -102   $\pm$ 30 & -180  $\pm$ 96   \\   
  $\frac{2}{3}$ & -76.5   $\pm$ 11.5 & -79.2  $\pm$ 37.6   \\   
\hline
\hline
\end{tabular*}
\end{table*} 

\subsubsection{Neutron skins}
\label{sec_ns_skin}

Here, we link the EoS described in Sec.~\ref{sec_snm} and \ref{sec_nm} to the neutron skin, defined as
\begin{equation}
S = <r^2>_n^{1/2} -  <r^2>_p^{1/2} \; .
\end{equation}
Relating the nucleus spatial extension as directly as possible to the microscopic EoS is best achieved by means of the droplet model~\cite{Swia05}. This also allows applications to heavy nuclei, which may be outside the reach of {\it ab initio} methods.
The predictions in Table~\ref{tab_skin} are taken from Ref.~\cite{Sam22}. Those are calculated from droplet model expressions where the symmetry energy and its density slope at saturation appear explicitely.

\begin{table*}
\caption{ The neutron skin of $^{208}$Pb, $S$, calculated as described in Ref.~\cite{Sam22} using the specified symmetry energy, $J$, and its slope at saturation, $L$. 
}
\label{tab_skin}
\begin{tabular*}{\textwidth}{@{\extracolsep{\fill}}cccc}
%\begin{tabular}{c c c c }
\hline
\hline
 $J$ (MeV)      & $L$ (MeV)  & $S$ (fm)  & source for $J$, $L$   \\
\hline    
\hline 
 31.3 $\pm$ 0.8  &  52.6 $\pm$ 4.0  & [0.13, 0.17] & \cite{SM22}   \\
 (31.1, 32.5)   &  [44.8, 56.2]  & [0.12, 0.17] &  \cite{DHS19}   \\
(28, 35)  &  [20, 72]  &     [0.078, 0.20] &     \cite{DHW21}    \\
(27, 43)                & [7.17, 135]  &  [0.055, 0.28]           &  \cite{LH19}    \\     
 38.29  $\pm$ 4.66  &  109.56  $\pm$ 36.41  &  [0.17, 0.31] &  \cite{Ree21}   \\              
\hline
\hline
\end{tabular*}
\end{table*}
\begin{table*}[hbt!]
\caption{Values of the neutron skins in $^{48}$Ca and in $^{208}$Pb from a variety of experimental methods. It is our understanding that the quoted errors refer to one standard deviation. For the values taken from Ref.~\cite{Tarb+2014} and Ref.~\cite{Bro+2007_2}, the first and second uncertainties are statistical and systematic errors, respectively. In Ref.~\cite{Bro+2007}, the first uncertainty is the experimental error, while the second originates from theoretical modeling of the experimental charge densities. Finally, in the result from Ref.~\cite{Roca+2013}, the first two uncertainties are the experimental and the theoretical error,  respectively, and the third one stands for an estimated uncertainty in the symmetry energy at saturation.}
\label{tab_exp_2}
\begin{tabular*}{\textwidth}{@{\extracolsep{\fill}}ccccc}
\hline
\hline
 Type of measurement & Extracted neutron skin in $^{48}$Ca & Extracted neutron skin in $^{208}$Pb  \\ 
\hline 
\hline
Proton-nucleus scattering~\cite{Clark+2003}  & 0.056 - 0.102 & 0.083 - 0.111 \\ 
 Proton-nucleus scattering~\cite{Zen+2010} &       &  0.211$^{+0.054}_{-0.063}$   \\
Proton-nucleus scattering~\cite{Staro+1994} &    &  0.20 $\pm$ 0.04  \\
Proton-nucleus scattering~\cite{Shlo+1979} &   0.10 $\pm$ 0.03 &      \\
Polarized proton-nucleus scattering~\cite{Ray+1979}  & 0.23 $\pm$ 0.05   &  0.16 $\pm$ 0.05  \\
Polarized proton-nucleus scattering~\cite{Zen+2018}  & 0.168$^{+0.025}_{-0.028}$   &    \\   
Polarized proton-nucleus scattering~\cite{Zen+2010}  &    &   0.211 $^{+0.054}_{-0.063}$  \\              
Pionic probes~\cite{Fried2012}    &0.13 $\pm$ 0.06   &  0.11 $\pm$ 0.06 \\
Pionic probes~\cite{Gib+1992}    &0.11 $\pm$ 0.04   &       \\
Coherent $\pi$ photoproduction~\cite{Tarb+2014} &     &  0.15 $\pm$ 0.03$^{+0.01}_{-0.03}$   \\
Coherent $\pi$ photoproduction~\cite{Zana+2015} &     &  0.20 $^{+0.01}_{-0.03}$   \\ 
Antiprotonic atoms~\cite{Bro+2007} &       &   0.20 ($\pm$ 0.04) ($\pm$ 0.05)    \\
Antiprotonic atoms~\cite{Bro+2007_2} &       &   0.16 ($\pm$ 0.02) ($\pm$ 0.04)    \\
Antiprotonic atoms~\cite{Trz+2001} &       &   0.15 $\pm$ 0.02    \\                         
Electric dipole polarizability~\cite{Roca+15} &    &  0.13 - 0.19 \\
Electric dipole polarizability~\cite{Roca+2013} &    &  0.165 ($\pm$0.09)($\pm$ 0.013) ($\pm$0.021) \\
Electric dipole polarizability &      &             \\
{\it via} polarized scattering at forward angle~\cite{Tam+11} &    &  0.156 $^{+0.025}_{-0.021}$ \\     
Electric dipole polarizability~\cite{Birk+2017} & 0.14 - 0.20   &   \\
Pygmy dipole resonances~\cite{Klim+2007} &      & 0.18 $\pm$ 0.035   \\
Interaction cross sections~\cite{Mats+2022} &   0.105 $\pm$ 0.06 &       \\
$(\alpha, \alpha')$ GDR 120 MeV~\cite{Kras+1994} &       & 0.19 $\pm$0.09  \\
$\alpha$-particle scattering~\cite{Gils+1984}  & 0.171 $\pm$ 0.05  &        \\
\hline
\hline
\end{tabular*}
\end{table*}

The first three entries in Table~\ref{tab_skin} are obtained from EoS based on chiral EFT, with chiral two- and three-nucleon interactions at N$^3$LO. One can see that they are relatively soft, cover a narrow range, and are in good agreement with one another. Consistent with earlier discussions,  the corresponding neutron skins are relatively small. Most recently,
ab initio predictions for the neutron skin of $^{208}$Pb have become
available~\cite{Hu22}. The reported range is between 0.14
and 0.20 fm, smaller than the values extracted from parity violating
electron scattering. 
The fourth line corresponds to an analysis based on current constraints from nuclear theory and experiment. 
In Ref.~\cite{LH19}, the authors utilized 48 phenomenological models, both relativistic mean field and Skyrme Hartree-Fock. 
The last line shows the values of $J$ and $L$ from the PREX II experiment. Note that the reported value for the skin of $^{208}$Pb  in Ref.~\cite{Ree21} is (0.283 $\pm$ 0.071) fm, giving a range between 0.21 and 0.35 fm.

\paragraph{Experiments and phenomenological analyses.} 
Indirect measurements of the neutron skin in $^{208}$Pb and $^{48}$Ca have been performed using a variety of techniques, such as those listed below.  Parity-violating electron scattering will be addressed separately. Some representative 
measurements are: Proton-nucleus elastic scattering~\cite{Clark+2003, Zen+2010, Staro+1994, Shlo+1979};
Polarized proton-nucleus elastic scattering~\cite{Ray+1979, Zen+2018};
Pionic probes~\cite{Fried2012, Gib+1992};
Coherent pion photoproduction~\cite{Tarb+2014};
 Antiprotonic atom data~\cite{Bro+2007,Bro+2007_2, Trz+2001};
 Electric dipole polarizability~\cite{Roca+15,Roca+2013,Tam+11, Birk+2017};
Pygmy dipole resonsnces~\cite{Klim+2007};                             
 Interaction cross sections with microscopic optical potentials~\cite{Mats+2022}; 
 $(\alpha, \alpha')$ giant dipole resonance (GDR)~\cite{Kras+1994};
 $\alpha$-particle scattering~\cite{Gils+1984};

In Table~\ref{tab_exp_2}, we 
summarize values for $^{208}$Pb and $^{48}$Ca neutron skins deduced from the indicated experiments. There is a considerable spread, as the result of a multitude of methods and theoretical input over decades. Analyses of hadronic scattering experiments, in particular, require modeling of the nuclear potential.

\begin{figure*}[!t] 
\centering
\hspace*{-3cm}
\includegraphics[width=7.0cm]{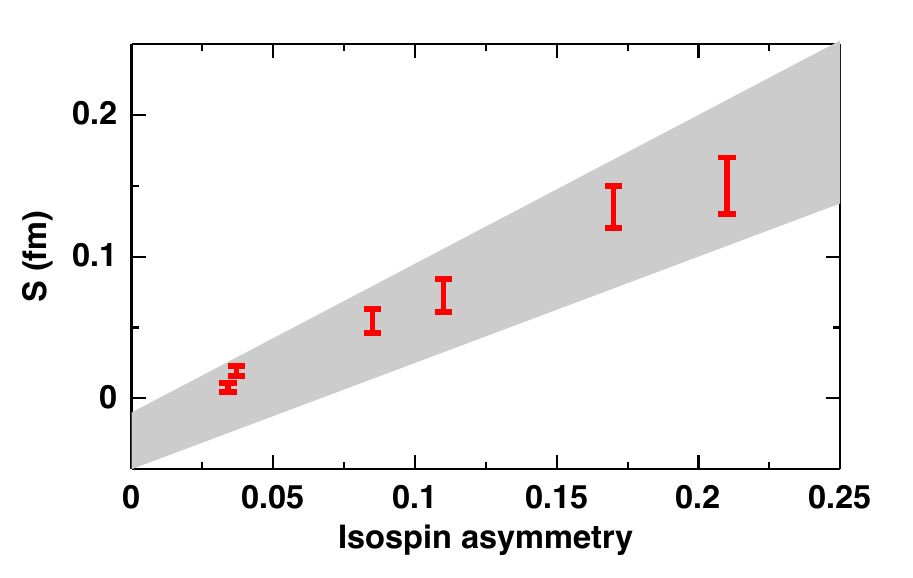}\hspace{0.01in} 
\vspace*{-0.05cm}
 \caption{(Color online) Red bars: Neutron skin of $^{58}$Ni,  $^{27}$Al, $^{59}$Co, $^{90}$Zr, $^{48}$Ca,
and $^{208Pb}$, in order of increasing isospin asymmetry, as predicted in
Refs.~\cite{Sam22}. The shaded area is bounded by linear fits to the data~\cite{Swia05}. 
}
\label{skin_fig}
\end{figure*}

The parity-violating experiment to measure the neutron skin of $^{48}$Ca was recently completed~\cite{RRN22}.The extracted value of the neutron skin, $S = 0.121 \pm 0.026 \pm 0.024$, would not be expected based on PREX-II. In the words of one of the CREX collaboration member, K.D. Paschke, ``{\it ...the contrast between the two measurements is a bit 
surprising and provides a challenge to the theoretical 
description of nuclei}." The results from CREX and PREX-II are not consistent with each other because the two nuclei are rather similar with regard to isospin asymmetry, see Fig.~\ref{skin_fig}. Our findings for $^{208}$Pb are
 consistent with the CREX reported value for $^{48}$Ca (both on the smaller side): S($^{48}$Ca)  = 0.12 -- 0.15 fm and  S($^{208}$Pb) = 0.13 -- 0.17 fm.  Based on previous measurements of the skin in $^{48}$Ca, we see no strong reasons to deem the CREX result surprising or unexpected, whereas the opposite is true for $^{208}$Pb. 
A more in-depth discussion follows below.

\paragraph{Parity-violating electron scattering.}
The parity-violating electron scattering asymmetry, $A_{PV}$, is defined for a spin-zero nucleus as
\begin{equation}
\label{apv}
A_{PV} = \frac{\sigma_R - \sigma_L}{\sigma_R + \sigma_L} \; ,
\end{equation}
where $\sigma_{R(L)}$ is the elastic cross section for right (left) handed electrons~\cite{Donn+1989}.
 $A_{PV}$ is proportional to the ratio of weak ($F_W(q)$) to charge ($F_{ch}(q)$) form factors, whith 
$q$ the four-momentum transfer. $F_{ch}(q)$ is taken from 
existing measurements and $F_W(q)$ is  extracted from the measured $A_{PV}$. We recall that the weak and charge form factors are the Fourier transforms of the weak charge density and the charge density, respectively:
\begin{equation}
\label{fw}
F_W(q) = \frac{1}{Q_W} \int d^3r j_0(qr) \rho_W(r) \; ,
\end{equation}
where $Q_W$ is the weak charge of the nucleus, and 
\begin{equation}
\label{fch}
F_{ch}(q) = \frac{1}{Z} \int d^3r j_0(qr) \rho_{ch}(r) \; .
\end{equation}
In Eqs.~(\ref{fw}) and~(\ref{fch}), $j_0(qr)$ is the zero$^{th}$ order spherical Bessel function.
In Eq.~(\ref{fw}), a form is assumed for $\rho_W(r)$ and the radius parameter of the density function is adjusted to reproduce the experimental $A_{PV}$. A form must also be assumed for $\rho_{ch}$ in Eq.~(\ref{fch}), but its parametrization relies upon well-established measurements from charged electron scattering. 
The CREX result is found insensitive to the assumed form for the weak charge density~\cite{RRN22}. 

\paragraph{Status of neutron skin predictions from {\it ab initio} theory.}

\begin{table*}[hbt!]
\caption{Status of {\it ab initio} predictions for the skins (in fm) in $^{48}$Ca and in $^{208}$Pb. The two results from Ref.~\cite{Hu+2022} for both $^{48}$Ca and in $^{208}$Pb show acceptable ranges within 68\% and 90\% of the credibility region (CR). The last two entries were obtained from density functional theory~\cite{DFT}.}
\label{tab_ab}
\begin{tabular*}{\textwidth}{@{\extracolsep{\fill}}cccc}
\hline
\hline
 Nucleus & Predicted skin &  Source   \\ 
\hline 
\hline
 $^{48}$Ca &  0.120 - 0.150 & Ref.~\cite{Hag+2016}       \\
$^{48}$Ca &  0.141 - 0.187 & Ref.~\cite{Hu+2022}   68\% CR   \\
$^{48}$Ca &  0.123 - 0.199 & Ref.~\cite{Hu+2022}    90\% CR  \\
$^{208}$Pb &  0.139 - 0.200 & Ref.~\cite{Hu+2022}  68\% CR     \\
$^{208}$Pb &  0.120 - 0.221 & Ref.~\cite{Hu+2022}   90\% CR   \\
$^{48}$Ca &  0.114 - 0.186 & Ref.~\cite{Nov+2023}       \\
$^{208}$Pb &  0.184 - 0.236 & Ref.~\cite{Nov+2023}      \\
$^{48}$Ca &  0.12 - 0.15 & Ref.~\cite{Sam22}       \\
$^{208}$Pb &  0.13 - 0.17 & Ref.~\cite{Sam22}   \\
\hline
$^{48}$Ca &  0.176 $\pm$ 0.018  & Ref.~\cite{DFT}       \\
$^{208}$Pb & 0.168 $\pm$ 0.022 & Ref.~\cite{DFT}   \\
\hline
\hline
\end{tabular*}
\end{table*}

We show in Table ~\ref{tab_ab} recent predictions for the skins in $^{48}$Ca and in $^{208}$Pb. Some comments are in place. In both Ref.~\cite{Hag+2016} and Ref.~\cite{Nov+2023}, the nature of the $NN$ chiral potentials, N$^2$LO$_{sat}$ (see Table~\ref{tab_pots} for characteristics of the various potentials), 
and $\Delta$N$^2$LO$_{GO}$ (see Table~\ref{tab_pots}  ).
is such that the results are not truly {\it ab initio}.
Note, also, that the value for $^{208}$Pb from Ref.~\cite{Nov+2023} is not a prediction, but was obtained using the linear regression the authors constructed from the skins of lighter nuclei.

\begin{figure*}[!t] 
\centering
\hspace*{-1cm}
\includegraphics[width=7.7cm]{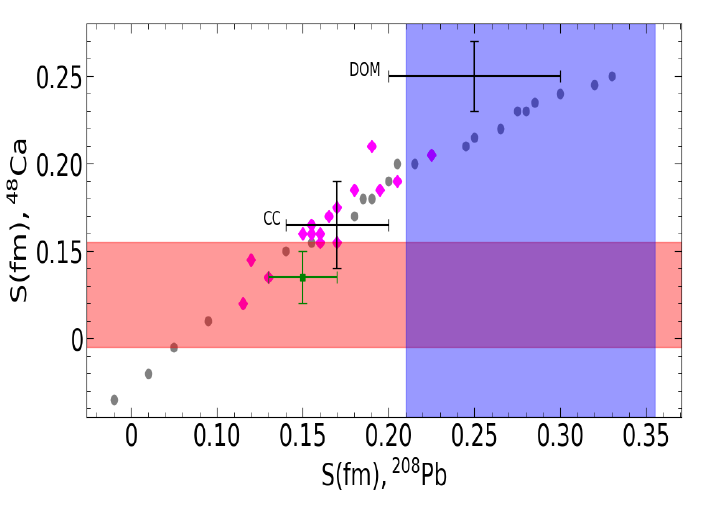}\hspace{0.01in} 
\vspace*{-0.05cm}
 \caption{(Color online) The neutron skin of $^{48}$Ca {\it vs.} the one of $^{208}$Pb. The red and blue bands show the results from CREX and combined PREX-I and PREX-II, respectively. The gray circles and pink diamonds are the result of relativistic and non-relativistic mean-field models, respectively.
Coupled cluster
and dispersive optical model predictions are indicated as CC and DOM, respectively.
}
\label{rrr}
\end{figure*}

\begin{figure*}[!t] 
\centering
\hspace*{-3cm}
\includegraphics[width=7.0cm]{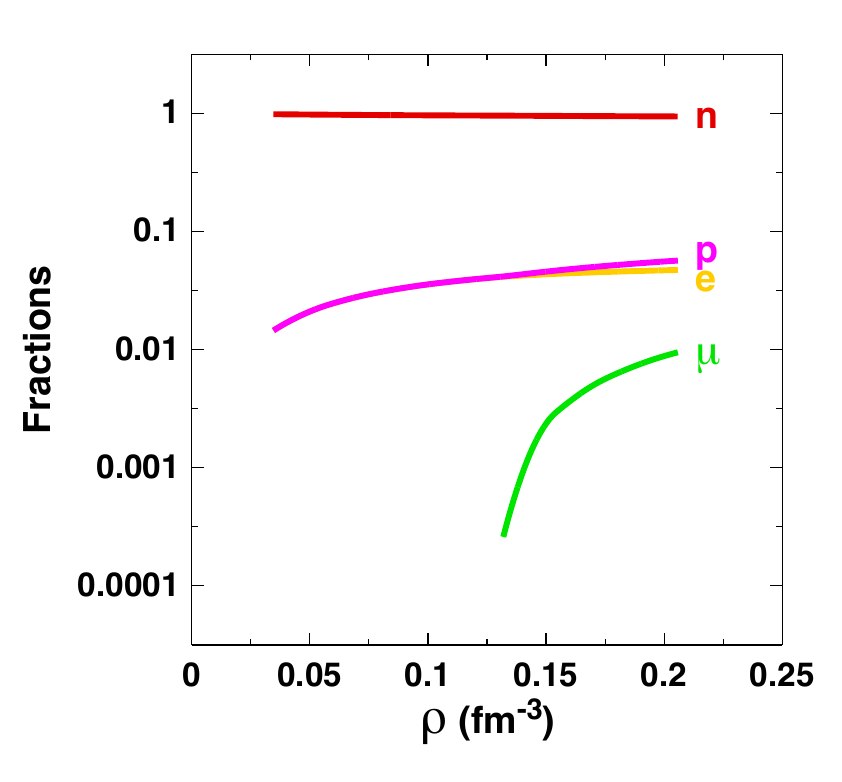}\hspace{0.01in} 
\vspace*{-0.5cm}
 \caption{(Color online) Particle fractions as a function of density in $\beta$-equilibrated matter with neutrons, protons, electrons, and muons.
}
\label{frac}
\end{figure*}

Figure~\ref{rrr} shows the neutron skin in $^{48}$Ca {\it vs.} the one in $^{208}$Pb. All values, except for the green bars, have been extracted from Fig.~5 of Ref.~\cite{RRN22}. The PREX-II and PREX-I combined experimental result is shown by the blue bar, while the red vertical bar is the CREX result. The
gray circles (pink diamonds) show results from a variety of relativistic
(non-relativistic) density functionals, which give values of the $L$ parameter ranging from small and negative to large and positive. 
Coupled cluster (CC)~\cite{Hag+2016, Nov+2023}
and dispersive optical model (DOM) predictions~\cite{Atk}  are also displayed. Our predictions are shown by the green bars~\cite{Sam22}. There exist, of course, RMF models that agree with the PREX result, and, correspondingly, generate values for $^{48}$Ca that are also on the larger side.

We conclude that a value between 0.212 fm and 0.354 fm (0.283 $\pm$ 0.071) for the skin of $^{208}$Pb is outside the boundaries set by microscopic theory. Simultaneous consistency with both CREX and PREX seems to be a challenge even for phenomenology.
%On the other hand, PREX aside, a small skin for $^{48}$Ca does not appear to be peculiar based on the facts we reviewed above, see also Fig.~\ref{Latt_fig}. At the same time, it's important to keep in mind that an experimental measurement has a probability of %approximately 68\% (95\%) to be within $\pm$ one standard deviation ($\pm$ two standard deviations). That is, it is possible that the PREX value is the result of a large statistical fluctuation.

We  look forward to hearing from MREX (Mainz Radius Experiment) at the MESA accelerator.  The MREX experiment promises to determine 
``{\it ...the neutron-skin thickness of $^{208}$Pb with ultimate precision.}"~\cite{Bec18}.
    
\subsubsection{Equation of state for stellar matter}
\label{sec_beta}

In this section we outline the main steps to construct the EoS for stellar matter in $\beta$-equilibrium. We define the total energy per baryon as:
\begin{equation}
\label{total_eng_per_prt}
\begin{aligned}
e_T (\rho) = e_0(\rho) + e_{sym} \ (Y_{n}-Y_{p})^2 +
e_e + e_\mu + \sum_{i = n,p} Y_i m_i  \; ,
\end{aligned}
\end{equation}
where $Y_{n,p}$ stands for the neutron or proton fraction. The last term accounts for the baryon rest energy, while $e_{e/\mu}$ are the energies (per baryon) of the electrons and muons, respectively. All terms are functions of density.

The energy density ($\epsilon_{i}$), pressure ($p_{i}$), number density ($\rho_{i}$) and chemical potential ($\mu_i$) for each lepton species, $i$, at a given Fermi momentum, $(k_{F})_{i}$, can be expressed as:
\begin{equation}
\label{eng_den}
\epsilon_i = \frac{\gamma}{2\pi^2} \int^{k_{F_{i}}}_{0} \sqrt{k^2 + m_{i}^{2}} \ k^2 \ dk  \; ,
\end{equation}
\begin{equation}
\label{prs}
p_i = \frac{1}{3} \frac{\gamma}{2\pi^2} \int^{k_{F_{i}}}_{0} \frac{k^2}{\sqrt{k^2 + m^2_{i}}} \ k^2 \ dk  \; ,
\end{equation}
\begin{equation}
\label{den}
\rho_i = \frac{\gamma}{2\pi^2} \int^{k_{F_{i}}}_{0} k^2 \ dk  \; ,
\end{equation}
\begin{equation}
\label{chempot}
\mu_{i} = \frac{\partial \epsilon_{i}}{\partial \rho_{i}} = \sqrt{k^{2}_{F_i}+m_{i}^{2}}  \; ,
\end{equation}
where $\gamma$ is the spin/isospin degeneracy factor.

The energy per volume is obviously related to the energy per particle:
\begin{equation}
\label{eps_eq_den_der}
\epsilon = \rho \; e(\rho)  \; ,
\end{equation}

Generally, the particle fraction,
\begin{equation}
\label{frac_expr}
Y_i=\frac{\rho_i}{\rho}  \; ,
\end{equation}
is related to the chemical potential through:
\begin{equation}
\label{chempot_frac}
\mu_{i} = \frac{\partial \epsilon_{i}}{\partial \rho_{i}} = \frac{\partial e_{i}}{\partial Y_{i}}  \; .
\end{equation}
From the condition of energy minimization, subjected to the constraints of conserved nucleon density and global charge neutrality,
\begin{equation}
\label{frac_bary}
\rho_p + \rho_n = \rho \quad \mbox{or} \quad Y_p + Y_n = 1  \; ,
\end{equation}
\begin{equation}
\label{frac_chrg}
\rho_p = \rho_e + \rho_{\mu} \quad \mbox{or} \quad Y_p = Y_e + Y_{\mu}  \; ,
\end{equation}
one can obtain all particle fractions.
The proton fraction plays a major role in the cooling process of neutron stars, therefore, cooling data have the potential to shed light on the symmetry energy, closely related to the 
  proton fraction in $\beta$-equilibrated matter. Below, we give a simple demonstration of this very direct relation.

Neglecting, for simplicity, the muon  contribution ($ Y_p$ = $Y_e$), the equations above give:
\begin{equation}
e_e = \frac{\hbar c}{4 \pi ^2} \rho ^{1/3}(3 \pi ^2 Y_p)^{4/3} \,
\end{equation}

\begin{equation}
 \frac{\partial e_{e}}{\partial Y_{p}} = \frac{\hbar c}{3 \pi ^2} (3 \pi ^2)^{4/3}  ( \rho Y_p) ^{1/3} \; ,
\end{equation}

\begin{equation}
   \frac{\partial e_{p}}{\partial Y_{p}} =    \frac{\partial e_{n}}{\partial Y_{n}}  -   \frac{\partial e_{e}}{\partial Y_{e}} \; ,
\end{equation} 
from which we obtain:
\begin{equation}
\label{esym_x}
\begin{aligned}
4 \ e_{sym} (\rho) \ (1 - 2\cdot Y_p) = \hbar c (3\pi^2)^{\frac{1}{3}} (\rho \ Y_{p})^{\frac{1}{3}} \; ,
\end{aligned}
\end{equation}
where we have neglected the mass difference between the proton and the neutron.

If $\rho_{DU}$ is the density at which $Y_p$ is equal to the value needed for direct Urca processes (DU), about $1/9$, the following relation holds:
\begin{equation}
\label{urca2}
\begin{aligned}
e_{sym} (\rho_{DU}) = \hbar c \frac{9}{28} (\pi^2/3)^{\frac{1}{3}}  (\rho_{DU})^{1/3} \; .
\end{aligned}
\end{equation}
This simple relation is quite insightful.
If, for example, $\rho_{DU}$ is close to saturation density, the symmetry energy would be over 50 MeV at that point. If $\rho_{DU}  = \frac{2}{3}\rho_0$, the symmetry energy at that density would be over 40 MeV. In our and other microscopic predictions, DU processes are more likely to open at a few to several times normal density (based on projections, since chiral EFT predictions cannot be extended to such high densities). Using PREX II constraints, the Urca threshold is found to be approximately 1.5$\rho_0$ or just above 0.2 $fm^{-3}$~\cite{Kum23}. From Eq.~(\ref{urca2}), the value of the symmetry energy at 1.5$\rho_0$ is then about 58 MeV, clearly indicationg an unusually steep density dependence, taking the PREX II value of 38.1 $\pm$ 4.7 MeV at saturation. 

As noted above, the CREX experiment~\cite{RRN22} is now completed and the results are not consistent with PREX II. The authors of Ref.~\cite{Kum23} developed different parametrizations of RMF models constrained by CREX results, PREX II results, or a combination of both, and report that the direct Urca threshold density decreases from 0.71 fm$^{-3}$ to 0.21 fm$^{-3}$ as the skin of $^{208}$Pb increases from 0.13 fm to 0.28 fm. 

At normal nuclear densities, typical predictions of the symmetry energy are consistent with a proton fraction approximately equal to 1/25~\cite{Latt91}, far from the DU threshold. On the other hand, including PREX-II constraints, the threshold density for DU is found to be as low as 1.47$\rho_0$, with a threshold mass of 0.736 solar masses~\cite{Thapa22}.

\subsection{Astrophysics implications}

 A neutron star is the remnant collapsed core of a giant star which has undergone a supernova explosion. A star must have sufficient mass, estimated to be between 8 and 25 $M_{\odot}$, to undergo a supernova event at the end of its life cycle~\cite{B_1_2013}.  Due to its extremely compact nature, the neutron star is supported against further gravitational collapse into a black hole by mechanisms of nuclear origin, which make these objects excellent natural laboratories for exploring the nuclear EoS.

We begin this section by reviewing the relevance of neutron stars for nuclear physics, from a historical perspective.
In 1934, just two years after the discovery of the neutron~\cite{C_1_1932}, Baade and Zwicky proposed the existence of a very dense stellar object, which they named neutron star, arising from the remnants of a supernova~\cite{BZ_1_1934, BZ_2_1934}. In 1939 Tolmann~\cite{Tol39}, and, independently, Oppenheimer and Volkoff~\cite{OV39}, estimated the mass--radius relationship of these neutron stars based on general relativity and crude models for the nuclear force, and developed the famous Tolmann--Oppenheimer--Volkoff (TOV) equation. The TOV equation allows to predict a theoretical upper limit on the possible mass of neutron stars. However, due to the very limited understanding of nuclear interactions at the time, their predictions were not accurate.
 With improved understanding of nucleonicEnergy emission from a neutron star interactions, a more realistic picture of neutron stars and their structure emerged~\cite{C_1_1959, AS_1_1960, S_1_1960, IK_1_1965, TC_1_1966, M_1_1971, S_1_1972, SS_1_1972, KN_1_1986, IK_1_1969}. 

 While the existence of neutron stars was theoretically possible, finding evidence of their existence remained a challenge. Initial attempts to compute and observe the thermal signature of neutron stars~\cite{C_1_1964, TC_1_1966, YLS_1_1999} were unsuccessful. In 1967, Pacini~\cite{P_1_1967} postulated that fast rotating neutron stars could generate electromagnetic emission from a powerful magnetic dipole. Soon after, Bell and Hewish~\cite{HBPSC_1_1968} discovered the first radio pulsar, characterized by a stable periodic electromagnetic signal. Later that year, Gold theorized that neutron stellar objects were suitable candidates to explain the unusual characteristics of the pulsar signal~\cite{G_1_1968}.

 The connection between supernovas and pulsars was firmly established in 1969, with the discoveries of the Vela~\cite{LVM_1_1968} and Crab Nebula~\cite{RBS_1_1969} pulsars. Hundreds of pulsars were discovered in the 1970s and 1980s using radio astronomy, while more recent developments have identified pulsars whose signals span the electromagnetic spectrum~\cite{G_1_1996}. To date, more than two and a half thousand pulsars have been discovered~\cite{M_1_2017}, in many configurations, such as binary pulsar systems~\cite{HT_1_1975}, main-sequence binary-pulsar systems~\cite{JMLBKQA_1_1992}, globular clusters~\cite{LBMKBC_1_1987} with orbiting exo-planets~\cite{WF_1_1992}, and displaying a wide variety of unusual periodic, signals~\cite{M_etal_1_2006, PTH_1_1995}. The relatively recent GW170817 neutron star merger event, detected through gravitational wave signatures by LIGO/Virgo~\cite{A_etal_1_2017} along with the accompanying gamma-ray burst~\cite{A_etal_2_2017}, generated additional observational data.

 While the neutron star radius cannot be measured directly, the mass of neutron stars in binary systems can be inferred from observation together with the application of gravitational theory. With constraints on the mass of a star, observational data allow for indirect inference of the radius, for instance through Doppler shift~\cite{G_1_1996}. Observation-based constraints consistently place the estimated radius of a neutron star in the range of 10--15 km. Using accreting and bursting sources, the radius of the canonical-mass neutron star was determined to be within a range of 10.4 to 12.9 km~\cite{SLB_1_2013}, while analysis from the LIGO/Virgo observations determined the radius to be between 11.1 and 13.4 km~\cite{AGKV_1_2018}. Upper limits on the neutron star radii, as determined from iron emission lines, were placed between 14.5 and 16.5 km~\cite{C_etal_1_2008}.

 The mass range deduced from observed neutron stars is around 1--2 $M_{\odot}$. To date, the smallest mass neutron star has been determined to be $\approx$ 1.17 $M_{\odot}$~\cite{M_etal_1_2015}, while the most massive observed neutron star is $\approx$ 2.14 $M_{\odot}$~\cite{CF_1_2020}. We recall that the Chandrasekhar mass limit of white dwarf stars is 1.4 $M_{\odot}$. If this mass is exceeded, electron degeneracy would no longer be able to support a white dwarf star from gravitational collapse. Observational constraints on neutron star masses yield values clustered around 1.4 $M_{\odot}$~\cite{Z_etal_1_2011} and for this reason this value is taken as the neutron star canonical mass. This also led to the idea that white dwarf collapse may be an additional  mechanism for the formation of neutron stars~\cite{TSYL_1_2013, SQB_1_2015, RFBSCK_1_2019}.

Neutron star models are generally in good agreement with observational constraints for the radius. For instance, the radius of the canonical-mass neutron star predicted  from the set of EoS applied in Ref.~\cite{LS_1_2014} is predicted to be in the range 10.45--12.66 km. From a variety of techniques, based on experimentally determined quantities correlated to symmetry energy parameters, the radius is determined to be between 10.7 and 13.1 km~\cite{T_etal_1_2012, LL_1_2013, LS_1_2014}, while using a range of theoretical models a limit of 9.7 to 13.9 km is obtained~\cite{SG_1_2012, HLPS_2_2013, LS_1_2014}. Recent surveys of neutron star physics and theoretical approaches include Refs.~\cite{OHKT_1_2017, BV_1_2020, V_1_2018}.

On theoretical grounds, the largest mass was predicted to be 3.2 $M_{\odot}$~\cite{RR_1_1974}, based on only three assumptions: (1) General Relativity is the appropriate theory to describe these massive stars;  (2) the EoS is constrained by Le Chatelier's principle ($\partial P$/$\partial \epsilon \geq$ 0);  (3) the causality condition, which constrains the speed of sound in dense matter to remain below the speed of light. While such massive neutron stars may be theoretically possible, none has been observed.

Over the past few years, interest in compact stars has increased considerably as we have entered the ``multi-messenger era'' of astrophysical observations. The GW170817 neutron star merger event has yielded new and independent constraints on the radius of the canonical mass neutron star~\cite{A_etal_1_2017}. Astronomy with gravitational waves provides additional opportunities to explore these exotic systems.

The remarkable point is that the EoS of neutron-rich matter essentially shapes the properties of neutron stars. The mass--radius relationship of neutron stars is uniquely determined by the star's pressure and energy density profile and thus reliable observational constraints can shed light on the EoS.

The structure of a neutron star probes a very large range of densities, from the density of iron up to several times the nuclear matter saturation density, and thus no single theory of hadrons can be considered reliable if extended to those regions. On the other hand, contemporary {\it ab initio} theories of nuclear and neutron matter at normal densities can be taken as the baseline for any extension or extrapolation method, which will unavoidably involve some degree of phenomenology. We recall that the radius of a 1.4 $M_{\odot}$ is sensitive to the pressure at normal densities, see Ref.~\cite{SM19} and references therein, and thus it can pose constraints on microscopic theories of the EoS at those densities where such theories are applicable.

In summary, neutron stars offer a unique view into the properties of dense matter, as discussed in the following sections.

\subsubsection{Polytropic continuation}
\label{sec_poly}

Chiral predictions have a limited domain of validity, which, in the previous section, we estimated to be about twice the saturation density. The densities within neutron stars can reach five to six times saturation density. Additionally, appearance of strange hadrons and/or phase transitions becomes likely at these extreme densities, and therefore an appropriate method for extrapolating the EoS to these densities must be employed.
 To accomplish this, we express the high density pressure through polytropes~\cite{Rea09}:
\begin{equation}
\label{prs_poly}
P(\rho) =  \rho^2 \frac{\partial e_T(\rho)}{\partial \rho} = \alpha \rho^{\Gamma} \; ,
\end{equation}
where $\alpha$ is chosen such as to ensure continuity at the matching density.  A comment is in place: while continuity of the pressure is of course preserved, additional considerations are necessary to ensure continuity of the derivative. The latter would be essential to implement thermodynamic consistency of the piecewise EoS, which is beyond our present scope. Note, further, that the presence of discontinuities in the polytropic index is not unusual for the purpose of describing the global features of the star~\cite{Rea09}.
Following Ref.~\cite{SM19},  we match piecewise polytropes to the {\it ab initio} predictions as explained next.

\begin{figure*}[!t]
\centering
\hspace*{-1cm}
\includegraphics[width=7.5cm]{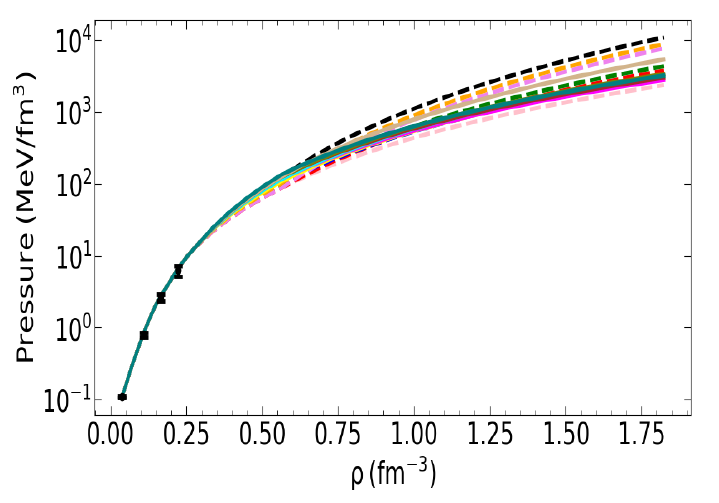}
\caption{Pressure in $\beta$-stable matter as a function of density.  The figure shows the spreading of the pressure values due to the matching of polytropes as explained in the text. Each curve is made up of the microscpic part (common to all cases) and twp polytropes with different adiabatic index, joined by smooth interpolation. The microscopic prediction are obtained at N$^3$LO and cutoff equal to 450 MeV. }
\label{pgam}
\end{figure*}

 We empirically estimate a maximum density for the microscopic EoS based on the following arguments. We translate some trial maximum density into the corresponding Fermi momentum in pure NM. Clearly,
 for the chiral expansion to make sense, the expansion parameter $p/ \Lambda$ must be less than 1, where $p$ is a characteristic momentum of the system under consideration, which we take to be the average momentum in a free neutron gas at that density. For $\rho$ = 0.277 fm$^{-3}$,
we obtain a value of 68\% for $p/ \Lambda$, which is actually a pessimistic estimate, since the average momentum in $\beta$-stable matter is smaller than in pure NM. 
 Having chosen the matching density, $\rho_1$, we join the pressure predictions with polytropes of different adiabatic index, ranging from 1.5 to 4.5. This range is chosen following guidelines from the literature, in particular Ref.~\cite{Rea09}, where constraints on phenomenologically parameterized neutron-star equations of state are investigated. To simulate a (likely) scenario where the pressure displays different slopes in different density regimes, we define a second matching density, $\rho_2$, approximately equal to 2$\rho_1$, at which point a set of polytropes covering the same range of $\Gamma$ is attached to each of the previous polytropes.  Figure~\ref{pgam} shows a subselection of EoS obtained in this way. 
It is important to emphasize that high-density EoS extrapolations are not meant to be a replacement for microscopic theoretical predictions~\cite{SM19} which, at this time, are not feasible at super-high densities. Instead, the spreading of the high-density pressure values from the piecewise variation of the polytrope index allows to probe the sensitivity of lower-density predictions to the much larger uncertainty at high density. 
To complete the EoS on the low-density side, we attach a crustal EoS~\cite{HW65, NV73}.

The extension of the EoS through polytropes is a quite general method. However, other options exist, such as the parametrizations used in Ref.~\cite{Kan+21}, or the iterative procedure to obtain the high-density EoS utilized in Ref.~\cite{Tew18}.

\subsubsection{The TOV equations}
\label{sec_tov}
With the EoS available over a full range of densities, we move to the mass--radius relation in a neutron star. In this section we will briefly review the relativistic equations for hydrostatic equilibrium, the TOV equations~\cite{Tol39, OV39} and how the mass--radius relationship emerges from them for a given input EoS. We will not go beyond the TOV equations to address, for instance, rotating stars, as our main purpose is to highlight the chief role of the EoS in the structure of neutron stars, which can be well demonstrated through static, cold stars.

The TOV equation describes a spherically symmetric inertial massive object composed of a perfect fluid in hydrostatic equilibrium. The equation relates the pressure within the star to the mass-energy density as functions of the radial distance from the star's center:
\begin{equation}
\label{TOV}
\frac{d P(r)}{d r} = - \frac{G}{c^{2}} \ \frac{(P(r) + \epsilon (r) ) \ (M(r) + 4 \pi r^3 \frac{P(r)}{c^{2}})}{r (r - \frac{2 G M(r)}{c^{2}})} \; .
\end{equation}
A spherical shell of material is related to the energy density at a distance $r$  from the star's center by:
\begin{equation}
\label{dMR}
\frac{d M(r)}{dr} = \frac{4 \pi}{c^{2}} r^{2} \epsilon(r) \; .
\end{equation}
The star's gravitational mass ($M$) is determined from the radius ($R$) and the mass-energy density ($\epsilon(r)$):
\begin{equation}
\label{MR}
M(R) = \int^{R}_{0} \frac{\epsilon(r)}{c^{2}} \ d^{3} r \; .
\end{equation}
Since the pressure and energy-density are functions of density, for a fixed central density the mass--radius of the star can be determined by Equations~(\ref{dMR}) and~(\ref{TOV}). Equation~(\ref{TOV}) can be integrated numerically by summing over shells of fixed width at incremented distance from the star's center so as to evaluate the total pressure as a function of radial distance. Equation~(\ref{dMR}) can be integrated in the same fashion, simultaneously, to determine the mass contained within each spherical shell. To accomplish this, we employed the fourth-order Runge--Kutta method.
The radial distance at which the pressure effectively vanishes corresponds to the star's radius. Then, Equation~(\ref{MR}) provides the total mass enclosed within such radius.

The $M(R)$ relation obtained from selected EoS extended with piecewise polytropes as explained above is shown in Fig.~\ref{MR}. We apply the constraint of causality and we only consider polytropes, which can support a maximum mass of at least 2.01 $M_{\odot}$, to be consistent with the lower limit of the (2.08 $\pm$ 0.07) $M_{\odot}$ observation reported in Ref.~\cite{Mil21} for the J0740 + 6620 pulsar along with a radius estimate of (\mbox{12.35 $\pm$ 0.75) km}. Of the EoS that are consistent with the maximum mass constraints, some are causal up to and beyond the maximum mass, others are displayed up to the point where causality is violated. Of course, we must keep in mind that, at and near the central densities corresponding to the heavier stars, the EoS are polytropic and not the result of microscopic theory. What we do observe is that the best combination of polytropes, which satisfies both the maximum mass constraint and causality up and beyond the maximum mass, is given by a first polytrope with adiabatic index equal to about 3.0 - 3.2, joined to a second polytrope with lower adiabatic index, approximately 2.7 - 2.8. This scenario is consistent with a softening of the EoS at the highest densities.

\begin{figure*}[!t]
\centering
\hspace*{-1cm}
\includegraphics[width=9.5cm]{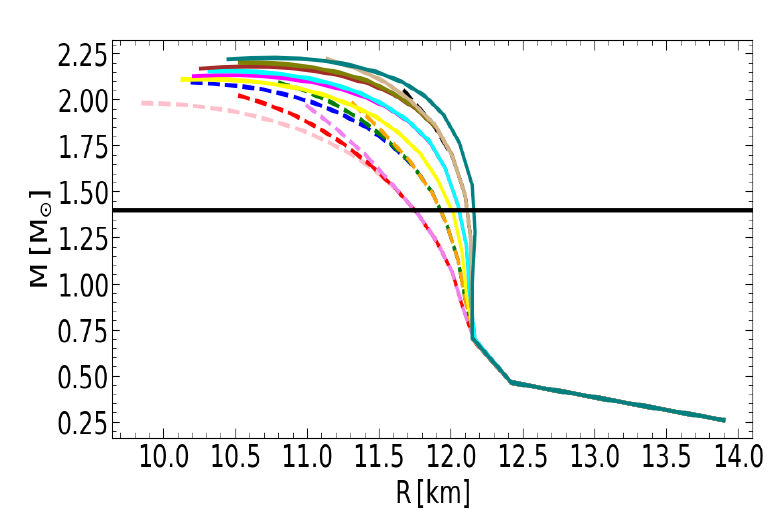}
%\vspace*{-2cm}
\caption{Mass-radius relation for a variety of EoS from Fig.~\ref{pgam}. Each color/pattern corresponds to the same color/pattern in Fig.~\ref{pgam}. Some curves are causal up and beyond the maximum mass, others are cut at the point where causality is violated. The horizontal line marks the 1.4 $M_{\odot}$ mass. }
\label{MR}
\end{figure*}

\subsubsection{The radius of the average-mass neutron star}
\label{sec_rad}

A fully microscopic EoS up to central densities of the most massive stars---potentially involving non-nucleonic degrees of freedom and phase transitions---is not within reach. Nevertheless, neutron stars are powerful natural laboratories for constraining theories of the EoS. One must be mindful about the theory's limitations and the best ways to extract useful information from the observational constraints. 
Constraints on the radius of a medium-mass neutron star, $R_{1.4}$, are becoming more stringent, with the current uncertainty reported at about 2 km. Furthermore, $R_{1.4}$ is known to be sensitive to the pressure in neutron-rich matter near normal densities, accessible to modern effective field theories of nuclear forces.  For these reasons, we focused on predicting $R_{1.4}$ with proper uncertainty quantification.
Our neutron star calculations employing the EoS described in Sec.~\ref{sec_snm} and \ref{sec_nm} are presented in detail in Ref.~\cite{SM22}.

Using the tools reviewed above, 
our estimate of the radius for the canonical-mass neutron star can be stated as:
\begin{equation}
R_{1.4} = (11.96 \pm 0.80) \; \mbox{km} \; ,
\label{R1.4}
\end{equation}
in excellent agreement with the LIGO/Virgo range of 11.1 to 13.4 km~\cite{Ann18}.

Based on our analysis in Section~\ref{sec_ns_skin}, we are confident that
the estimate given in Equation~(\ref{R1.4}), approximately (12 $\pm$ 1) km, is characteristic of EFT predictions based on high-quality 2NF and properly calibrated (leading and subleading) 3NFs. The range currently cited for chiral EFT-based predictions of $R_{1.4}$ is between 10 km and 14 km, accounting for additional theoretical uncertainties.  In fact, it is interesting to notice that the extensive analysis from Ref.~\cite{LH19}, where 300,000 possible EoS were generated, provides a range for $R_{1.4}$ between 10.0 and 12.7 km, with 12.0 km being the most probable value. We recall that the outcome of PREX II gives 13.33 km as the lower limit for the radius, which is problematic to reconcile with a multitude of microscopic predictions~\cite{Ree21}.

Gravitational wave astronomy offers new exciting opportunities for nuclear astrophysics.  Even though chiral EFT cannot reach out to the extreme-density and yet unknown regimes at the core of these remarkable stars,
continuously improved {\it ab initio} calculations of the nuclear EoS are an essential foundation for interpreting current and future observations in terms of microscopic nuclear forces.

\section{Conclusions}
\label{sec_concl}

The past 20 years have seen great progress in our understanding of nuclear forces
in terms of low-energy QCD. Key to this development was the realization that
low-energy QCD is equivalent to an effective field theory (EFT) which              
has become known as chiral perturbation theory (ChPT) or chiral EFT.
In fact, it can be argued that the chiral EFT approach to nuclear forces has considerable advantages over any earlier attempts, because 
  chiral EFT 
\begin{itemize}
\item
is rooted in low-energy QCD;
\item
is, in principal, model-independent;
\item
occurs with an organizational scheme (`power counting') that allows to quantify the uncertainty of  predictions;
\item
generates two- and many-body forces on an equal footing .
\end{itemize}

But the strength of chiral EFT extends beyond a purely formal level.
{\it Ab inito} calculations applying chiral nuclear forces have been, in part, very successful.
One example is the
explanation of nuclear matter saturation. Though it was speculated early on
that the addition of 3NFs might solve the problem, nonrelativistic calculations with phenomenological
3NFs failed to do so~\cite{Day83,CPW83,APR98}.
In contrast, the chiral 3NF at only leading order (NNLO) has the ability to solve that 
problem~\cite{Sam12,Cor14,Sam15,MS16,Sam18,Heb11}.
The $N$-$d$ $A_y$ puzzle, which could never be resolved with phenomenological
3NFs~\cite{Kie10},
is another example for the success of
EFT-based 3NFs~\cite{Gir19}.

Nevertheless problems remain. Just to mention one important issue:
 The simultaneous {\it ab initio} description of
the bulk properties of medium-mass nuclei und nuclear matter saturation
remains an outstanding challenge.

We have also reviewed the equations of state, for both SNM and pure NM, based upon chiral forces. We surveyed a large cross section of representative literature addressing the nuclear symmetry energy, with the objective to identify and discuss current gaps or problems, and provide recommendations for future research. There has been enormous progress in nuclear theory
since the days of one-boson-exchange or phenomenological $NN$
potentials and phenomenological 3NFs, selected with no clear scheme or guidance. 
Predictions from state-of-the-art nuclear theory favor a softer density dependence of the symmetry energy -- on the low-to-medium end of the spectrum found in the literature -- and, naturally, smaller neutron skins and radii of canonical-mass neutron stars.
To advance our understanding of these intriguing systems, it is important to build on that progress.

Last but not least, it needs to be pointed out that
there are still problems at the foundation of the approach:
 Since EFT is a {\it field theory},
the standards to which it must measure up are higher
than for a model. A sound EFT must be renormalizable and 
allow for a proper power counting (order-by-order arrangement).
The presently used chiral nuclear potentials are based on naive dimensional analysis 
(`Weinberg counting') and apply a
cutoff regularization scheme.
In that scheme, one wishes to see cutoff independence of the results.
Such independence is seen to a good degree below the 
breakdown scale~\cite{Mar13,Cor14,Cor13},
but to which extent that is satisfactory is controversial.
The problem is due to the nonperturbative
resummation necessary for typical nuclear physics problems (bound states).
However, there is hope that from the current 
discussion~\cite{HKK19,NTK05,Bir06,LY12,Lon16,Val11,Val11a,Val16,Val17,EGM17,Kon17,Epe18}
constructive solutions may emerge~\cite{Kol20,Val19,EO21}.

In conclusion, considering both formal aspects and evidence of successful applications, 
one may say that chiral EFT has substantially advanced
 the field of theoretical nuclear physics~\cite{MS20}.
 But, at the same time, the community needs to be aware of the
 open challenges and 
 work together building on the progress of the past 20 years.

	%end of the core of the manuscript

	\section*{Acknowledgements}
         
	It is a great pleasure to thank D.~R.~Entem  for his continuous collaboration on topics reviewed in this article.
	This work was supported in part by the US Department of Energy under Grant No.\ DE-FG02-03ER41270. 
%	\section*{Author's contributions \textit{(optional section)}}
%	Detailing here the contributions of the authors of the review.

\bibliography{bibRM}
		
\end{document}